\documentclass[aps,prd,amsmath,floats,onecolumn,floatfix,nofootinbib,showpacs]{revtex4}
\usepackage{amssymb,graphicx} 
\usepackage{bm}
\usepackage{mathrsfs}
\usepackage[usenames]{color}
\usepackage[normalem]{ulem}

\newcommand{\Complex}{\mathbb{C}}

\newcommand{\ScriPlus}{\mathscr{I}^+}

\newcommand{\hz}{\mbox{\em \r{h}\hspace{0.3mm}}} 
\newcommand{\Gammaz}{\Gamma\hspace{-0.25cm}{}^{\mbox{\r{~}}}{}\hspace{-0.12cm}}

\begin{document}
\title{Tetrad formalism for numerical relativity on conformally
  compactified constant mean curvature hypersurfaces}

\author{James M. Bardeen$^1$, Olivier Sarbach$^2$, Luisa T. Buchman$^3$}
\affiliation{${}^1$Physics Department, University of Washington,
  Seattle, Washington 98195 USA}
\affiliation{$^2$Instituto de F\'{\i}sica y Matem\'aticas,
Universidad Michoacana de San Nicol\'as de Hidalgo,
Edificio C-3, Ciudad Universitaria, 58040 Morelia, Michoac\'an, M\'exico}
\affiliation{${}^3$Theoretical Astrophysics, California Institute of
  Technology, Pasadena, California 91125 USA}
\date{\today}

\begin{abstract}

  We present a new evolution system for Einstein's field equations
  which is based on tetrad fields and conformally compactified
  hyperboloidal spatial hypersurfaces which reach future null
  infinity. The boost freedom in the choice of the tetrad is fixed by
  requiring that its timelike leg be orthogonal to the foliation,
  which consists of constant mean curvature slices. The rotational
  freedom in the tetrad is fixed by the 3D Nester gauge. With these
  conditions, the field equations reduce naturally to a first-order
  constrained symmetric hyperbolic evolution system which is coupled
  to elliptic equations for the gauge variables. The conformally
  rescaled equations are given explicitly, and their regularity at
  future null infinity is discussed.  Our formulation is potentially
  useful for high accuracy numerical modeling of gravitational
  radiation emitted by inspiraling and merging black hole binaries and
  other highly relativistic isolated systems.
\end{abstract}


\maketitle

\section{Introduction}
\label{Sec-Intro}

The calculation of gravitational waveforms generated by the inspiral
and coalescence of binary black holes is now a routine matter for
numerical relativists (see Refs.~\cite{fP09,jCjBbKjvM10} for recent
reviews). One of the most pressing goals for those working in the
field today is to compute these waveforms as accurately as possible,
without losing computational efficiency. Optimally, this goal is
accomplished by including future null infinity ($\ScriPlus$) in the
computational grid because then the Bondi news
function~\cite{hBmBaM62,rS62,jS89}, which contains all the gravitational
wave information, can be read off without the need for
extrapolation. One way to include $\ScriPlus$ in the numerical grid is
Cauchy characteristic matching (see Ref.~\cite{jW09} for a review), which
joins a standard $3+1$ Cauchy code in an interior region with a
characteristic code in an outer region. However, arranging stable data
transfer in both directions at the interface between the two regions
is difficult in general; hence, a modified approach called Cauchy
characteristic extraction (CCE)~\cite{jW09,cRnBdPbS10,mBbSjWyZ10}, in
which data is extracted from the interior of a standalone Cauchy
development and extrapolated to $\ScriPlus$ using a characteristic
code, is now often used to extract binary black hole inspiral and
merger waveforms at $\ScriPlus$.  While CCE is certainly a significant
improvement over wave extraction at the finite boundary of a
conventional Cauchy code, potential sources of error remain. Initial
conditions for the characteristic code cannot easily be made
compatible with the Cauchy time development.  Also, either errors in
boundary conditions for the Cauchy code can affect the CCE or there is
insufficient time for all the junk radiation present in the initial
data of the Cauchy code to escape. Furthermore, the extraction imposes
a gauge on the characteristic code in which the expansion of
$\ScriPlus$ generally does not vanish, making it somewhat of a challenge
to obtain the Bondi news accurately~\cite{mBbSjWyZ10}.

An approach which, we argue, has greater potential for high accuracy
waveform calculations is to conformally compactify the spatial
geometry on hyperboloidal spacelike hypersurfaces, as originally
advocated by Friedrich~\cite{hF83}, but with elliptic equations
determining the gauge degrees of freedom.  Hyperboloidal hypersurfaces
behave like conventional time slices near the compact object sources,
but smoothly transition to asymptotically approach null slices in the
physical spacetime at $\ScriPlus$. For a wide class of problems of
physical interest, the compactified conformal geometry is regular, and
the entire space section can be represented on a finite resolution and
relatively small coordinate grid, enabling (we expect) efficient
simulations and accurate waveform calculations.  We can choose
boundary conditions on the elliptic gauge equations at $\ScriPlus$
which make the expansion of $\ScriPlus$ vanish, and thereby simplify
reading off the asymptotic gravitational wave amplitudes.  No outer
boundary conditions are required on the dynamical variables, since
$\ScriPlus$ is an ingoing null hypersurface in the conformally
compactified geometry.

Friedrich~\cite{hF83} derived a first-order symmetric fully hyperbolic
system evolved on compactified hyperboloidal hypersurfaces, with
tetrad components of the Weyl tensor as the fundamental dynamical
variables and with the Bianchi identities providing the evolution
equations for these variables.  The system is manifestly regular at
$\ScriPlus$, thanks to the conformal invariance of the Weyl tensor.
We refer the reader to the review article \cite{jF04} by Frauendiener
for further details, including references to numerical work based on
this formulation. The system is very useful for obtaining analytic
results on existence and uniqueness of radiative spacetimes 
(e.g.~\cite{hF86b}).

Metric-based proposals for solving the Einstein equations on
conformally compactified hyperboloidal hypersurfaces include the
completely hyperbolic generalized harmonic scheme of
Zengino\u{g}lu~\cite{aZ08a} and the mixed hyperbolic-elliptic
formulation of Moncrief and Rinne~\cite{vMoR09}.  Both of these are
second order in spatial derivatives.  Rinne~\cite{oR10a} has
implemented a simplified version of the Moncrief-Rinne formalism in an
axisymmetric finite difference code, and demonstrated long-term stable
dynamical evolution of the Einstein equations. Zengino\u{g}lu and
Tiglio~\cite{aZmT09b} and Zengino\u{g}lu and Kidder~\cite{aZlK10b}
have successfully evolved wave equations for test fields on
compactified hyperboloidal hypersurfaces on fixed black hole
backgrounds.  Conformally flat hyperboloidal Bowen-York binary black
hole initial data have been constructed in~\cite{lBhPjB09}.

The tetrad framework we present in this paper uses the dyadic notation
of the Einstein-Bianchi system of Estabrook and
Wahlquist~\cite{fEhW64}, modified to allow for conformal rescaling.
Instead of basing the dynamical system on the Weyl tensor and the
Bianchi identities, our fundamental variables are the 24 tetrad
connection coefficients (i.e., Ricci rotation coefficients) evolved
using the Einstein equations.  In the context of certain dynamical
gauge conditions, this system can be put in a first-order symmetric
hyperbolic form, the ``WEBB'' equations~\cite{lBjB03,ERW97}. The
mathematical structure of these equations is analyzed in
Refs.~\cite{fE05,fE06}, and 1D numerical tests are presented in
Ref.~\cite{lBjB05b}.  A more general discussion of symmetric
hyperbolic systems for the Einstein equations and their suitability
for numerical relativity is given by Friedrich in~\cite{hF96}.  Purely
hyperbolic numerical evolution in general relativity, particularly in
a first-order scheme, needs to accurately preserve a fairly large
number of constraints.

Our approach here, like that of Moncrief and Rinne~\cite{vMoR09}, is a mixed
hyperbolic-elliptic system based on a special class of hyperboloidal hypersurfaces called constant mean curvature (CMC) hypersurfaces, for which the trace of
the extrinsic curvature is uniform in both space and time.  Some of
the elliptic equations are singular at $\ScriPlus$, where the
conformal factor vanishes, which forces the solutions to have a highly
constrained asymptotic behavior.  This is a feature, not a
bug, since it enables a detailed analysis of the
asymptotic regularity conditions which must be imposed in order to
obtain non-singular evolution of the conformal geometry. We differ
from Moncrief and Rinne in that our evolution equations are first
order in space as well as time, and include connection coefficients of
the conformal spatial geometry in our fundamental set of dynamical
variables.  Also, instead of the coordinate metric and coordinate
components of tensors, our approach is based on orthonormal tetrad
basis vectors. The tetrad components of all tensors are coordinate
scalars, and the metric for raising and lowering tetrad indices is
trivial. Consequently, the resulting equations are simpler in form
and in interpretation than those of Moncrief and Rinne.

The extra gauge degrees of freedom in our tetrad formalism
corresponding to local Lorentz boosts and rotations of the tetrad
frames are fixed in a natural way which simplifies the equations and
reduces the number of variables compared to a general tetrad
formalism. The time-like vector of the tetrad is chosen to be the
unit normal to the CMC hypersurfaces, which constrains the triad of
spatial vectors to be tangent to the CMC hypersurfaces. The rotational
degrees of freedom in the conformal spatial triad vectors are fixed by
imposing the 3D Nester gauge~\cite{jN89}. In a certain global sense,
the Nester gauge makes the triad vectors at different points in the
hypersurface as nearly related by parallel transport as possible. The
Nester gauge leads naturally~\cite{jN91} to a choice of conformal factor, which satisfies an elliptic equation derived from the Hamiltonian constraint and is enforced by  elliptic equations for potentials that determine the conformal angular velocity of the triad frames and the trace of the conformal extrinsic curvature. Of the 24 tetrad
connection coefficients, only 10 are dynamical and form a Maxwell-like
symmetric hyperbolic system. The remaining evolution equations are
the nine linear advection equations for the coordinate components of
the three spatial triad vectors.

As in Ref.~\cite{vMoR09}, we put considerable effort into analyzing
the properties of the equations and their solutions in the
neighborhood of $\ScriPlus$.  The evolution equations for the
dynamical variables contain terms which have a conformal factor in the
denominator and therefore are potentially singular at $\ScriPlus$.
Certain well-known regularity conditions must be imposed to make these
terms finite, such as the ``zero-shear'' condition on the null
generators of $\ScriPlus$ (see Ref.~\cite{lApC94}).  Once imposed in
the initial conditions, they are preserved by the evolution equations.
An additional and stronger regularity condition is also usually
assumed, that the conformally invariant Weyl tensor vanish at
$\ScriPlus$.  We refer to this as the ``Penrose regularity
condition'', and it is also automatically preserved by the evolution
equations.  The review of Frauendiener~\cite{jF04} puts the various
regularity and smoothness assumptions at $\ScriPlus$ in context.

The question of how much smoothness in the conformal geometry at
$\ScriPlus$ is, in some sense, physically generic has been the matter
of considerable debate in the literature. The original analyses of
Bondi et al.~\cite{hBmBaM62} and Sachs~\cite{rS62} assumed analyticity,
but the more systematic conformal analysis of Penrose~\cite{rP65} 
reduced this to $C^3$ smoothness.  Later, Chru\'sciel 
et al.~\cite{pCmMdS95} showed that ``polyhomogeneous'' terms in 
the expansion of the angular part of the metric in Bondi-Sachs 
coordinates, terms of the form $x^n(\log x)^m$, where $x \equiv 1/r$, 
are consistent with the Einstein equations and the ability to define a 
physically suitable Bondi energy.  The coefficient of the ``leading log'' 
term is a constant during evolution, so if these polyhomogeneous terms are 
ever present they were always present back to past null infinity, where 
they are associated with incoming radiation.  

We conclude from this that in the context of typical
astrophysical problems, where one can assume there is no or negligible
incoming radiation at past null infinity, it is perfectly appropriate to 
exclude polyhomogeneous terms of the type
contemplated in Ref.~\cite{pCmMdS95} from ever being present.  
Similarly, violations of the regularity conditions
described above, including the Penrose regularity condition, can only 
be present if they were present at past null infinity, and cannot be 
generated by any dynamics of the physical system.
However, as was shown in Ref.~\cite{lApC94} and
as we derive in Appendix~\ref{App-AsymptoticSolns}, polyhomogeneous
terms in both the conformal and physical extrinsic curvature and the
conformal factor are generically present when the spacetime is 
foliated on CMC hypersurfaces.
In fact, we show that they {\em must} be
present whenever outgoing radiation is present at $\ScriPlus$, and are 
not present otherwise.  These terms
arise from the special character of the momentum constraint equation
and the elliptic equation for the conformal factor on a CMC slice as
it becomes asymptotically null at $\ScriPlus$. In a more flexible 
hyperboloidal slicing, where the trace of the extrinsic
curvature only approaches a constant at $\ScriPlus$, the polyhomogeous
terms need not be present.  In fact, they are clearly not present when
the spacetime is evolved using the regular conformal field equations
of Friedrich~\cite{hF83}, which allows the foliation to evolve as part
of the hyperbolic system.

There are some significant advantages to our mixed hyperbolic-elliptic
system.  The boundary conditions on the gauge variables determined by
the elliptic equations can be adjusted to simplify and control the
geometric properties of and coordinates on $\ScriPlus$.  For instance,
the boundary condition on the elliptic equation for the conformal
lapse, together with the boundary condition on the trace of the
conformal extrinsic curvature, can keep the intrinsic conformal
geometry of the intersection of the CMC hypersurface with $\ScriPlus$
that of a two-sphere of constant area and the time coordinate on the
CMC hypersurfaces equal to the retarded Minkowski time at $\ScriPlus$.
The shift vector controls the evolution of the spatial coordinates.
We propose a few alternative vector elliptic equations for the shift,
all with a boundary condition chosen to keep $\ScriPlus$ at a fixed
coordinate radius.

The final conformally rescaled system is a symmetric hyperbolic
evolution scheme which is coupled to an elliptic system for the
conformal factor and the gauge variables.  Hyperbolic-elliptic
formulations have been discussed in the literature, both from
mathematical and numerical viewpoints. A 
hyperbolic-elliptic formulation of Einstein's equations in spatial
harmonic coordinates on CMC hypersurfaces, very similar to
the Moncrief-Rinne formulation but without the conformal
compactification, has been proven well-posed by Andersson and
Moncrief~\cite{lAvM03}. Note that the Andersson-Moncrief formulation
is not constrained -- elliptic equations are solved only for the lapse
and shift. The Meudon group proposed a fully constrained,
conformally rescaled system~\cite{sBeGpG04} for which both numerical
tests and mathematical analyses have been
performed~\cite{sBeGpG04,iCjIeGjJjN08,iCpChDjJjNeG09}.  Finally,
elliptic equations have been periodically enforced in a constraint
projection scheme \cite{mHlLrOhPmSlK04} for a first-order reduction of
a scalar field equation on a black hole background, and shown not to
be prohibitively expensive. Therefore, we hope that a numerical
implementation of the proposal presented in this paper with efficient
elliptic solvers such as those in the Spectral Einstein Code {\tt
  SpEC}~\cite{SpECwebsite,hPlKmSsT03} will be possible with reasonable
computational effort, especially since we would need relatively small
grid sizes due to conformal compactification and the relatively small
number of wave oscillations out to $\ScriPlus$ on a CMC hypersurface.

Our paper is organized as follows. In Sec.~\ref{Sec:3+1}, we specify
our choice of variables and notation and, based on the
hypersurface-orthogonal gauge, we present the $3+1$ decomposition of
the triad and connection coefficients as originally developed in the
dyadic formalism of Ref.~\cite{fEhW64}. In Sec.~\ref{Sec-BasicEqns},
we discuss the evolution and constraint equations of the WEBB tetrad
formulation when specialized to a hypersurface-orthogonal gauge, and
we decompose both the extrinsic curvature and the dyadic form of the
spatial connection coefficients into symmetric trace-free,
antisymmetric, and trace parts. The antisymmetric and trace parts are
fixed by our gauge conditions which are discussed in
Sec.~\ref{Sec:GaugeConds}. While the antisymmetric part of the
extrinsic curvature is identically zero, its trace is fixed to a
constant by the CMC slicing condition, and the trace and antisymmetric
parts of the dyadic form of the spatial connection coefficients are
determined by the 3D Nester gauge condition (the latter up to a
gradient). The remaining symmetric trace-free parts are governed by a
nice Maxwell-like symmetric hyperbolic evolution system which is
coupled to advection equations for the triad vectors and elliptic
equations for the gauge variables. The conformal decomposition is
performed in Sec.~\ref{Sec:Conformal}. The Nester gauge is covariant
with respect to conformal transformations, and this property motivates
a natural choice for the conformal factor. With this choice, the
antisymmetric part of the dyadic representation of the
conformal connection coefficients is zero, which further simplifies
the equations. Also in Sec.~\ref{Sec:Conformal}, we derive manifestly
regular elliptic equations guaranteeing that these gauge conditions
are preserved throughout evolution, and we discuss possible elliptic
equations for the shift. In Sec.~\ref{Sec:RegularityConds}, we analyze
necessary conditions to impose on the fields at $\ScriPlus$ in order
to have a regular time evolution. Furthermore, we evaluate explicitly
the apparently singular terms at $\ScriPlus$ in the right-hand side of
the evolution equations. Our evolution system, together with its
constraints and elliptic equations for the gauge variables including
boundary conditions, is summarized in Sec.~\ref{Sec:Summary}, and
issues such as comparisons with other approaches and extraction of
gravitational waveforms are discussed in Sec.~\ref{Sec:Discussion}.
More technical aspects relevant to our formulation are given in
Appendices~\ref{App-AsymptoticSolns}, \ref{App-ConstraintPropagation}
and \ref{App-NesterDirac} where we discuss, respectively, asymptotic
expansions of the fields near $\ScriPlus$, the constraint propagation
system, and the relation of the 3D Nester gauge condition to the 3D
Dirac equation.

Throughout this paper, we use the sign convention $(-,+,+,+)$ for the
metric.  We denote by $\dot{\ScriPlus}$ the 2D intersection of the
$\ScriPlus$ null hypersurface with a hyperboloidal CMC spatial
hypersurface, and by the $\doteq$ sign an equality that holds only at
$\ScriPlus$. We adopt the Wald~\cite{Wald84} sign convention for the extrinsic curvature of a spatial hypersurface, so it is positive for expansion toward the future.

\section{$3+1$ Decomposition}
\label{Sec:3+1}

The tetrads in the spinor formulation of Newman and
Penrose~\cite{eNrP62} are aligned along null congruences and are
therefore conducive to understanding radiative properties of a
spacetime. In contrast, in the tetrad formulation presented by
Estabrook and Wahlquist~\cite{fEhW64}, the tetrad is oriented with
respect to a preferred timelike congruence. This can be useful if a
physically preferred timelike direction, such as a fluid 4-velocity,
exists.  However, this is not the case in vacuum.  Our choice is to
orient the timelike leg of the tetrad orthogonal to the CMC
hypersurfaces of constant coordinate time $\Sigma_t$. It follows then
that the spacelike legs of the tetrad are tangent to $\Sigma_t$. This
is called a {\em hypersurface-orthogonal} gauge condition, and we
assume this gauge condition throughout the paper.

Our notation follows that of Ref.~\cite{lBjB03}.  Spacetime indices
$(0-3)$ are Greek and spatial indices $(1-3)$ are Latin.  The
beginning of the alphabet is used for tetrad indices, and the middle
of the alphabet is used for coordinate indices.  Spatial tetrad
indices will not be raised and lowered, since the spatial metric in
the tetrad basis is trivial. In Ref.~\cite{lBjB03}, there was a
distinction between the tetrad vectors with components
$\lambda_{\alpha}^{\;\mu}$ and the spatial triad vectors ${\bf B}_a$
defined as projections of the spatial tetrad vectors tangent to
$\Sigma_t$, with only spatial components $B_a^{\;k}$.  In a
hypersurface-orthogonal gauge, the spatial tetrad vectors have zero
time components and are identical to the ${\bf B}_a$.  The directional
derivatives along the three spatial legs of the tetrad are
then\footnote{The vector $A_a$ of Ref.~\cite{lBjB03}, which is the
  velocity of the tetrad congruence relative to a hypersurface rest
  frame, is identically zero.}
\begin{equation} 
  D_a = B_a^{\;k} \: \frac{\partial}{\partial x^k}.
\end{equation}
The Lie derivative along the timelike leg of the tetrad is written in
terms of the lapse function ($\alpha$) and the shift vector
($\beta^k$):
\begin{equation} 
  D_0 =  \frac{1}{\alpha} \: \left(\frac{\partial}{\partial t} - \pounds_\beta \right).
\end{equation} 
Let us stress that we regard variables with pure tetrad indices (such
as $K_{ab}$, $N_{ab}$, $a_b$ and $\omega_b$ below) as scalar fields on
$\Sigma_t$. In this sense, the Lie derivative operator $\pounds_\beta$
with respect to the shift acts as the directional derivative on such
variables. In contrast to this, we have $\alpha D_0 B_a^{\;i} =
(\partial_t - \beta^j\partial_j) B_a^{\;i} + D_a\beta^i$ which differs
from the definition in~\cite{lBjB03}, where $\alpha D_0 B_a^{\;i} =
(\partial_t - \beta^j\partial_j) B_a^{\;i}$.

The coordinate basis spacetime metric $g_{\mu \nu}$ and the spatial
metric $h_{ij}$ can be calculated from the tetrad and triad vectors as
\begin{equation} 
  g^{\mu \nu} = \eta^{\alpha \beta} \: \lambda_{\alpha}^{\;\mu} \:
  \lambda_{\beta}^{\;\nu},\qquad
  h^{ij} =B_a^{\;i} \: B_a^{\;j},
\end{equation} 
where $\eta^{\alpha \beta}$ is the Minkowski metric. The connection
coefficients with respect to the Levi-Civita connection $\nabla$ of $g_{\mu\nu}$
in the tetrad basis are calculated from the commutators,
\begin{equation} 
  \Gamma_{\alpha \beta \gamma} = {\bf \lambda}_{\alpha}
  \cdot \nabla_{\gamma} \: {\bf \lambda}_{\beta} = -\Gamma_{\beta \alpha
    \gamma} = \frac{1}{2} \: \left({\bf \lambda}_{\beta} \cdot \left[{\bf \lambda}_{\alpha},
    {\bf \lambda}_{\gamma} \right] \: + \: {\bf \lambda}_{\gamma} \cdot 
    \left[{\bf \lambda}_{\alpha}, {\bf \lambda}_{\beta} \right] \: - \: 
    {\bf \lambda}_{\alpha} \cdot \left[{\bf \lambda}_{\beta}, {\bf \lambda}_{\gamma} \right] 
  \right).
\label{Eq:GammaDefComm}
\end{equation} 

In the dyadic formalism~\cite{fEhW64}, the $24$ independent $\Gamma_{\alpha \beta
  \gamma}$ are decomposed into two $3 \times 3$ matrices
\begin{equation} 
  K_{ab} \equiv \Gamma_{b0a},\qquad
  N_{ab} \equiv \frac{1}{2} \: \varepsilon_{bcd} \: \Gamma_{cda},
\end{equation} 
and two vectors
\begin{equation} 
  a_{b} \equiv \Gamma_{b00}, \qquad
  \omega_{b} \equiv -\frac{1}{2} \: \varepsilon_{bcd} \: \Gamma_{cd0}.
\end{equation} 
In the hypersurface-orthogonal gauge, the $K_{ab}$ are the triad
 components of the symmetric
extrinsic curvature tensor of $\Sigma_t$. 
The sign of $K_{ab}$ is opposite to that of Misner, Thorne, and
Wheeler \cite{MTW73}, in that positive values imply expansion of the
normals, not contraction. The $N_{ab}$ describe the induced connection
coefficients of $\Sigma_t$.  It is often convenient to write the
anti-symmetric part of $N_{ab}$ as the vector
\begin{equation} 
n_b \equiv \frac{1}{2} \: \varepsilon_{bcd} \: N_{cd}.
\end{equation}
The vectors $a_b$ and $\omega_b$ are, respectively, the acceleration
and the angular velocity relative to Fermi-Walker transport of the
tetrad frame. The acceleration vector field is given by the gradient 
of the logarithm of the lapse,
\begin{equation} 
  a_b = D_b(\log\alpha). \label{Eq:HypSurfOrtho}
\end{equation}

\section{Basic Equations}
\label{Sec-BasicEqns}

Here we write the basic Einstein evolution and constraint equations of
the WEBB tetrad formulation (these are derived in Ref.~\cite{lBjB03}
from the Einstein equations and the Riemann constraints, see also 
Appendix~\ref{App-ConstraintPropagation}), specialized
to a hypersurface-orthogonal gauge. Note that $K_{ab}$ is explicitly
symmetric. The resulting evolution equations for $K_{ab}$ and
$N_{ab}$, respectively, are
\begin{eqnarray}
  D_0 K_{ab} - \varepsilon_{acd} D_c N_{db} &=& (D_a + a_a) a_b
  - \varepsilon_{bcd} N_{ac} \, a_d + 2 \, \varepsilon_{cd(a} K_{b)c} \, \omega_d
  + N N_{ab} \nonumber \\
  &+& \frac{1}{2} \varepsilon_{adf} \, \varepsilon_{bce} \left(K_{dc} K_{fe}
    -N_{dc} N_{fe} \right)  - K_{ac} K_{cb} - N_{ca}
  N_{cb} + 8\pi G\left[ \sigma_{ab} - \frac{1}{2}\delta_{ab}(\sigma_{cc} + \rho) \right],
\label{Eq:OrigKabEvoln} \\
  D_0 N_{ab} + \varepsilon_{acd} D_c K_{db} &=& -(D_a + a_a)\omega_b
  + \varepsilon_{bcd} \left(K_{ac} \, a_d +N_{ac} \, \omega_d \right)
  + \varepsilon_{acd} N_{cb} \, \omega_d \nonumber \\
  &-& N K_{ab} + 2 N_{c[a} K_{b]c} +\varepsilon_{adf} \varepsilon_{bce} N_{dc} K_{fe}
   - 8\pi G\varepsilon_{abc} j_c,
\label{Eq:OrigNabEvoln}
\end{eqnarray}
where $N$ is the trace of $N_{ab}$, $K$ is the trace of $K_{ab}$, and
$a_b = D_b(\log\alpha)$. Here, we have also
defined the energy density $\rho = \tau_{00}$, the energy flux $j_b =
-\tau_{0b}$, and the stress tensor $\sigma_{ab} = \tau_{ab}$, where
$\tau_{\mu\nu}$ denotes the stress-energy tensor.

The Hamiltonian and momentum constraint equations are, respectively,
\begin{eqnarray}
  2 D_a n_a &=& N_{ab} N_{ab} + \frac{1} {2}\left( K_{ab} K_{ab} 
    - N_{ab} N_{ba} - K^2 - N^2 \right) + 8\pi G\rho,
\label{Eq:OrigHamConstraint} \\
  D_b K_{ab} - D_a K &=& - \varepsilon_{abc} \, K_{bd} N_{dc} 
  + 2 K_{ac}\, n_c - 8\pi G j_a. 
\label{Eq:OrigMomConstraint} 
\end{eqnarray}
There is also a constraint for $N_{ab}$ analogous to the momentum
constraint equation, which is obtained from the Riemann identities:
\begin{equation}
  D_b N_{ab} - D_a N = - \varepsilon_{abc} \, N_{bd} N_{dc}. 
 \label{Eq:OrigNabConstraint} 
\end{equation}

Evolution and constraint equations for the spatial triad vectors
$\bf{B}_a$ are derived from the commutators of the orthonormal basis
vectors and are
\begin{eqnarray}
  D_0 B_a^{\;k} &=& -\varepsilon_{abc} \,  \omega_b \, B_c^{\;k} - K_{ac} B_c^{\;k},\label
{Eq:OrigBakEvoln}\\
  \varepsilon _{abc} \,D_b \,B_c^{\;k} &=& N_{ba} \, B_b^{\;k} - N B_a^{\;k}. \label
{Eq:OrigBakConstraint}
\end{eqnarray}

In the following, it will be useful to separate out the symmetric,
trace-free parts of $K_{ab}$ and $N_{ab}$, denoted by
$\hat{K}_{ab}$ and $\hat{N}_{ab}$, so that
\[ 
K_{ab} = \hat K_{ab} + \frac{\delta _{ab}} {3}\,K,\qquad
N_{ab} = \hat N_{ab} + \varepsilon _{abc} n_c + \frac{\delta _{ab}}{3}\,N.
\]
From Eqs.~(\ref{Eq:OrigKabEvoln}) and~(\ref{Eq:OrigNabEvoln}), we
obtain the following evolution equations for $\hat{K}_{ab}$ and
$\hat{N}_{ab}$:
\begin{eqnarray}
  D_0 \hat K_{ab} - D_c \hat N_{d(a} \varepsilon_{b)cd}
  &=& \left\{ (D_{(a} + a_{(a}) a_{b)} - (D_{(a} - a_{(a}) n_{b)}
  - K\hat K_{ab} + \frac{2N}{3}\hat N_{ab} - 2\hat
  N_{ac} \,\hat N_{cb} \right.\nonumber\\ 
  &+& \left. \varepsilon_{cd(a} \left[\hat N_{b)c} \, (2n_d - a_d) + 2 \hat K_{b)c} \,\omega _d 
\right] + 8\pi G\sigma_{ab} \right\}^{TF}
\label{Eq:EvolnHatKab}\\
  D_0 \hat N_{ab} + D_c \hat K_{d(a} \varepsilon_{b)cd}
  &=& \left\{ - (D_{(a} + a_{(a})\omega _{b)} 
   - \frac{4N}{3} \hat K_{ab} -\frac{K}{3} \hat N_{ab} + 2\hat K_{c(a} \hat N_{b)c}  
   + \varepsilon_{cd(a} \left[{\hat K_{b)c}\, a_d 
   + 2 \hat N_{b)c} \,\omega _d } \right] \right\}^{TF}
  \label{Eq:EvolnHatNab} 
\end{eqnarray}
where $\{ ... \}^{TF}$ denotes the trace-free part of the expression
inside the parenthesis. The evolution equation for the trace of the
extrinsic curvature, from which an elliptic equation for the lapse
will be derived in Sec.~\ref{SubSec:CMC}, is
\begin{equation}
D_0 K = (D_b + a_b - 2 n_b)a_b - \hat{K}_{ab}\hat{K}_{ab} - \frac{1}{3} K^2
 - 4\pi G(\sigma_{aa} + \rho).
\label{Eq:EvolnK}
\end{equation}
We have used the Hamiltonian constraint,
Eq.~(\ref{Eq:OrigHamConstraint}), to eliminate the scalar 3-curvature.
The evolution equations for the trace and the antisymmetric part of
$N_{ab}$, from which we will derive elliptic equations for $\omega_a$
and the time derivative of the conformal factor in
Sec.~\ref{SubSec:Ktilde}, are
\begin{eqnarray}
 D_0 N &=& -(D_a + a_a)\omega_a - \hat{N}_{ab}\hat{K}_{ab} - \frac{1}{3} K N,
 \label{Eq:EvolnNSubstHamCon}\\
  2 \, D_0 \, n_a &=& -\varepsilon_{abc} \left(D_b + a_b - 2 n_b
  \right) \omega_c - \frac{2K}{3}(a_a + n_a) - \varepsilon_{abc} \hat{K}_{bd} \hat{N}_{dc}
  +\hat{K}_{ab} \left( a_b + n_b \right) - 8\pi G j_a.
 \label{Eq:EvolnLittlenSubstMomCon} 
\end{eqnarray}
The second equation has been simplified by applying the momentum
constraint equation~(\ref{Eq:OrigMomConstraint}).

The Hamiltonian and momentum constraint
equations~(\ref{Eq:OrigHamConstraint})
and~(\ref{Eq:OrigMomConstraint}), expressed in terms of the symmetric
trace-free and antisymmetric parts of $K_{ab}$ and $N_{ab}$, are
\begin{eqnarray}
  D_a n_a &=& \frac{1} {4} \left( \hat K_{ab} \hat K_{ab} +\hat N_{ab}
    \hat N_{ab} \right) + \frac{3} {2} \,n_a n_a
  -\frac{\left(K^2+N^2\right)}{6} + 4\pi G \rho, 
\label{Eq:HamConstraint} \\
  D_b \hat K_{ab} &=& \frac{2}{3}D_a K - \varepsilon_{abc} \,\hat K_{bd} \,\hat N_{dc}
  +3 \, \hat K_{ab} \, n_b - 8 \pi G j_a. 
 \label{Eq:MomConstraint} 
\end{eqnarray}
The $N_{ab}$ constraint equation~(\ref{Eq:OrigNabConstraint}) becomes
\begin{equation}
  D_b \hat{N}_{ab} = \frac{2}{3} D_a N - \varepsilon_{abc} \, D_b \, n_c -
  \frac{4}{3}N n_a +2 \hat{N}_{ab} n_b.
  \label{Eq:NabConstraint}
\end{equation}

Finally, the spatial triad variables $B_a^{\;k}$ are evolved via
\begin{equation}
  \label{Eq:EvolnSpatialTriad}
  D_0 B_a^{\;k} = -\varepsilon _{abc} \,\omega _b \, B_c^{\;k} - \hat K_{ac}\, B_c^{\;k} - 
\frac{K} {3} B_a^{\;k},
\end{equation} 
and are constrained by
\begin{equation}
  \label{Eq:ConstraintSpatialTriad}
  \varepsilon _{abc} \,D_b \,B_c^{\;k} = \hat N_{ab}\, B_b^{\;k} +
  \varepsilon_{abc} \,n_b\,B_c^{\;k} - \frac{2N}{3}B_a^{\;k} .
\end{equation}

\section{Gauge Conditions}
\label{Sec:GaugeConds}

In the WEBB tetrad formalism, in addition to specifying the usual
lapse function and shift vector in order to control the evolution of
the coordinates, gauge conditions are required to specify the
evolution of the tetrad vectors.  In our hypersurface-orthogonal
implementation, the acceleration vector is given in terms of the lapse
by Eq.~(\ref{Eq:HypSurfOrtho}).  The remaining freedom is the
evolution of the angular velocity (with respect to Fermi-Walker frame 
dragging) 3-vector $\omega_a$.  In this section, we derive the
equation for the lapse function appropriate to our CMC hypersurface
condition and the elliptic equations which preserve the 3D Nester 
gauge conditions on the spatial triad
during the evolution. We assume for the present that the shift vector
is given; discussion and derivation of the shift equations is deferred
to Sec.~\ref{SubSec:ShiftCond}.

\subsection{CMC slicing}
\label{SubSec:CMC}

In order to fix the lapse and the corresponding foliation of
spacetime, we require that the trace of the extrinsic curvature be a
positive constant in both space and time,
\begin{equation}
K = {\rm const.} > 0,
\end{equation}
which with our sign convention means that the spatial hypersurfaces go to 
future, rather than past, null infinity.  Then
Eq.~(\ref{Eq:EvolnK}) becomes the following linear, elliptic equation
for the lapse:
\begin{equation}
  \left[ -(D_a - 2n_a) D_a + \hat{K}_{ab}\hat{K}_{ab} + \frac{1}{3} K^2
   + 4\pi G(\rho + \sigma_{aa}) \right]\alpha = 0.
\label{Eq:CMCAlpha}
\end{equation}
We notice that the operator $(D_a - 2n_a)D_a = \nabla_a\nabla_a$ is
equal to the covariant Laplacian on the three-geometry, where here and
in the following, $\nabla_a$ denotes the covariant derivative with
respect to the three-geometry. It follows that the operator inside the
square parenthesis is (formally) positive if the strong energy
condition, $\rho + \sigma_{aa} \geq 0$, holds.

\subsection{3D Nester gauge}
\label{SubSec:3DNester}

After imposing hypersurface orthogonality, 
 the remaining freedom in the choice of the tetrad is
described by local rotations of the spacelike legs. Nester \cite{jN89}
proposed a set of natural gauge conditions in order to fix this
freedom up to a global rotation. These conditions are conveniently
formulated using differential forms. Let $\theta^a$, $a = 1,2,3$, be
the co-frame associated with ${\bf B}_a$; that is, $\theta^1$,
$\theta^2$, $\theta^3$ are the basis of one-forms defined by
\begin{displaymath} 
  \theta^a({\bf B}_b) = \delta^a{}_b.
\end{displaymath} 
The Cartan structure equations (see for instance~\cite{Straumann})
imply the following relation between $\theta^a$ and the connection
coefficients $\Gamma_{acd}$:
\begin{displaymath} 
  d\theta^a = \Gamma^a{}_{cd} \theta^c\wedge \theta^d.
\end{displaymath} 
The 3D Nester gauge condition is\footnote{This condition has a natural
  generalization to higher dimensions, see \cite{jN92}.}
\begin{equation} 
  d\Xi + \delta\Pi = 0,
\label{Eq:3DNesterCondition}
\end{equation} 
where the one-form $\Xi$ and the three-form $\Pi$ are defined by
\begin{eqnarray}
  \Xi &\equiv& i_{{\bf B}_a} d\theta^a =
  \Gamma^a{}_{ab}\theta^b, \nonumber\\
  \Pi &\equiv& \frac{1}{2}\delta_{ab}\theta^a\wedge d\theta^b =
  \frac{1}{2}\Gamma_{bcd}\theta^b\wedge\theta^c\wedge\theta^d,\nonumber
\end{eqnarray}
respectively. Here, $\delta$ denotes the co-differential which is
defined as the formal adjoint of the exterior derivative operator $d$
with respect to the inner product $(\nu,\eta)\equiv\int_{\Sigma_t} \nu
\wedge *\eta$ for $p$-forms $\nu$ and $\eta$. In terms of the Hodge
dual $*$, it is explicitly given by $\delta = (-1)^p * d *$ when
acting on a $p$-form. Provided suitable boundary conditions are
specified at the boundary of $\Sigma_t$, the two terms arising in
Eq.~(\ref{Eq:3DNesterCondition}) are mutually orthogonal with respect
to the product $(\cdot,\cdot)$, and the 3D Nester gauge condition is
equivalent to
\begin{equation}
  d\Xi = 0, \qquad d*\Pi = 0.
\end{equation}
In our notation, we have $\Xi = 2n_a\theta^a$ and $*\Pi =
N$. Therefore, assuming that $\Sigma_t$ is connected and that its first
cohomology vanishes, the 3D Nester gauge implies
\begin{equation}
n_a = D_a\Phi,\qquad N = {\rm const}.
\label{Eq:Integrated3DNesterCondition}
\end{equation}
for a function $\Phi$ on $\Sigma_t$. We choose the constant value of
$N = 0$, since this tends to minimize unnecessary twisting of the
triad vectors from one spatial point to another, which is particularly
appropriate in the context of asymptotic flatness. The existence and
uniqueness of a set of orthonormal three-frames globally satisfying
the 3D Nester gauge can be related to the existence and uniqueness of
a nowhere vanishing spinor field satisfying the 3D Dirac equation, see
Ref.~\cite{aDfM89} and Appendix~\ref{App-NesterDirac}.

\subsection{Properties of the evolution system in the CMC and 3D Nester gauge}
\label{SubSec:System}

The gauge conditions we have adopted have the following nice
properties: they fix the gauge variables $K$, $N$ and $n_a$
(the last up to a gradient). As a consequence, the evolution system
reduces to the advection equation~(\ref{Eq:EvolnSpatialTriad}) for the
triad vectors ${\bf B}_a$ and the Maxwell-like equations~(\ref{Eq:EvolnHatKab},~\ref{Eq:EvolnHatNab}) for the symmetric,
traceless parts of $K_{ab}$, $N_{ab}$. This system constitutes a
quasilinear, symmetric hyperbolic evolution system, assuming that the
gauge variables $a_b = D_b(\log\alpha)$, $n_b = D_b\Phi$, and
$\omega_b$ are known. The gauge variables are determined by a set of
coupled elliptic equations obtained by requiring that the gauge
conditions be preserved during the evolution.  The lapse is obtained
by solving the elliptic equation~(\ref{Eq:CMCAlpha}), which is a
consequence of setting the time derivative of $K$ to zero, preserving
CMC slicing.  Next, we note that in the 3D Nester gauge, the
Hamiltonian constraint~(\ref{Eq:HamConstraint}) yields an elliptic
equation for the scalar potential $\Phi$ which might be used to obtain
$n_b$. Finally, the requirement that the time evolution preserve the
3D Nester gauge condition yields an elliptic system of equations for
$2\alpha D_0\Phi$ and $\alpha\omega_b$. This system can
be derived as follows. Using the commutation relation $[\partial_t -
\pounds_\beta,D_a]\Phi = -\alpha(K_{ac} +
\varepsilon_{abc}\omega_b)D_c\Phi$ in
Eq.~(\ref{Eq:EvolnLittlenSubstMomCon}), with $n_a = D_a\Phi$, we
obtain
\begin{equation} 
  D_a(2\alpha D_0\Phi) + \varepsilon_{abc} D_b(\alpha\omega_c) 
  = -\alpha\left[ \varepsilon_{abc}\hat{K}_{bd}\hat{N}_{dc} - \hat{K}_{ab}(3n_b + a_b)
 + \frac{2}{3} K a_a + 8\pi G j_a \right].
\label{Eq:GradD0Phi}
\end{equation}
In addition, with $a_a = D_a (\log\alpha)$ and $N=0$,
Eq.~(\ref{Eq:EvolnNSubstHamCon}) can be rewritten as
\begin{equation}
    D_a(\alpha\omega_a) = -\alpha \hat{N}_{ab}\hat{K}_{ab}.
\label{Eq:DivOmega}
\end{equation}
As shown in Appendix~\ref{App-NesterDirac},
Eqs.~(\ref{Eq:GradD0Phi},~\ref{Eq:DivOmega}) are equivalent to the
inhomogeneous 3D Dirac equation, and they can be cast as a
second-order elliptic system for appropriate potential fields. This
elliptic formulation will be presented explicitly in the next
section, after the conformal decomposition of the fields has been
performed.

Summarizing, in the CMC and 3D Nester gauge conditions, our system
reduces naturally to a symmetric hyperbolic evolution problem for the
fields ${\bf B}_a$, $\hat{K}_{ab}$, $\hat{N}_{ab}$ which is coupled to
an elliptic system which determines the gauge variables $\alpha$,
$\Phi$ and $\omega_a$.  Appropriate conditions on the evolution of the
spatial coordinates, and the corresponding equations for the shift
vector, will be discussed in Sec.~\ref{SubSec:ShiftCond}.  The
well-posedness of the full, coupled hyperbolic-elliptic system depends
on the structure of the elliptic equations for the gauge variables,
and the choice of the shift. A well-posed Cauchy formulation has been
obtained by Andersson and Moncrief~\cite{lAvM03} for a somewhat
similar metric-based system.

\section{The conformally rescaled equations}
\label{Sec:Conformal}

With compactified spatial coordinates and asymptotically null spatial
hypersurfaces, so that $\ScriPlus$ is at a finite coordinate radius
$R_+$, the physical spacetime (coordinate) metric $g_{\mu \nu}$ is
singular at $\ScriPlus$.  The spatial metric must generate infinite
spatial distances over a finite range of coordinates, and the lapse
becomes infinite as the hypersurface becomes null.  In the conformal
approach to asymptotically flat spacetimes, pioneered by
Penrose~\cite{rP64}, these infinities are absorbed into a conformal
factor $\Omega$ such that the conformal metric $\tilde{g}_{\mu \nu}$,
defined by
\begin{displaymath}
  g_{\mu \nu} = \Omega^{-2} \tilde{g}_{\mu \nu},
\end{displaymath}
is finite at $\ScriPlus$, with $\Omega = 0$ at $\ScriPlus$ and
positive everywhere in the interior.  The physical manifold $M$ can be
considered as part of a larger conformal manifold $\tilde{M}$
containing the boundary of the physical manifold $\partial M$ as a
regular null hypersurface and in which the CMC hypersurfaces are
completely spacelike.  The conformal metric is taken to be regular at
$\ScriPlus$, with the precise amount of smoothness there open to some
debate (see Sec.~\ref{Sec-Intro}).

In the $3+1$ tetrad formalism, the above scaling of the coordinate
metric under a conformal transformation corresponds to the lapse and
the coordinate components of the spatial triad vectors scaling as
\begin{displaymath}
  \alpha = \Omega^{-1} \tilde{\alpha}, \qquad  B_a{}^{i} = \Omega\tilde{B}_a{}^{i}.
\end{displaymath}
The coordinate components of the shift vector are unchanged by the
conformal transformation.  We can then write for the directional
derivatives
\begin{displaymath}
  D_0 = \Omega \tilde{D}_0, \qquad D_a = \Omega \tilde{D}_a.
\end{displaymath}
Note, however, that the triad components of the shift vector transform
as $\tilde{\beta}^a = \Omega \beta^a$.

From the definition of the tetrad connection coefficients,
Eq.~(\ref{Eq:GammaDefComm}),
\begin{displaymath}
  \Gamma_{\alpha \beta \gamma}  = \Omega \: \tilde{\Gamma}_{\alpha \beta \gamma} 
  \: + \: \tilde{D}_{\alpha} \Omega \: \eta_{\beta \gamma} \: - \: \tilde{D}_{\beta} 
  \Omega \: \eta_{\alpha \gamma}.
\end{displaymath}
This translates into the following conformal transformations of the
dyadic decompositions of the connection coefficients:
\begin{displaymath}
  N_{ab} = \Omega\tilde{N}_{ab} + \varepsilon_{abc}\tilde{D}_c\Omega,
  \qquad K_{ab} = \Omega \tilde{K}_{ab} \: - \: \tilde{D}_0 \Omega \: \delta_{ab},
\end{displaymath}
from which
\begin{eqnarray} 
&&  \hat{N}_{ab} = \Omega\hat{\tilde{N}}_{ab}, 
  \qquad N = \Omega\tilde{N}, 
  \qquad n_b = \Omega\tilde{n}_b + \tilde{D}_b\Omega, \label{Eq:NRescale} \\
&&  \hat{K}_{ab} = \Omega\hat{\tilde{K}}_{ab},
  \qquad K = \Omega\tilde{K} - 3\tilde{D}_0\Omega,\label{Eq:KRescale}
\end{eqnarray}
and 
\begin{equation} 
a_b = \Omega\tilde{a}_b \: - \: \tilde{D}_b \Omega, \qquad
\tilde{a}_b = \tilde{D}_b \left(\log{\tilde{\alpha}} \right), \qquad
\omega_b = \Omega \tilde{\omega}_b.
\end{equation}

The choice of conformal scaling for the energy-momentum tensor is
somewhat arbitrary. The physical tetrad components of a radiation
energy-momentum tensor are expected to scale as $\Omega^4$, two powers
from the inverse $r^2$ scaling of divergent radiation in a Minkowski
frame, and two redshift-time dilation factors due to our tetrad frame
becoming asymptotically null. Therefore, we will define the conformal
energy-momentum tensor by
\begin{displaymath}
  \tau_{\alpha  \beta} = \Omega^4 \, \tilde{\tau}_{\alpha  \beta},\qquad
  \tilde{\rho} \equiv \tilde{\tau}_{00},\qquad
  \tilde{j}_b\equiv -\tilde{\tau}_{0b},\qquad
  \tilde{\sigma}_{ab} \equiv \tilde{\tau}_{ab}.
\end{displaymath}

In principle, there is quite a bit of freedom in the choice of the
conformal factor $\Omega$.  Any function which is positive over the
physical manifold and vanishes in a suitable way at $\ScriPlus$ would
do, since a change in $\Omega$ can be compensated for by a change in
the determinant of the conformal metric.  One possible choice would be
to demand that $\Omega^{-6}$ equal the determinant of the physical
spatial metric, so the determinant of the conformal spatial metric is $1$.  However,
such a choice would be highly coordinate dependent.  What seems to us
to be a much better choice, given our use of the 3D Nester gauge, is
$\Omega = e^{\Phi}$, where $\Phi$ is the Nester potential for the
physical space, see Eq.~(\ref{Eq:Integrated3DNesterCondition}).  This
has the great simplifying consequence that $\tilde{n}_a \equiv 0$
\cite{jN91}, and together with $N=0$ this means that the conformal
connection coefficients $\tilde{N}_{ab} = \hat{\tilde{N}}_{ab}$ are
symmetric and traceless.  Another consequence is that the triad vector
fields, as vector fields in the conformal 3-space, are divergenceless:
\begin{equation}
  \tilde{\nabla}_k \tilde{B}_a{}^k = -2\tilde{n}_a = 0.
    \label{Eq:divB}
\end{equation}
The covariant conformal derivative of an arbitrary vector field ${\bf
  v} = v_a {\bf B}_a$ is
\begin{displaymath} 
  \tilde{\nabla}_a v_b = \tilde{D}_a v_b - \hat{\tilde{N}}_{ac} \, v_d \, \varepsilon_{cdb},
\end{displaymath}
and $\tilde{\nabla}_a v_a = \tilde{D}_a v_a$. Therefore, in
the conformal Laplacian of a scalar or the conformal divergence of a
vector field, the covariant derivatives can be replaced by directional
derivatives. Furthermore, the conformal Ricci tensor is 
\begin{equation}
    \tilde{R}_{ab} 
    = -\tilde{D}_c\hat{\tilde N}_{d(a}\varepsilon_{b)cd} + 
    2\hat{\tilde N}_{ac}\hat{\tilde N}_{bc} - \delta_{ab}\hat{\tilde N}_{cd}\hat{\tilde N}_{cd} 
    \label{Eq:ConfRab}.
\end{equation} 
The Ricci scalar has the simple form $\tilde{R} =
-\hat{\tilde{N}}_{ab}\hat{\tilde{N}}_{ab}$ and is negative definite.

The Hamiltonian constraint equation~(\ref{Eq:HamConstraint}) then
becomes a relatively simple elliptic equation for $\Omega$:
\begin{equation}
  \Omega \: \tilde{D}_c \tilde{D}_c \Omega = \frac{3}{2} \left[ \left(\tilde{D}_c \Omega \right) 
    \left(\tilde{D}_c \Omega \right) - \left( \frac{K}{3} \right)^2\right] + \frac{\Omega^2}{4} 
  \left( \hat{\tilde{K}}_{cd}  \hat{\tilde{K}}_{cd} \: + \: 
    \hat{\tilde{N}}_{cd}  \hat{\tilde{N}}_{cd} \right) \: + \: 4\pi
  G \Omega^4 \tilde{\rho}. 
 \label{Eq:HamRescaled}
\end{equation}

With the Nester-gauge inspired choice of $\Omega$, and with the time
derivatives of $\Omega$ expressed in terms of $\tilde{K}$, the
evolution equations for the conformal quantities are
\begin{eqnarray}
  \tilde{D}_0 \Omega &=& -\frac{1}{3} \left(K - \Omega \tilde{K} \right),
\label{Eq:OmegaKt}\\
  \tilde{D}_0\tilde{B}_a{}^k &=&-\hat{\tilde{K}}_{ab}\tilde{B}_b{}^k -
  \varepsilon_{acd} \, \tilde{\omega}_c \tilde{B}_d{}^k -\frac{\tilde{K}}{3}
  \tilde{B}_a{}^k,
  \label{Eq:BaRescaled}\\
  \tilde{D}_0 \hat{\tilde{N}}_{ab} + \tilde{D}_c \hat{\tilde{K}}_{d(a}
  \varepsilon_{b)cd}
  &=& \left\{ 2 \hat{\tilde{K}}_{c(a} \hat{\tilde{N}}_{b)c} - \frac{\tilde{K}}{3}\hat{\tilde{N}}_{ab}
    + 2 \varepsilon_{cd(a} \hat{\tilde{N}}_{b)c} \, \tilde{\omega}_d 
    + \frac{1}{\tilde{\alpha}} \left[\varepsilon_{cd(a} \hat{\tilde{K}}_{b)c} 
      {\tilde{D}_d} {\tilde{\alpha}} - \tilde{D}_{(a}
      \left({\tilde{\alpha}} \, {\tilde{\omega}}_{b)}\right)
    \right] \right\}^{TF},
  \label{Eq:NabRescaled} \\
  \tilde{D}_0\hat{\tilde{K}}_{ab} - \tilde{D}_c \hat{\tilde{N}}_{d(a} \varepsilon_{b)cd} 
  &=& \left\{ - 2 \hat{\tilde{N}}_{ac} \hat{\tilde{N}}_{bc} - \frac{\tilde{K}}{3}\hat{\tilde{K}}_{ab}
    + 2\varepsilon_{cd(a} \hat{\tilde{K}}_{b)c} \, \tilde{\omega}_d \right.
  \nonumber\\
  &+& \left.  \frac{1}{\tilde{\alpha}} \tilde{\nabla}_a \tilde{\nabla}_b \tilde{\alpha} 
    - \frac{2}{\Omega} \left[\tilde{\nabla}_a \tilde{\nabla}_b \, \Omega 
      + \frac{K}{3}\hat{\tilde{K}}_{ab} \right] + 8\pi G \Omega^2 \tilde{\sigma}_{ab}
  \right\}^{TF}.
\label{Eq:KabRescaled}
 \end{eqnarray} 
 This system is subject to the constraint equations
\begin{eqnarray} 
  \varepsilon_{abc}\tilde{D}_b\tilde{B}_c{}^k -
  \hat{\tilde{N}}_{ab}\tilde{B}_b{}^k &=& 0,
\label{Eq:CurlBRescaled}\\
  \tilde{D}_b\hat{\tilde{N}}_{ab} &=& 0,
  \label{Eq:DivNRescaled}\\
  \tilde{D}_b\hat{\tilde{K}}_{ab} +\varepsilon_{abc}\hat{\tilde{K}}_{bd}
  \hat{\tilde{N}}_{dc} - \frac{2}{\Omega}(\tilde{D}_b\Omega)\hat{\tilde{K}}_{ab}
  &=& -8\pi G \Omega^2 \tilde{j}_a.
\label{Eq:MomRescaled}
\end{eqnarray} 

The conformal lapse $\tilde{\alpha}$ is determined by transforming the
CMC hypersurface condition, Eq.~(\ref{Eq:CMCAlpha}), to conformal
variables and then using the Hamiltonian constraint,
Eq.~(\ref{Eq:HamRescaled}), to eliminate the most singular terms at
$\ScriPlus$, giving
\begin{equation}
  \Omega\tilde{D}_a\tilde{D}_a\tilde{\alpha} -
  3(\tilde{D}_a\Omega)\tilde{D}_a\tilde{\alpha}
  + (\tilde{D}_a\tilde{D}_a\Omega)\tilde{\alpha} - \frac{\Omega}{2}\left(
    \hat{\tilde{N}}_{ab} \hat{\tilde{N}}_{ab} + 3\hat{\tilde{K}}_{ab}
    \hat{\tilde{K}}_{ab} \right) \tilde{\alpha}
  = 4\pi G \Omega^3 \left(3\tilde{\rho} + \tilde{\sigma}_{cc} \right) 
  \tilde{\alpha}.
\label{Eq:CMCRescaled}
\end{equation} 

We see that in principle, $\Omega$ can be found in two ways: by
solving the elliptic equation~(\ref{Eq:HamRescaled}) of the
Hamiltonian constraint or by integrating the evolution
equation~(\ref{Eq:OmegaKt}).  The elliptic equation is degenerate at
$\ScriPlus$ due to the hypersurface becoming null.  The apparently
singular terms constrain the asymptotic behavior of the solution, but
in practice cause no numerical problems~\cite{lBhPjB09, oR10a}.
Likewise, we do not expect the apparently singular terms in the lapse
equation~(\ref{Eq:CMCRescaled}) to impede the use of standard elliptic
solvers.  The time evolution equation for $\Omega$ could prove useful
in reducing the frequency of solving its elliptic equation.  However,
as yet we have no equations for $\tilde{K}$ and $\tilde{\omega}_a$.
Elliptic equations for these quantities come from requiring
preservation of the Nester gauge during the evolution, as shown in the
following subsection.

\subsection{Determination of $\tilde{K}$ and $\tilde{\omega}_a$}
\label{SubSec:Ktilde}

Now we derive the elliptic equations for the gauge fields $\tilde{K}$
and $\tilde{\omega}_a$.  These equations are based on
Eqs.~(\ref{Eq:DivOmega}) and~(\ref{Eq:GradD0Phi}) from Sec.~\ref{SubSec:System} which yield, in terms of the rescaled quantities,
\begin{eqnarray}
  -\tilde{D}_a(\tilde{\alpha}\tilde{\omega}_a) 
  &=& \tilde{\alpha} \hat{\tilde{N}}_{ab}\hat{\tilde{K}}_{ab},
 \label{Eq:DivergenceOmega}\\
 \tilde{D}_a\left( \frac{2}{3}\tilde{\alpha}\tilde{K}\right) 
 + \varepsilon_{abc}\tilde{D}_b(\tilde{\alpha}\tilde{\omega}_c)
 &=& \tilde{\alpha}\tilde{\mu}_a,
\label{Eq:GradKTilde}
\end{eqnarray}
where 
\begin{equation}
\tilde{\mu}_a \equiv \hat{\tilde{K}}_{ab}\tilde{D}_b\left(
  \log\tilde{\alpha} + 2\log\Omega \right) -
\varepsilon_{abc}\hat{\tilde{K}}_{bd}\hat{\tilde{N}}_{dc} - 8\pi
G\Omega^2\tilde{j}_a.
\label{Eq:Defmutilde}
\end{equation}
While Eq.~(\ref{Eq:DivergenceOmega}) depends only on the rescaled
quantities, $\tilde{\mu}_a$ in Eq.~(\ref{Eq:GradKTilde}) contains the
apparently singular term involving the gradient of
$\log{\Omega}$. However, $\tilde{\mu}_a$ is, in fact, finite at
$\ScriPlus$. Indeed, using the rescaled momentum constraint,
Eq.~(\ref{Eq:MomRescaled}), we may rewrite
\begin{equation}
\tilde{\alpha}\tilde{\mu}_a = \tilde{D}_b(\tilde{\alpha}\hat{\tilde{K}}_{ab})
\label{Eq:mutildeRegular}
\end{equation}
which does not involve the conformal factor. However, we prefer
Eq.~(\ref{Eq:Defmutilde}), since it does not contain derivatives of
dynamical fields.  In Sec.~\ref{Sec:RegularityConds}, we derive
an explicit regular expression for the term
$\hat{\tilde{K}}_{ab}\tilde{D}_b\log{\Omega}$ at $\ScriPlus$.  In
Appendix~\ref{App-AsymptoticSolns}, we show that {\em second} radial
derivatives of $\tilde{K}_{ab}$ are logarithmically singular at
$\ScriPlus$ whenever outgoing radiation is present there.

In order to cast Eqs.~(\ref{Eq:DivergenceOmega},~\ref{Eq:GradKTilde})
into a second-order elliptic system, we make the following ansatz
which is motivated by the considerations at the end of
Appendix~\ref{App-NesterDirac}:
\begin{displaymath}
  \frac{2}{3}\tilde{\alpha}\tilde{K} = -\tilde{D}_a\zeta_a,\qquad
  \tilde{\alpha}\tilde{\omega}_a = \tilde{D}_a\psi + \varepsilon_{abc}\tilde{D}_b\zeta_c,
\end{displaymath}
where $\psi$ and $\zeta_a$ are a scalar function and a vector field,
respectively. Plugging this into
Eqs.~(\ref{Eq:DivergenceOmega},~\ref{Eq:GradKTilde}) and using the
commutation relation $[\tilde{D}_a,\tilde{D}_b] =
\varepsilon_{abd}\hat{\tilde{N}}_{cd}\tilde{D}_c$, we obtain the
elliptic system
\begin{eqnarray}
&& -\tilde{D}_a\tilde{D}_a\psi - \hat{\tilde{N}}_{ab}\tilde{D}_a\zeta_b = \tilde{\alpha}
\hat{\tilde{N}}_{ab}\hat{\tilde{K}}_{ab},
\label{Eq:EllipticPsi}\\
&& -\tilde{D}_b\tilde{D}_b\zeta_a + \hat{\tilde{N}}_{ab}\tilde{D}_b\psi
- \varepsilon_{abc}\hat{\tilde{N}}_{cd}\tilde{D}_d\zeta_b = \tilde{\alpha}\tilde{\mu}_a.
\label{Eq:EllipticZeta}
\end{eqnarray}
When supplemented with appropriate boundary conditions at $\ScriPlus$,
this system yields a unique solution for $(\psi,\zeta_a)$. We choose
homogeneous Dirichlet boundary conditions for $\psi$ and the
tangential components of $\zeta_a$. The normal component of $\zeta_a$,
however, is determined by the boundary condition from
Eq.~(\ref{Eq:BCKtilde}) which guarantees that the cross sections
$\dot{\ScriPlus}$ remain metric two-spheres. This translates into a 
mixed boundary condition for the normal component of $\zeta_a$
(see Sec.~\ref{Sec:Summary} below).

\subsection{Shift Condition}
\label{SubSec:ShiftCond}

The shift governs the evolution of the spatial coordinates. Therefore,
keeping the coordinate location of $\ScriPlus$, where $\Omega=0$, at
some fixed coordinate radius $R=R_+$ imposes a boundary condition on
the shift. The equation for the evolution of the conformal factor is
\begin{equation}
\label{Eq:EvolnOmega}
  \tilde \alpha \tilde D_0 \Omega 
  = \left( {\partial _t - \beta^k \partial _k } \right)\Omega 
  = -\tilde \alpha \left( {\frac{K}{3} - \frac{\Omega } {3}\tilde K} \right),
\end{equation}
with $\tilde{K}$ finite at $\ScriPlus$. The requirement that the
conformal factor remain zero at $\ScriPlus$, $\partial_t
\Omega\doteq 0$, implies that $\beta ^k \partial _k \Omega = \tilde
\beta ^a \tilde D_a \Omega\doteq K \tilde{\alpha}/3$, which gives a
constraint on the normal component of the shift. The outward normal
vector field $\tilde{s}_a$ with respect to the conformal geometry is
proportional\footnote{The minus sign in the expression
  $-\tilde{D}_a\Omega$ comes from the fact that $\Omega$ is strictly
  positive in the interior of $\Sigma_t$, hence, its gradient points
  inward from $\dot{\ScriPlus}$.} to $-\tilde{D}_a\Omega$, and the
normalization factor can be determined by evaluating the Hamiltonian
constraint, Eq.~(\ref{Eq:HamRescaled}), at $\ScriPlus$. This gives
\begin{equation}
\tilde{s}_a \doteq -\frac{3}{K}\tilde{D}_a\Omega,
\label{Eq:sa}
\end{equation}
so that the normal component of the shift at $\ScriPlus$ is
$\tilde{s}_a \tilde{\beta}^a\doteq - \tilde{\alpha}$.  It also makes
sense to require that the components of the shift tangent to
$\ScriPlus$ vanish. This means the coordinates on $\ScriPlus$ are
propagated along the null generators of $\ScriPlus$. The complete
shift boundary condition is then
\begin{equation}
    \tilde{\beta}_a \doteq -\tilde{\alpha}{\tilde s}_a, \qquad 
    \beta^i \doteq \frac{3\tilde{\alpha}}{K} \tilde{h}^{ij} \partial_j\Omega = 
    \frac{3\tilde{\alpha}}{K} \tilde{B}_a{}^i \tilde{B}_a{}^j \partial_j \Omega.
\label{Eq:ShiftBC}
\end{equation}
Satisfying this boundary condition is most straightforward if it is
imposed as a boundary condition on a vector elliptic equation for the
shift. This spatial coordinate gauge fixing may need to be done
carefully, and here we mention a few possibilities and their main features.

The first possibility, the {\em spatial harmonic} gauge, is motivated
by the metric-based formulation proved well posed by Andersson and
Moncrief~\cite{lAvM03}. With CMC foliations, this gauge casts the 
York version~\cite{jY79} of the standard Arnowitt-Deser-Misner (ADM) evolution equations~\cite{ADM62} into a nonlinear system of wave equations which
is coupled to elliptic equations for lapse and shift. Besides bringing
the evolution equations into hyperbolic form, the
spatial harmonic gauge has additional nice properties which were noted
in~\cite{lAvM03} and exploited in their proof. Namely, the terms
involving derivatives of the three-metric in the elliptic equation for
the lapse are eliminated. Similarly, the terms involving second
derivatives of the three-metric in the vector elliptic equation for
the shift cancel out. These properties were important for consistently
solving the elliptic equations to a given degree of smoothness in the
rigorous proof of well-posedness for the mixed hyperbolic-elliptic
system in Ref.~\cite{lAvM03}. While their analysis was carried out on
compact CMC slices in the absence of conformal rescaling, it motivated
the choice of the spatial harmonic gauge for the conformal spatial
geometry in the conformally compactified scheme of Moncrief and
Rinne~\cite{vMoR09}.

In terms of the conformal metric $\tilde{h}_{ij}$ the spatial harmonic gauge sets
\begin{equation}
 \tilde{V}^k \equiv \tilde{h}^{ij} \left[ \tilde{\Gamma}^k{}_{ij} - \Gammaz^k{}_{ij}  \right] = 0,
\label{Eq:SpatHarmGauge}
\end{equation}
where $\tilde{\Gamma}^k{}_{ij}$ and $\Gammaz^k{}_{ij}$ are,
respectively, the Christoffel symbols of the conformal three metric
$\tilde{h}_{ij}$ and of a given, possibly time-dependent, reference
metric $\hz_{ij}$ on $\Sigma_t$. Requiring this gauge condition to
remain satisfied during the evolution implies an elliptic equation for
the shift, derived from $\partial_t\tilde{V}^k = 0$.

In general, there is no simple way to express the Christoffel symbols
in terms of the triad connection coefficients. However, in the
conformal Nester gauge, there is a very simple expression for the
trace $\tilde{\Gamma}^k \equiv \tilde{h}^{ij}\tilde{\Gamma}^k{}_{ij}$
in terms of the conformal triad vectors. A straightforward calculation
gives
\begin{displaymath}
    \tilde{\Gamma}^k = -\partial_i \tilde{h}^{ki} - \tilde{h}^{ki} \left(
      \partial_i \sqrt{\tilde{h}}\,\right) / \sqrt{\tilde{h}} = -\tilde{D}_a 
    \tilde{B}_a{}^k - (\tilde{\nabla}_i \tilde{B}_a{}^i)\tilde{B}_a{}^k.
\end{displaymath}
In the conformal Nester gauge, $\tilde{\nabla}_i \tilde{B}_a{}^i =  0$
(see Eq.~(\ref{Eq:divB})), so
\begin{equation}
  \tilde{\Gamma}^k = - \tilde{D}_a  \tilde{B}_a{}^k.
    \label{Eq:Gammak}
\end{equation}
The elliptic equation for the shift comes from substituting
Eq.~(\ref{Eq:BaRescaled}) into
\begin{eqnarray}
  (\partial_t - \beta^\ell \partial_\ell)\tilde{V}^k 
  &=& -\tilde{D}_a \left[ (\partial_t - \beta^\ell\partial_\ell) \tilde{B}_a{}^k \right] 
  - (\tilde{D}_a\beta^i)\partial_i\tilde{B}_a{}^k
  \nonumber\\
  &&- \left[ (\partial_t - \beta^\ell\partial_\ell) \tilde{B}_a{}^{j} \right]
  \left( \partial_j \tilde{B}_a{}^k + 2\tilde{B}_a{}^i\Gammaz^k{}_{ij} \right)
  - \tilde{B}_a{}^i\tilde{B}_a{}^j(\partial_t - \beta^\ell\partial_\ell)
  \Gammaz^k{}_{ij} \nonumber\\
  &=& 0.
  \nonumber
\end{eqnarray}
After using Eqs.~(\ref{Eq:GradKTilde} - \ref{Eq:mutildeRegular}) to
eliminate derivatives of $\tilde{\alpha} \tilde{\omega}_a$ and
$\tilde{\alpha} \hat{\tilde{K}}_{ab}$, the result is
\begin{eqnarray}
  \tilde{D}_a\tilde{D}_a \beta^k
  &+& 2\tilde{B}_a{}^j\Gammaz^k{}_{ij}\tilde{D}_a\beta^i 
  + \left[ 2\tilde{\alpha} \tilde{\mu}_a - \frac{1}{3} \tilde{D}_a (\tilde{\alpha} 
    \tilde{K}) \right] \tilde{B}_a{}^k 
  \nonumber\\
  &+& 2\tilde{\alpha}\tilde{K}_{ab} \left( \tilde{D}_a\tilde{B}_b{}^k 
    + \tilde{B}_a{}^i \tilde{B}_b{}^j \Gammaz^k{}_{ij} \right)
  - \tilde{B}_a{}^i \tilde{B}_a{}^j (\partial_t - \beta^\ell\partial_\ell)
  \Gammaz^k{}_{ij} \nonumber\\
  &=& 0.
\label{Eq:SHShift}
\end{eqnarray}

It may be significant that the spatial harmonic gauge eliminates
derivatives of the triad vectors from the triad ``Laplacian'' operator
$\tilde{D}_a \tilde{D}_a$, which is the true scalar Laplacian operator
in the conformal Nester gauge. In the spatial harmonic gauge,
\begin{displaymath}
  \tilde{D}_a \tilde{D}_a = \tilde{B}_a{}^i \tilde{B}_a{}^j
  \frac{\partial^2} {\partial x^i \partial x^j} + \left(\tilde{D}_a
    \tilde{B}_a{}^k\right) \frac{\partial} {\partial x^k} = \tilde{B}_a{}^i
  \tilde{B}_a{}^j \left[ \frac{\partial^2} {\partial x^i \partial x^j}
 + \Gammaz^k{}_{ij} \frac{\partial} {\partial x^k} \right].
\end{displaymath}
In all of our elliptic equations except the shift equation, this
completely eliminates derivatives of the conformal triad vectors.
Notice also that the elliptic equation for the shift required for
preserving the spatial harmonic gauge, Eq.~(\ref{Eq:SHShift}), does
not contain second or higher derivatives of $\tilde{B}_a{}^k$.  An
appropriate choice for the reference metric of the spatial harmonic
gauge may well be the trivial one $\hz_{ij} = \delta_{ij}$, for which
the Christoffel symbols vanish.  This can accommodate multiple black
holes, as shown in the conformally flat initial data calculations of
Ref.~\cite{lBhPjB09}.

We now consider alternative coordinate gauges, based on the motivation
of choosing the shift such that the spatial metric components become
time independent as the physical and presumably the conformal
spacetime become stationary. The spatial harmonic gauge
may well have this property, but a more direct way of achieving the
goal is to choose the shift to minimize a functional quadratic in time
derivatives of the conformal metric. The two options of this type presented here are contingent on establishing analytically, or
by successful numerical trials, that they do not lead to
instabilities. We focus on minimizing the time dependence of the
conformal metric, rather than the physical metric, because our
conformal factor is a coordinate scalar determined by a coordinate
scalar equation.

The time dependence of our conformal metric is determined by, though
not equivalent to, the evolution equation~(\ref{Eq:BaRescaled}) for
the conformal triad fields, which is regular at $\ScriPlus$. Using the
constraint equation~(\ref{Eq:CurlBRescaled}), we can rewrite 
Eq.~(\ref{Eq:BaRescaled}) as
\begin{equation}
  \partial _t \tilde B_a{}^i = -\left[ \tilde D_a \tilde \beta_b 
    + \varepsilon _{acd} \tilde \beta_c\hat{\tilde{N}}_{db} 
    + \tilde \alpha \left( \tilde K_{ab} + \varepsilon _{acb} 
      \tilde{\omega} _c \right) \right] \tilde B_b{}^i.
\label{Eq:partialtB}
\end{equation}
The triad components of the shift depend on the conformal rescaling,
hence the $\tilde{\beta}_a$. Eq.~(\ref{Eq:partialtB}) should supersede
Eq.~(\ref{Eq:BaRescaled}) as the evolution equation for the triad
vectors when the shift is calculated as the $\tilde{\beta}_a$ rather
than the $\beta^i$. Contracting with the dual one-forms 
$\tilde \theta ^b{}_i$ gives
\[
\tilde F_{ba} \equiv \tilde \theta^b{}_i \,\partial _t \tilde B_a{}^i 
 = - \tilde D_a \tilde \beta_b - \varepsilon _{acd} \, \tilde\beta_c \hat{\tilde{N}}_{db} 
 - \tilde \alpha \left( {\tilde K_{ab} + \varepsilon _{acb} \, \tilde \omega _c } \right).
\]	 

One should bear in mind that the determinant of the conformal metric
is not time-independent in general for the Nester-based choice of
conformal factor. Therefore, the straightforward way to minimize the
time dependence of the conformal metric is the ``minimal conformal
strain'' condition minimizing $\int [(\partial_t \tilde h^{ij})
(\partial_t \tilde h^{k\ell}) \, \tilde h_{ik} \tilde h_{j\ell} \, w\, \sqrt
{\tilde h} \, ] \, d^3 x$, where $w$ is a weight function.  Since
$\tilde{h}^{ij} = \tilde{B}_a{}^i\tilde{B}_a{}^j$, this is equivalent
to minimizing the functional
\begin{equation}
I_1[\tilde{\beta}_a] = \int\tilde{F}_{(ab)}\tilde{F}_{(ab)} w \, \tilde\eta,
    \label{Eq:I1shiftaction}
\end{equation}
where $\tilde \eta = \sqrt {\tilde h} \, d^3 x$ is the natural volume
element of the conformal geometry.  The trivial choice of weight 
function, $w=1$, seems as good as any.  The
symmetrization of $\tilde{F}_{ab}$ in Eq.~(\ref{Eq:I1shiftaction}) is
what makes it essentially equivalent to the conventional minimal
strain condition in terms of $\tilde{h}^{ij}$.

With
\[
\tilde F_{(ab)} = - \tilde \nabla _{(a} \tilde \beta _{b)} - \tilde
\alpha \tilde K_{ab} ,
\]	 
where $\tilde \nabla_a$ is the conformal covariant derivative operator, 
minimizing $I_1$ gives the elliptic equation
\begin{equation}
- \tilde \nabla _b \tilde F_{(ab)} 
 = \frac{1} {2}\left( \tilde \nabla_b \tilde \nabla_b \tilde \beta_a 
  + \tilde \nabla_b \tilde \nabla_a \tilde \beta_b  \right) 
  + \tilde \nabla_b ( \tilde \alpha \tilde K_{ab} ) = 0,
\label{Eq:MinimalConformalStrain}
\end{equation}
and the boundary variation $\delta \tilde \beta_a \tilde F_{(ab)}
\tilde s_b$ vanishing is consistent with the Dirichlet boundary
condition~(\ref{Eq:ShiftBC}) at $\ScriPlus$ discussed above.  As in
the previous subsection, we prefer avoiding radial derivatives of the
fields in the source terms of the elliptic equations. For this reason,
we use the momentum constraint equation~(\ref{Eq:MomRescaled}), and
substitute
\[
\tilde \nabla_b ( \tilde \alpha \tilde K_{ab} )
 = \tilde{\alpha}\left[ \hat{\tilde{K}}_{ab}\tilde{D}_b(\log\tilde{\alpha} + 2\log\Omega)
  - 8\pi G\Omega ^2 \tilde{j}_a \right] + \frac{1}{3}\tilde D_a \left( \tilde{\alpha}\tilde{K} \right).
\]
As mentioned before, a regular expression for the term
$\hat{\tilde{K}}_{ab}\tilde{D}_b\log\Omega$ at $\ScriPlus$ will be
derived in Sec.~\ref{Sec:RegularityConds}.  We note that expanding
the covariant derivatives gives
\[
\tilde \nabla_b \tilde \nabla_b \tilde \beta_a 
 = \tilde D_b \tilde D_b \tilde \beta_a 
 + 2\varepsilon_{acd} \hat{\tilde{N}}_{bd} \tilde D_b \tilde{\beta}_c
 + \hat{\tilde{N}}_{ab} \hat{\tilde{N}}_{bc}\tilde{\beta}_c 
 - \hat{\tilde{N}}_{bc} \hat{\tilde{N}}_{bc}\tilde{\beta}_a
\]
and 
\[
\tilde \nabla_b \tilde \nabla_a \tilde \beta_b 
 = \tilde \nabla_a\left( \tilde \nabla _b \tilde \beta_b \right) + \tilde R_{ab}\tilde \beta_b 
 = \tilde D_a \left( \tilde D_b \tilde \beta_b \right) + \tilde R_{ab}\tilde \beta_b,
\]
where the constraint Eq.~(\ref{Eq:DivNRescaled}) has been used and
$\tilde{R}_{ab}$ is given by Eq.~(\ref{Eq:ConfRab}).

Smarr and York \cite{SY78} argue that a different shift condition, the
``minimal distortion'' condition, most cleanly suppresses that part of
the time dependence of the spatial metric associated with
diffeomorphisms of the spatial coordinates. This minimizes the time
dependence of what they call the conformal spatial metric, $\hat
h^{ij} \equiv (h^{1/3}) \, h^{ij} $, by minimizing the action $\int
[(\partial_t \hat h^{ij}) (\partial_t \hat h^{k\ell}) \hat h_{ik} \hat
h_{j\ell} \sqrt h\,] \, d^3 x$.  Their rationale for minimal distortion
rather than minimal strain is based on using the Hamiltonian
constraint to determine the determinant of the spatial metric, but we
make a quite different kind of conformal decomposition.  For
completeness, we note that the minimal distortion shift can be
implemented in our formalism by defining $\hat B_a{}^i \equiv B_a{}^i
/(\det B) = \tilde B_a{}^i /(\det \tilde B)$ and $\hat{\tilde{F}}_{ab}
\equiv \tilde F_{ab} - \frac{1} {3}\delta _{ab} \, \tilde F_{cc} $.
The minimal distortion version of the action is
\begin{equation}
I_2[\tilde{\beta}_a] = \int \hat{\tilde{F}}_{(ab)} \hat{\tilde{F}}_{(ab)} \tilde \eta. 
    \label{Eq:Iqshiftaction}
\end{equation}
The differential equation for the shift is then
\begin{eqnarray}
  0 = - \tilde \nabla_b \hat{\tilde{F}}_{(ab)} &=& \frac{1}{2}\left(
    \tilde \nabla_b \tilde \nabla_b \tilde \beta_a 
    + \tilde \nabla_b \tilde \nabla_a \tilde \beta_b \right)
  - \frac{1}{3}\tilde \nabla_a \tilde \nabla_b \tilde \beta_b  
  + \tilde \nabla_b \left( \tilde \alpha \hat{\tilde{K}}_{ab}  \right)
  \nonumber \\
  &=&  \frac{1} {2}\left( \tilde \nabla_b \tilde \nabla_b \tilde \beta_a 
    + \frac{1}{3}\tilde \nabla_a \tilde \nabla_b \tilde \beta_b 
    + \tilde R_{ab} \tilde \beta_b  \right) 
  + \tilde \nabla_b \left( \tilde \alpha \hat{\tilde{K}}_{ab}  \right),
 \label{Eq:EllipticBeta}
\end{eqnarray} 
and the boundary variation is $\delta \tilde
\beta_a\hat{\tilde{F}}_{(ab)} \tilde s_b$, which vanishes for the
Dirichlet boundary condition~(\ref{Eq:ShiftBC}).

All of the above forms of the shift equation are regular at
$\ScriPlus$ and should give rise to a well-posed elliptic boundary value problem on a CMC hypersurface (with other variables fixed) as long as the only boundary is at
$\dot{\ScriPlus}$. If black holes are present, there should also be
inner boundaries at excision surfaces inside or at the apparent
horizons, but we defer discussion of inner boundary conditions to a
future paper.

\section{Regularity Conditions at $\ScriPlus$}
\label{Sec:RegularityConds}

The evolution
equations~(\ref{Eq:OmegaKt},~\ref{Eq:BaRescaled},~\ref{Eq:NabRescaled})
for $\Omega$, $\tilde{B}_a{}^k$ and $\hat{\tilde{N}}_{ab}$ are
completely regular at $\ScriPlus$. However, Eq.~(\ref{Eq:KabRescaled})
for the evolution of $\hat{\tilde K}_{ab}$ has the apparently singular
term
\begin{displaymath}
  S_{ab} \equiv \frac{1}{\Omega}\left( \tilde \nabla_a \tilde \nabla_b \Omega 
    + \frac{K} {3}\hat{\tilde K}_{ab} \right)^{TF}
\end{displaymath}
at $\ScriPlus$. In this section, we derive the appropriate regularity
conditions which ensure that $S_{ab}$ is finite at $\ScriPlus$.  To
analyze the limit of the covariant Hessian $\tilde \Xi _{ab} \equiv \tilde
\nabla _a \tilde \nabla _b \Omega = \tilde \nabla _{(a} \tilde \nabla
_{b)} \Omega $ at $\ScriPlus$, we introduce the tensor $\tilde \gamma
_{ab} \equiv \delta _{ab} - \tilde s_a \tilde s_b$ projecting tangent
to $\dot{\ScriPlus}$ and the 2D extrinsic curvature of
$\dot{\ScriPlus}$
\begin{equation}
    \tilde \kappa _{ab} \equiv \tilde \gamma _{ac}\tilde \gamma_{bd} \tilde \nabla _c \tilde s_d  
    = \tilde \gamma_{ac} \tilde \nabla_c \tilde s_b,
\end{equation}
with a similar sign convention to that of the 3D extrinsic curvature, such that
$\tilde \kappa \equiv \tilde \kappa_{cc}$ is positive for expansion of the  
outward normals.  With $\tilde s_b\doteq -(3/K)\tilde \nabla _b \Omega$ and
smoothly extending $\tilde s_b$ as the tangent vector to the normal
spacelike geodesic in a neighborhood of $\dot{\ScriPlus}$, decomposing
$\tilde{\Xi}_{ab}$ into components along and perpendicular to
$\tilde{s}_a$ gives
\begin{eqnarray}
\tilde \Xi _{ab}   
  &\doteq& \tilde s_a \tilde s_b \tilde s_c \tilde s_d \tilde \nabla_c\tilde \nabla_d \Omega  
  + \tilde s_a \tilde s_c \tilde \gamma _{bd} \tilde \nabla_d \tilde\nabla_c \Omega  
  + \tilde s_b \tilde s_d \tilde \gamma_{ac} \tilde \nabla_c\tilde \nabla_d \Omega  
  + \tilde \gamma_{ac}\tilde \gamma_{bd}\tilde \nabla_c\tilde \nabla_d \Omega
\nonumber\\
  &\doteq& \tilde s_a \tilde s_b \tilde s_c \tilde \nabla_c 
  \left( \tilde s_d\tilde \nabla _d \Omega \right) - \frac{K} {3}\tilde \kappa_{ab},
\label{Eq:HessianAtScriPlus} 
\end{eqnarray} 
since $\tilde s_c \tilde \nabla_c \tilde s_d = 0$, $\tilde \kappa_{bc} \tilde s_c = 0$ and $\Omega\doteq 0$. Take the trace of Eq.~(\ref{Eq:HessianAtScriPlus}) to get
\[
\tilde \nabla_c \tilde \nabla_c \Omega
 \doteq \tilde s_c \tilde \nabla_c\left( \tilde s_d \tilde \nabla_d \Omega \right)
  - \frac{K}{3}\tilde \kappa,
\]
and the gradient of Eq.~(\ref{Eq:HamRescaled}) gives
\[
\tilde s_b \tilde \Xi _{ab} 
 \doteq \frac{1}{3}\tilde s_a \tilde D_b \tilde D_b \Omega 
 \doteq \frac{1}{3}\tilde s_a \tilde s_c \tilde \nabla_c 
 \left( \tilde s_d \tilde\nabla_d \Omega \right) - \frac{K}{9}\tilde \kappa \tilde s_a,
\]	 
from which 
\begin{displaymath}
    \tilde s_c\tilde \nabla_c(\tilde s_d\tilde \nabla_d \Omega) = \tilde{s}_c \tilde{s}_d 
    \tilde{\Xi}_{ab} \doteq -\tilde \kappa K/6 
\end{displaymath} 
and
\begin{equation}
    \tilde \Xi _{ab}\doteq - \frac{K} {3}\left(
    \tilde \kappa _{ab} + \frac{1} {2}\tilde \kappa \tilde s_a \tilde s_b \right).
    \label{Eq:HessianAtScri}
\end{equation}
Therefore, we find for the singular term in Eq.~(\ref{Eq:KabRescaled})
\begin{displaymath}
\Omega S_{ab} = (\tilde{\Xi}_{ab})^{TF} + \frac{K}{3}\hat{\tilde K}_{ab}
 \doteq -\frac{K}{3}\left( 
 \tilde \kappa_{ab} - \frac{1}{2}\tilde \gamma_{ab} \tilde \kappa - \hat{\tilde K}_{ab} \right).
\end{displaymath}
To ensure that $S_{ab}$ is finite, we see that the condition
\begin{equation}
\hat{\tilde \kappa} _{ab} 
 \equiv \tilde \kappa _{ab} - \frac{1}{2}\tilde \gamma_{ab} \tilde\kappa 
 \doteq \hat{\tilde K}_{ab}
\label{Eq:RegularityConditions}
\end{equation}
must hold at $\dot{\ScriPlus}$, and both
\begin{equation}
    \hat{\tilde K}_{ab} \tilde s_b = O(\Omega)
    \label{Eq:TransverseK}
\end{equation}
and
\begin{equation}
    \hat{\tilde \kappa}_{ab} - \tilde \gamma_{ac}\tilde \gamma_{bd}\hat{\tilde K}_{cd}
    = O(\Omega)
\label{Eq:ZeroShear}
\end{equation}
at $\dot{\ScriPlus}$. 

The first of these conditions, that only the transverse-traceless part
of $\tilde K_{ab}$ can be non-zero at $\dot{\ScriPlus}$, is what is
required to resolve the apparently singular terms in the sources for
the elliptic equations discussed in Sec.~\ref{Sec:Conformal}.  As we
show at the end of this section, the limit of $\Omega^{-1}\hat{\tilde
  K}_{ab}\tilde{s}_b $ at $\ScriPlus$ can be derived from the momentum
constraint equation~\cite{lApC94}. The second condition,
Eq.~(\ref{Eq:ZeroShear}), is known as the zero-shear condition, and
physically is a constraint on the initial data that limits the
amplitude of incoming waves in the neighborhood of $\ScriPlus$ and
ultimately at past null infinity. There is an extensive literature
(see~\cite{lApC94}) showing how these conditions relate to the
existence of $\ScriPlus$ as a regular null hypersurface in the
conformal spacetime.

While Eq.~(\ref{Eq:HessianAtScri}) can be inverted to get an
expression for $\tilde{\kappa}_{ab}$ at $\dot{\ScriPlus}$ in terms of
directional derivatives of $\Omega$, it is better to evaluate
$\tilde{\kappa}_{ab}$ from its defining expression
 \begin{equation}
     \tilde{\kappa}_{ab} = \tilde{\gamma}_{ac} \tilde{\nabla}_c \tilde{s}_b = 
     \tilde{\gamma}_{ac} \left( \tilde{D}_c \tilde{s}_b 
      + \varepsilon_{bde} \tilde{s}_d \hat{\tilde{N}}_{ce} \right),
      \label{Eq:kappaEval}
 \end{equation}
 which does not contain any second normal derivatives of $\Omega$.
 Since the directional derivative of $\tilde{s}_b$ is tangent to
 $\dot{\ScriPlus}$, $\tilde{\kappa}_{ab}$ at $\dot{\ScriPlus}$ is
 obtained by substituting $\tilde{s}_b = -(3/K) \tilde{D}_b \Omega$
 into Eq.~(\ref{Eq:kappaEval}).  The trace of the 2D extrinsic
 curvature is then
\begin{equation}
    \tilde{\kappa} = \tilde{\gamma}_{ab} \tilde{D}_b \tilde{s}_a  \doteq 
    -\frac{3}{K} \tilde{\gamma}_{ab} \tilde{D}_b \tilde{D}_a \Omega.
\end{equation}
Away from $\dot{\ScriPlus}$, $\tilde{s}_b$ is found by integrating the
geodesic equation.

In the following, we compute an explicitly finite expression at
$\ScriPlus$ for the apparently singular term $S_{ab}$ when the
regularity conditions~(\ref{Eq:TransverseK}) and~(\ref{Eq:ZeroShear})
are satisfied.  The direct approach using l'H\^opital's rule would
require going to higher order in the asymptotic expansions, which is
awkward in the context of the tetrad equations.  The Nester gauge
conditions, as elliptic equations which are regular at $\ScriPlus$,
impose no local constraints on the behavior of the
$\hat{\tilde{N}}_{ab}$ near or at $\ScriPlus$.  On the other hand, the
asymptotic expansions are rather straightforward when dealing with
coordinate components in certain special coordinate systems. We derive
in Appendix~\ref{App-AsymptoticSolns} the asymptotic expansion of the
conformal factor in Gaussian normal coordinates, and from this,
expressions for the {\em coordinate} components $S_{ij}$.  Since $S$
is a 3-tensor, it is then trivial to obtain the corresponding
expression for the triad components, though evaluating this expression
is not so easy, since it requires knowing $\tilde{s}_a$ away from
$\dot{\ScriPlus}$.

An alternative approach is to impose the Penrose regularity condition
that the Weyl tensor vanish at $\ScriPlus$. In
Appendix~\ref{App-AsymptoticSolns}, we use the asymptotic expansions
in Gaussian normal coordinates to evaluate the Weyl tensor and
determine the additional constraint on the asymptotic behavior of the
extrinsic curvature beyond the zero-shear condition required to
satisfy the Penrose condition. While it is still somewhat
controversial, we believe that the Penrose condition is reasonable, in
that it is only a slightly stronger restriction on the amplitude of
incoming waves near $\ScriPlus$ in the initial data than the
zero-shear condition, and is preserved in the subsequent evolution.

A standard expression for the electric part of the physical Weyl
tensor, in which time derivatives have been eliminated using the
evolution equations (as in Ref.~\cite{vMoR09}, Eq.~(5.1)), is
\begin{equation}
  E_{ab} = C_{0a0b} = \left[ R_{ab} - K_{ac} K_{cb} + KK_{ab} - 4\pi G\sigma_{ab} \right]^{TF}
  = \left[ R_{ab} - \hat{K}_{ac}\hat{K}_{cb} + \frac{K}{3}\hat{K}_{ab} - 4\pi G\sigma_{ab}
  \right]^{TF}.
\label{Eq:EabNoTimeDerivs}
\end{equation}
The conformal invariance of the Weyl tensor means that the conformal
Weyl tensor, i.e., the Weyl tensor calculated from the full
space-time conformal metric, is $\tilde E_{ab} = \Omega^{-2}
E_{ab}$. It is the coordinate components $E_{ij}$ which are truly
invariant, and the scaling relation between physical and conformal
triad vectors introduces the $\Omega^{-2}$ factor. In our conformal
variables, and taking advantage of the simplifications of the Nester
triad gauge, we have
\begin{eqnarray}
\tilde E_{ab} &=& \Omega^{-2} E_{ab}
\nonumber\\
 &=& \frac{1}{\Omega}\left[ \tilde \nabla_a \tilde \nabla_b \Omega
  + \frac{K}{3}\hat{\tilde K}_{ab}  \right]^{TF}
  + \left[ \tilde{R}_{ab} - \hat {\tilde K}_{ac} \hat {\tilde K}_{bc} 
  - 4\pi G\Omega^2\tilde{\sigma}_{ab} \right]^{TF}
\nonumber\\
 &=& S_{ab} - \tilde{D}_c \hat{\tilde N}_{d(a}\varepsilon_{b)cd}
  + \left[ 2\hat{\tilde N}_{ac}\hat{\tilde N}_{bc} - \hat{\tilde K}_{ac}\hat{\tilde K}_{bc}
        - 4\pi G\Omega^2\tilde{\sigma}_{ab} \right]^{TF}.
\nonumber
\end{eqnarray}
Note that the expression in Eq.~(\ref{Eq:EabNoTimeDerivs}) is not
conformally invariant, since the Einstein equations used to transform
the electric tensor to that form are not conformally invariant. The
Penrose regularity condition implies that $\tilde E_{ab} \doteq 0$,
such that
\begin{equation}
S_{ab} \doteq \tilde{D}_c \hat{\tilde N}_{d(a}\varepsilon_{b)cd}
 - \left[ 2\hat{\tilde N}_{ac}\hat{\tilde N}_{bc} - \hat{\tilde K}_{ac}\hat{\tilde K}_{bc} 
\right]^{TF}.
\end{equation}
Introduced into the evolution equation~(\ref{Eq:KabRescaled}) for
$\hat{\tilde K}_{ab}$, the result is
\begin{eqnarray}
\tilde D_0 \hat{\tilde K}_{ab}  &\doteq&  -\tilde D_c \hat{\tilde N}_{d(a} \varepsilon _{b)cd} 
  + \left[ 
  2\left( \hat{\tilde N}_{ac} \hat{\tilde N}_{bc}  - \hat{\tilde K}_{ac} \hat{\tilde K}_{bc} \right)
  - \frac{\tilde K}{3}\hat{\tilde K}_{ab} 
  + 2\varepsilon _{cd(a} \hat{\tilde K}_{b)c} \tilde \omega_d  
  + \frac{1}{\tilde \alpha} \tilde \nabla_a \tilde \nabla_b \tilde \alpha \right]^{TF}.
\label{Eq:PenroseBdaryEvol}
\end{eqnarray} 

The conformal momentum constraint, Eq.~(\ref{Eq:MomRescaled}), can be
used to obtain more explicit information on the asymptotic behavior of
the extrinsic curvature. Start from the equation in the form
$\Omega\tilde \nabla_b \left( \Omega^{-1}\hat{\tilde K}_{ab} \right) -
\Omega^{-1}(\tilde\nabla_b\Omega)\hat{\tilde K}_{ab} = -8\pi
G\Omega^2\tilde{j}_a$. Decompose $\hat{\tilde K}_{ab}$ into
longitudinal and transverse parts relative to the normal vector
$\tilde{s}_a$ at $\ScriPlus$,
\[
\hat{\tilde K}_{ab} = \tilde s_a \tilde s_b \left( \tilde s_c \tilde
  s_d \hat{\tilde K}_{cd} \right) + 2\tilde s_{(a} \left(\tilde \gamma
  _{b)d} \tilde s_c \hat{\tilde K}_{cd} \right) + \left(
  \tilde\gamma_{ac}\tilde \gamma_{bd} \hat{\tilde K}_{cd} \right).
\]
Substituting into the momentum constraint and using the properties of
$\tilde s_a$ gives
\begin{eqnarray}
  && \tilde s_a \left[ \left( \Omega\tilde s_b \tilde \nabla_b 
  +  \Omega\tilde{\kappa} - \tilde{s}_b\tilde{D}_b \Omega \right)  
  \left( \frac{\tilde s_c \tilde s_d \hat{\tilde K}_{cd} }{\Omega} \right) 
  + \left( \Omega \tilde \nabla_b - \tilde{D}_b \Omega \right) 
  \left( \frac{ \tilde s_c \tilde \gamma _{bd} \hat{\tilde K}_{cd}}{\Omega} \right) \right]
\nonumber \\
 &&+\; \left( \Omega \tilde s_b \tilde \nabla_b + \Omega\tilde{\kappa} - 
  \tilde{s}_b\tilde{D}_b \Omega \right) 
  \left( \frac{\tilde s_c \tilde \gamma_{ad}\hat{\tilde K}_{cd}}{\Omega} \right)
   + \tilde \kappa _{ab} \left( \tilde s_c \tilde \gamma _{bd}\hat{\tilde K}_{cd} \right)       
  + \left(\tilde \nabla_b - 2\frac{\tilde{D}_b \Omega}{\Omega} \right)
  \left( \tilde \gamma_{ac} \tilde\gamma_{bd}\hat{\tilde K}_{cd} \right) \nonumber \\
 && = -8\pi G\Omega ^2 \tilde j_a .
\nonumber
\end{eqnarray}
The purely or partially longitudinal parts of $\tilde{K}_{ab}/\Omega$
are finite at $\dot{\ScriPlus}$, though we shall see in
Appendix~\ref{App-AsymptoticSolns} that their radial derivatives are
logarithmically divergent. Despite this, those derivative terms vanish
at $\dot{\ScriPlus}$, since they contain an $\Omega$ factor in
front. Projecting along $\tilde{s}_a $ and evaluating at
$\dot{\ScriPlus}$ yields
\begin{equation}
  \Omega ^{-1} \tilde s_c \tilde s_d \hat{\tilde K}_{cd} \doteq
  \frac{3}{K}\tilde \kappa_{ab} \left( 
  \tilde \gamma_{ac} \tilde \gamma_{bd} \hat{\tilde K}_{cd} \right) 
= \frac{3}{K}\tilde \kappa_{cd} \hat{\tilde K}_{cd}.
\label{Eq:MomRescaledProjOnNormalAtScriPlus}
\end{equation}
Since $\tilde \gamma_{cd} \hat{\tilde K}_{cd} = - \tilde s_c \tilde
s_d \hat{\tilde K}_{cd}$, which is zero at $\ScriPlus$, $\tilde
\kappa_{ab}$ can be replaced by its traceless part $\hat{\tilde
  \kappa}_{ab} = \tilde \kappa_{ab} - \frac{1}{2}\tilde \gamma_{ab}
\tilde \kappa$ in
Eq.~(\ref{Eq:MomRescaledProjOnNormalAtScriPlus}). The projection
perpendicular to $\tilde s_a $ gives at $\ScriPlus$
\[
\Omega ^{-1} \tilde s_c \tilde \gamma _{ad} \hat{\tilde K}_{cd}
\doteq - \frac{3}{K}\tilde{\gamma }_{ae}\tilde \gamma _{bd}
\tilde\nabla_b \left( \tilde \gamma _{ce} \hat{\tilde K}_{cd} \right).
\]
Combining these results, the $\Omega ^{-1}\tilde s_b \hat{\tilde
  K}_{ba}$ term that appears in various equations can be explicitly
evaluated as
\[
\Omega ^{-1} \tilde s_b \hat{\tilde K}_{ba} 
\doteq - \frac{3}{K}\tilde{\gamma}_{ae} \tilde \gamma _{bd} \tilde \nabla _b 
\left( \tilde \gamma_{ec} \hat{\tilde K}_{cd} \right)
 + \frac{3}{K}\tilde s_a \hat{\tilde \kappa}_{bc} \hat{\tilde K}_{bc} .
\]

Based on the asymptotic expansions in Gaussian normal coordinates, we
show in Appendix~\ref{App-AsymptoticSolns} that the regularity
conditions $\hat{\tilde K}_{ab} \tilde s_b\doteq 0$ and $\tilde
\gamma_{ac}\tilde \gamma_{bd}\hat{\tilde K}_{cd}\doteq \hat{\tilde
  \kappa}_{ab}$ are preserved by the time evolution.

\section{Summary}
\label{Sec:Summary}

The mixed hyperbolic-elliptic system proposed in this paper for
solving the Einstein equations on CMC hypersurfaces in a conformally
compactified asymptotically flat spacetime is, we believe, worth
testing as a way of improving the accuracy of numerical calculations
of gravitational radiation emitted in mergers of black holes. There is no need to use
artificial boundary conditions at a finite radius when the
computational domain extends all the way to future null infinity, and
asymptotic gravitational wave amplitudes are calculated directly,
without the need for any extrapolation procedure.  A purely hyperbolic
system would possibly be more computationally efficient, since there
are no elliptic equations to solve, but requires extra care in
constructing a numerical scheme to suppress errors in the additional
constraints among the additional variables such a system entails.

Our representation of tensors by tetrad, rather than coordinate,
components is perhaps more controversial.  We fix the timelike vector
of the tetrad by requiring it to be orthogonal to the CMC
hypersurfaces of constant coordinate time, but there is gauge freedom
in the orientation of the triad of spatial vectors within the
hypersurface, in addition to the gauge freedom in the evolution of the
spatial coordinates.  A key part of our scheme is combining what we
consider to be the ``natural'' choice of triad gauge, the 3D Nester
gauge, with the choice of conformal gauge (definition of the conformal
factor), such that the non-trivial spatial triad connection
coefficients for the conformal geometry of the CMC hypersurfaces are
equivalent to the five independent components of the traceless
symmetric $\hat{\tilde{N}}_{ab}$. The hyperbolic part
of our evolution system for the conformally rescaled variables
consists of the coupled evolution
Eqs.~(\ref{Eq:NabRescaled},~\ref{Eq:KabRescaled}) for the
$\hat{\tilde{N}}_{ab}$ and the traceless part of the conformal
extrinsic curvature $\hat{\tilde{K}}_{ab}$, respectively, plus the
simple advection Eq.~(\ref{Eq:BaRescaled}) (or~(\ref{Eq:partialtB}) 
when $\tilde{\beta}_a$ is used)
for the nine coordinate
components of the conformal triad vectors, $\tilde{B}_a{}^{i}$.  This
is a total of nineteen evolution equations.

The remaining six tetrad connection coefficients -- the triad
components of the acceleration and angular velocity (relative to
Fermi-Walker transport) of the tetrad frames -- are determined by
elliptic equations: Eq.~(\ref{Eq:CMCRescaled})
for the conformal lapse, whose logarithmic gradient is the conformal
acceleration, and Eqs.~(\ref{Eq:EllipticPsi})
and~(\ref{Eq:EllipticZeta}) for a scalar potential $\psi$ and a vector
potential $\zeta_a$, which together determine the conformal angular
velocity $\tilde{\omega}_a$ and the trace of the conformal extrinsic
curvature $\tilde{K}$.  The conformal gauge is enforced by the
Hamiltonian constraint Eq.~(\ref{Eq:HamRescaled}) for the conformal
factor $\Omega$, but some interpolation may be possible using the
evolution Eq.~(\ref{Eq:OmegaKt}).  The coordinate gauge conditions are
fixed by elliptic equations for the triad components of the
conformal shift.  We suggest three options,
Eqs.~(\ref{Eq:SHShift}),~(\ref{Eq:MinimalConformalStrain}),
and~(\ref{Eq:EllipticBeta}), but some experimentation will be
necessary to determine which of these, or perhaps some quite different
scheme, works best in practice.  We do not propose explicitly imposing
the momentum constraints during the evolution, since this does not
seem to be necessary for stability \cite{oR10a}, and we similarly
allow the $\tilde{D}_b \hat{\tilde{N}}_{ab} = 0$ constraint to evolve
freely.  We analyze the propagation of errors in these and other
constraints in Appendix~\ref{App-ConstraintPropagation}.

No boundary conditions are necessary at $\ScriPlus$ for the evolution
equations, which is a major advantage of the CMC hypersurface
condition. Boundary conditions are required for the elliptic
equations.  Fundamental to the conformal decomposition is the boundary
condition $\Omega \doteq 0$ on the conformal factor.  As discussed in
Sec.~\ref{SubSec:ShiftCond}, the boundary condition at
$\dot{\ScriPlus}$ on the shift vector is $\tilde{\beta}_a \doteq
-\tilde{\alpha}\tilde{s}_a$, where $\tilde{s}_a =
-(3/K)\tilde{D}_a\Omega$ is the unit normal to $\dot{\ScriPlus}$ in
the conformal 3-space.  This ensures that $\ScriPlus$ is at a fixed
coordinate radius $R_+$ and that the angular coordinates in
$\dot{\ScriPlus}$ are propagated along the null generators of
$\ScriPlus$.  These angular coordinates are related to the nominally
Cartesian computational coordinates in the same way as in flat space.

The boundary conditions for the Nester gauge potentials $\psi$ and
$\zeta_a$ introduced in Sec.~\ref{SubSec:Ktilde} are naturally chosen
to be homogeneous Dirichlet conditions for $\psi$ and the part of
$\zeta_a$ tangent to $\dot{\ScriPlus}$:
\begin{equation}
  \psi \doteq 0, \qquad \tilde{\gamma}_{ab}\zeta_b \doteq 0,
\end{equation}
since we do not want the conformal triad angular velocity $\tilde{\omega}_a$ to
be unnecessarily non-zero.  For a flat conformal spatial geometry,
$\hat{\tilde{N}}_{ab}$ and therefore the source terms for $\psi$ are
zero, which should make $\psi = 0$ globally.  Furthermore, in a
spherically symmetric spacetime, the boundary conditions should allow
$\zeta_a$ to be spherically symmetric, which means it does not
contribute to $\tilde{\omega}_a$.

In order to keep the intrinsic geometry of the $\dot{\ScriPlus}$
2-surface that of a true sphere, with an inverse circumferential radius 
$\xi_0$ constant in time, as discussed in Appendix~\ref{App-AsymptoticSolns}, 
the boundary condition~(\ref{Eq:BCKtilde}) on $\tilde{K}$ must be satisfied. 
In terms of the Nester vector potential, this becomes 
$\tilde{D}_a \zeta_a \doteq -\tilde{\alpha} \tilde{\kappa}$, 
or
\begin{equation}
  \tilde{s}_b \tilde{D}_b \left( \tilde{s}_a \zeta_a \right)
  - \tilde{\kappa}\tilde{s}_a \zeta_a \doteq -\tilde{\alpha} \tilde{\kappa}.
\end{equation}
While we use Gaussian normal coordinates to derive this condition, it
is valid regardless of the computational coordinate system since it is
expressed in terms of spatial coordinate scalars. The area of the
$\dot{\ScriPlus}$ 2-sphere is then constant in time, which means that
the expansion of $\ScriPlus$ vanishes.  Note that $\xi_0$ is {\em not}
necessarily equal to the inverse of the coordinate radius of
$\ScriPlus$, $R_+$.  The geometrical definition of $\xi_0$ is that the
2D scalar curvature of $\dot{\ScriPlus}$ is $2\xi_0^{\,2}$.  In
Sec.~\ref{Sec:Discussion}, we use the zero expansion result, plus the
fact that our boundary conditions on the shift maintain a simple
relation between our computational coordinates and standard polar
angle coordinates on $\dot{\ScriPlus}$, to obtain a simple expression
for the Bondi news function in terms of our variables.

Finally, we consider the boundary condition on the conformal lapse.  
We can make our time coordinate correspond to retarded Minkowski 
time at $\ScriPlus$ by imposing the boundary condition 
\begin{equation}
    \tilde{\alpha} \doteq \frac{K}{3\xi_0} \equiv \tilde{\alpha}_0
\end{equation}
on Eq.~(\ref{Eq:CMCRescaled}) for the conformal lapse, since CMC slicing 
of Minkowski space has $\alpha \sim K r /3$ as $r \rightarrow \infty$.  
Since $\xi_0$ is uniform on all of the $\ScriPlus$ hypersurface to the future 
of the initial hypersurface, so is $\tilde{\alpha}_0$.

The question of what boundary conditions to impose on an excision
surface on or inside the apparent horizon of a black hole is left open
for the present.

\section{Discussion}
\label{Sec:Discussion}

An important issue when formulating the equations on CMC hypersurfaces
is dealing with singularities, real or apparent, which are associated
with the hypersurfaces becoming asymptotically null as they approach
$\ScriPlus$.  The elliptic equations for the conformal factor $\Omega$
and the conformal lapse $\tilde{\alpha}$ are genuinely singular
there. However, the singular terms in these equations do not cause any
singularity in the solutions; instead they force any solutions which
satisfy the boundary conditions at $\dot{\ScriPlus}$ to have very
particular asymptotic behaviors.  These asymptotic behaviors,
additionally constrained by the usual ``zero-shear'' and Penrose
regularity conditions, were discussed in
Sec.~\ref{Sec:RegularityConds}.  We also found it useful to explore
the asymptotic behavior at greater depth in a special coordinate
system, Gaussian normal coordinates
(Appendix~\ref{App-AsymptoticSolns}).  These coordinates are {\em not}
suitable as global coordinates in an actual numerical calculation, but
their simplicity in a neighborhood of $\dot{\ScriPlus}$ facilitates
drawing general conclusions about the asymptotic behavior of scalars
and complete tensors.  In this way, we are able to analyze the
apparently singular terms that appear in some of the evolution
equations, show that they are in fact finite at $\ScriPlus$, and
evaluate their limits at $\dot{\ScriPlus}$.  However, the conformal
factor, the conformal lapse, and the conformal extrinsic curvature are
not completely smooth at $\dot{\ScriPlus}$ on a CMC hypersurface when
any outgoing radiation is present.  The fourth normal derivative of
$\Omega$ and the second normal derivative of the conformal extrinsic
curvature are logarithmically infinite there. This is a gauge
artifact of the CMC slicing condition, has nothing to do with our
spatial gauge conditions, and is not prevented by any physically
allowable regularity conditions.  In principle, it should slow the
convergence of a pseudo-spectral numerical scheme or a high-order
finite difference scheme, but the impact may not be severe. The only
apparent way of avoiding this potential problem is to make the lapse
evolve dynamically as part of the hyperbolic system.

It is instructive to give a more detailed comparison of our system
with the coordinate-based system of Moncrief and
Rinne~\cite{vMoR09}. They have a set of eleven evolution equations for the
six components of the conformal spatial metric and the five components
of the traceless part of the canonical momentum $\pi^{\rm{tr} \, ij}$.
The traceless canonical momentum is, within a factor of the square
root of the determinant of the conformal metric, equivalent to our
$\hat{\tilde{K}}_{ab}$.  The Moncrief-Rinne evolution equations are
second order in spatial derivatives of the metric and much more
complicated than our evolution equations, but there are fewer of them,
both because our system is first-order and because the spatial triad
vectors have three more degrees of freedom than the conformal metric.
Since Moncrief and Rinne also impose the CMC hypersurface condition, their equation
for the lapse is essentially the same as ours, and there is the same
breakdown in smoothness due to polyhomogeneous terms.  Their conformal
factor is defined differently by demanding that the conformal scalar
curvature be uniform in both space and time, but it satisfies a
similar elliptic equation. The difference in conformal gauge does not
affect the leading behavior of the conformal factor at
$\dot{\ScriPlus}$, and the basic structure of the apparently singular
terms in the equations is the same.  Since the triad angular velocity
plays no role in a coordinate-based system, Moncrief and Rinne have
three fewer elliptic equations than we do. Their variable $\Gamma$
defined in their Eq.~(2.11) is equivalent to our
$\tilde{\alpha}\tilde{K}$, but is determined by a quite different
elliptic equation.  They note the advantages of imposing a boundary
condition on $\Gamma$ equivalent to our Eq.~(\ref{Eq:BCKtilde}). Their
elliptic equation for the shift preserves the spatial harmonic
condition on the spatial coordinates.

The tradeoff between complicated spatially second-order evolution
equations and more, but simpler, first-order evolution equations will
probably make little difference in computational efficiency; however,
high accuracy may be easier to achieve with our first-order equations.
The more serious numerical efficiency issue in our formalism is the
three extra elliptic equations. All of our variables except for the
triad vectors and the shift vector are spatial coordinate scalars,
which with the globally optimized triad orientations provided by the
Nester gauge, may be better behaved in complicated dynamic scenarios
than the coordinate components of the metric tensor and the conformal
extrinsic curvature.  In both approaches, it is critical to use highly
optimized elliptic solvers.

Our asymptotic wave amplitude is $\hat{\tilde{\kappa}}_{ab}
\doteq \hat{\tilde{K}}_{ab}$, whose coordinate components are 
$\breve{\chi}_{\mbox{\tiny{AB}}}$ in the asymptotic expansion of the 
spatial metric in Gaussian normal coordinates developed in 
Appendix~\ref{App-AsymptoticSolns}.  To see how this relates to 
the Bondi news, we solve asymptotically for the coordinate 
transformation between our Gaussian normal coordinates 
defined on CMC hypersurfaces and Bondi coordinates.  The inverse 
metric in Bondi coordinates $u,\,x,\, \bar{x}^{\mbox{\tiny{A$'$}}}$, 
with $u$ the retarded time, $x$ the inverse of the Bondi radius $r$, 
and $\bar{x}^{\mbox{\tiny{A$'$}}}$ the angular coordinates, is
\[
\bar{g}^{uu} = \bar{g}^{u \mbox{\tiny{A$'$}}} = 0, \quad \bar{g}^{ux}
= x^2 e^{-2\beta}, \quad \bar{g}^{xx} = x^4 W e^{-2\beta}, \quad
\bar{g}^{x \mbox{\tiny{A$'$}}} = x^2 U^{\mbox{\tiny{A$'$}}}
e^{-2\beta}, \quad \bar{g}^{\mbox{\tiny{A$'$}} \mbox{\tiny{B$'$}}} =
x^2 \bar{h}^{\mbox{\tiny{A$'$}} \mbox{\tiny{B$'$}}},
\]
with the determinant of $\bar{h}^{\mbox{\tiny{A$'$}} \mbox{\tiny{B$'$}}}$ 
equal to $(\sin{\theta})^{-2}$.
For a certain class of solutions of the Einstein equations, the
Bondi-Sachs metric functions $\beta$, $W$, $U^{\mbox{\tiny{A$'$}}}$,
and $\bar{h}^{\mbox{\tiny{A$'$}} \mbox{\tiny{B$'$}}}$ have expansions
in powers of $x$ away from future null infinity at $x = 0$, provided
that the Penrose regularity condition is satisfied (see
\cite{pCmMdS95}).  

For the special class of {\em inertial} Bondi coordinates (see Sec.~6
of Ref.~\cite{jW09} and Ref.~\cite{mBbSjWyZ10}), the angular
metric $\bar{h}_{\mbox{\tiny{A$'$}}\mbox{\tiny{B$'$}}}$ has the
expansion
\[
\bar{h}_{\mbox{\tiny{A$'$}} \mbox{\tiny{B$'$}}} =
\breve{h}_{\mbox{\tiny{A$'$}} \mbox{\tiny{B$'$}}} +
\bar{\chi}_{\mbox{\tiny{A$'$}} \mbox{\tiny{B$'$}}} x + O(x^3),
\]
where as in Appendix~\ref{App-AsymptoticSolns},
$\breve{h}_{\mbox{\tiny{A$'$}} \mbox{\tiny{B$'$}}}$ is the unit sphere
metric, $\bar{\chi}_{\mbox{\tiny{A$'$}} \mbox{\tiny{B$'$}}}$ is a
traceless, symmetric tensor on the unit sphere, and the absence of an
$O(x^2)$ term is equivalent to the Penrose regularity condition in
this context.  Also, $\beta \doteq 0$ and $U^{\mbox{\tiny{A$'$}}}
\doteq 0$.  The Bondi mass aspect $M$, whose average over solid angle
is the Bondi energy, is defined by $W = 1 - 2
M(u,\bar{x}^{\mbox{\tiny{A$'$}}}) x + O(x^2)$. The Bondi news is
basically the time derivative of $\bar{\chi}_{\mbox{\tiny{A$'$}}
  \mbox{\tiny{B$'$}}}$, often represented as a complex scalar $c$
whose real part is $\bar{\chi}^{\theta'}_{~\theta'} = -
\bar{\chi}^{\varphi'}_{~\varphi'}$ and imaginary part is
$\bar{\chi}^{\varphi'}_{~\theta'}\,\sin{\theta'} =
\bar{\chi}^{\theta'}_{~\varphi'}/\sin{\theta'}$. The square of the Bondi 
news at a given physical point on $\ScriPlus$ is invariant under 
transformations of the Bondi coordinates from one inertial Bondi 
frame to another~\cite{rS62}.

In Appendix~\ref{App-AsymptoticSolns}, we discuss the asymptotic
conformal spatial metric in Gaussian normal coordinates
$z,\,x^{\mbox{\tiny{A}}}$ and obtain asymptotic expressions for the
lapse and shift which preserve the CMC hypersurface and Gaussian
normal coordinate conditions.  While these spatial coordinates are not
the computational coordinates, the shift satisfies the same boundary
condition at $\ScriPlus$.  The boundary condition on the conformal
lapse, as discussed in Sec.~\ref{Sec:Summary}, ensures that the
coordinate time $t$ on the CMC hypersurfaces asymptotically
corresponds to Minkowski retarded time.  The coordinate transformation
from these coordinates to the Bondi-Sachs coordinates can be
constructed as a power series in the distance $z$ from $\ScriPlus$,
starting from $u \doteq t$ and $\bar{x}^{\mbox{\tiny{A$'$}}} \doteq
x^{\mbox{\tiny{A}}}$, by requiring that the transformed metric have
the Bondi-Sachs form, with $\bar{g}^{uu} = \bar{g}^{u
  \mbox{\tiny{A$'$}}} = 0$.  The result is
\[
u = t + \frac{z}{2\tilde{\alpha}_0} + \frac{\tilde{\kappa}_0}
{8\tilde{\alpha}_0} z^2 + O(z^3), \quad
\bar{x}^{\mbox{\tiny{A$'$}}} = x^{\mbox{\tiny{A}}} + O(z^3). 
\]
Then 
\[
    x^2 \bar{h}^{\mbox{\tiny{A$'$}} \mbox{\tiny{B$'$}}} = 
    \Omega^2 \tilde{h}^{\mbox{\tiny{CD}}} 
    \frac{\partial \bar{x}^{\mbox{\tiny{A$'$}}}}{\partial x^{\mbox{\tiny{C}}}} 
    \frac{\partial \bar{x}^{\mbox{\tiny{B$'$}}}}{\partial x^{\mbox{\tiny{D}}}} 
    + O(z^6),
\]
and equating the determinants of the two sides gives 
\[
    x = \Omega \xi \left( 1 + O(z^3) \right) = \frac{K \xi_0}{3} z + O(z^2).  
\]
Note that $\partial x/\partial t = O(z^2)$, since our boundary
conditions keep $\partial \xi/\partial t \doteq 0$. From the
expansions of $\bar{h}^{\mbox{\tiny{A$'$}} \mbox{\tiny{B$'$}}}$ and
$\tilde{h}^{\mbox{\tiny{CD}}}$ (Eq.~\ref{Eq:Invhdecomp}), we see that
\begin{equation}
    \bar{\chi}_{\mbox{\tiny{A$'$}} \mbox{\tiny{B$'$}}} \doteq 
    -\frac{6}{K\xi_0} \breve{\chi}_{\mbox{\tiny{AB}}}.
\end{equation}

The expansion of $\ScriPlus$ clearly vanishes.  From 
\[
    x^2 e^{-2\beta} = \Omega^2 \frac{\beta^z}{\tilde{\alpha}^2} 
    \frac{\partial x}{\partial z} \frac{\partial u}{\partial t} + O(z^4),
\]
$\beta = O(z)$, and from 
\[
    x^2 U^{\mbox{\tiny{A$'$}}} e^{-2\beta} = \Omega^2 \left(
    - \frac{\beta^{\mbox{\tiny{B}}} \beta^z}{\tilde{\alpha}^2}
    \frac{\partial x}{\partial z} \frac{\partial \bar{x}^{\mbox{\tiny{A$'$}}}}
    {\partial x^{\mbox{\tiny{B}}}} + O(z^3) \right),
\]
$U^{\mbox{\tiny{A$'$}}} = O(z^2)$.  Therefore, all of the conditions for 
an inertial Bondi frame are satisfied, and
the Bondi-Sachs ``news'' is $\dot{\bar{\chi}}_{\mbox{\tiny{A$'$}}
  \mbox{\tiny{B$'$}}} \equiv \partial_u \,\bar{\chi}_{\mbox{\tiny{A$'$}}
  \mbox{\tiny{B$'$}}}$.  The vacuum Einstein equations in
Bondi-Sachs coordinates give, since $\partial_u \doteq \partial_t$,
\begin{eqnarray}
  \frac{\partial M}{\partial u} &=&
  -\frac{1}{8} \breve{h}^{\mbox{\tiny{A$'$}} \mbox{\tiny{C$'$}}} \breve{h}^{\mbox{\tiny{B$'$}} \mbox{\tiny{D$'$}}} 
  \dot{\bar{\chi}}_{\mbox{\tiny{A$'$}} \mbox{\tiny{B$'$}}}
  \dot{\bar{\chi}}_{\mbox{\tiny{C$'$}} \mbox{\tiny{D$'$}}} 
  + \frac{1}{4} 
  \dot{\bar{\chi}}^{\mbox{\tiny{A$'$}} \mbox{\tiny{B$'$}}}_{~~~~ 
    \breve{\scriptscriptstyle |}\mbox{\tiny{A$'$}} \mbox{\tiny{B$'$}}}
  \nonumber\\
  &=& -\frac{1}{2}\left[ 
    \left( \frac{3}{K \xi_0} \right)^2  \breve{h}^{\mbox{\tiny{AC}}} \breve{h}^{\mbox{\tiny{BD}}} 
    \dot{\breve{\chi}}_{\mbox{\tiny{AB}}} \dot{\breve{\chi}}_{\mbox{\tiny{CD}}} + \left( \frac{3}{K \xi_0}
    \right) \dot{\breve{\chi}}^{\mbox{\tiny{AB}}}_{~~~ 
      \breve{\scriptscriptstyle |} \mbox{\tiny{AB}}} \right],
    \label{Eq:BondiMassEvol}
\end{eqnarray}
with $\breve{\scriptscriptstyle |}$ denoting covariant derivatives on
the unit sphere~\cite{ChrJezMac98}.  The monopole and dipole parts of the term linear 
in $\breve{\chi}^{\mbox{\tiny{AB}}}$ as integrated over the unit sphere vanish identically, 
so this term contributes to neither the Bondi energy nor the 
Bondi momentum. Eq.~(\ref{Eq:BondiMassEvol}) defines the precise
physical meaning of $\breve{\chi}_{\mbox{\tiny{AB}}}$ as the 2-tensor
describing the asymptotic amplitude of the gravitational waves.  This
2-tensor is equivalent to the 3-tensor whose triad components are
$\tilde{\kappa}_{ab} \doteq\hat{\tilde{K}}_{ab}$, as projected into
$\dot{\ScriPlus}$.

Our extraction of the Bondi news is actually quite a bit simpler than
it is in Cauchy-characteristic matching or Cauchy-characteristic
extraction formalisms~\cite{jW09}.  In the characteristic methods, the
matching or extraction is done on a world tube in the interior, and
the constraints controlling the conformal Bondi frame are integrated
outward on the null hypersurface, generically resulting in a
non-inertial Bondi frame at $\ScriPlus$.  The subsequent
transformation from the computational frame to the inertial Bondi
frame requires considerable numerical effort~\cite{jW09}, but a more
direct way of calculating the Bondi news from an arbitrary
characteristic frame is presented and tested in Ref.~\cite{mBbSjWyZ10}.

Although we contend that the approach we have presented holds much
promise for high accuracy calculations of gravitational waveforms from
3D numerical simulations of the Einstein equations, it remains to be
seen whether or not a stable and efficient numerical implementation
can be achieved. A good place to begin is to fine tune the numerical
solution of the elliptic equations of our method in the context of the
initial value problem and then evaluate whether or not the computation
time required is prohibitive. Further analysis of the well-posedness
of the entire hyperbolic-elliptic system would provide a theoretical
foundation for a more extensive numerical effort.

\begin{acknowledgments}
  We thank Frank Estabrook, Luis Lehner, Anil Zengino\u{g}lu, and
  particularly Vincent Moncrief for a number of discussions and
  helpful suggestions. OS was supported in part by grants CONACyT
  61173 and CIC 4.19 to Universidad Michoacana. LB held a Visitor in
  Physics appointment at the California Institute of Technology during
  the course of this work and is grateful to Mark Scheel and Kip
  Thorne for their hospitality.
\end{acknowledgments}

\appendix
\section{Asymptotic Expansions in Gaussian Normal Coordinates}
\label{App-AsymptoticSolns}

Quite a bit can be learned about the asymptotic behavior of the
conformal factor and the regularization of the apparently singular
terms in the evolution equations at $\ScriPlus$ by developing an
explicit asymptotic solution. This is awkward to do in our
implementation of the tetrad formalism because in the elliptic
equations of the Nester gauge, there is no local control over how the
orientation of the spatial triad vectors and therefore how
$\hat{\tilde N}_{ab}$ evolves in a neighborhood of $\dot{\ScriPlus}$.
On the other hand, by working directly with the coordinate components
of the spatial metric and other tensors in the CMC hypersurfaces, and
by making a special choice of spatial coordinates (Gaussian normal
coordinates based on $\dot{\ScriPlus}$ as a two-surface embedded in
the 3D conformal space~\cite{lApC94}) in a neighborhood of
$\ScriPlus$, we can straightforwardly obtain explicit asymptotic
solutions for the conformal factor and extrinsic curvature in terms of
quantities defined from the asymptotic expansion of the conformal
spatial metric as a power series in the conformal proper distance from
$\dot{\ScriPlus}$.  Moncrief and Rinne~\cite{vMoR09} have done a
similar asymptotic analysis in a more general coordinate system, but
at the cost of substantial additional complexity.

The Gaussian normal coordinates are constructed by starting from
angular coordinates on the $\dot{\ScriPlus}$ 2-surface, which can always be
chosen so that its metric is conformal to the standard unit sphere
metric, with coordinates $x^{\mbox{\tiny{A}}} = \left( {\theta ,\phi }
\right)$ . These angular coordinates are propagated inward from
$\ScriPlus$ along normal spacelike geodesics in the CMC hypersurface,
and a ``radial'' coordinate $z$ is defined to be equal to the proper
distance from $\dot{\ScriPlus}$ along these geodesics.  This construction
gives a conformal spatial metric of the form
\begin{equation}
  \label{Eq:ConformalSpatialMetric}
  d\tilde s^2  = dz^2  + \tilde h_{\mbox{\tiny{AB}}} \, dx^{\mbox{\tiny{A}}} dx^{\mbox{\tiny{B}}},
\end{equation}
since the geodesics normal to the $\dot{\ScriPlus}$ 2-surface are perpendicular 
to the 2-surfaces of constant proper distance.
Furthermore, since any metric on a 2-surface which has the topology
of a two-sphere is conformal to the standard unit sphere metric 
$\breve{h}_{\mbox{\tiny{AB}}}$, we can decompose 
\begin{equation}
  \tilde{h}_{\mbox{\tiny{AB}}} = \xi^{-2} \breve{h}_{\mbox{\tiny{AB}}},
 \label{Eq:hdecompScri}
\end{equation}
defining a 2D conformal factor $\xi$ such that 
\begin{equation}
    \tilde h = \det \tilde h_{\mbox{\tiny{AB}}} = \xi^{-4} \det 
    \breve{h}_{\mbox{\tiny{AB}}} = \xi^{-4} \sin^2{\theta}.
    \label{Eq:hdeterminant}
\end{equation}

Assuming a power series expansion of the 2D conformal metric about $\ScriPlus$ 
through order $z^2$, we then have the form 
\begin{equation}
  \tilde{h}_{\mbox{\tiny{AB}}} = \xi^{-2} \left[ \breve{h}_{\mbox{\tiny{AB}}} 
  - 2\breve{\chi}_{\mbox{\tiny{AB}}} \,z + \left( \breve{\chi}^{\mbox{\tiny{CD}}} 
  \breve{\chi}_{\mbox{\tiny{CD}}} \breve{h}_{\mbox{\tiny{AB}}} - 
  \breve{\psi}_{\mbox{\tiny{AB}}} \right) z^2 + O\left( z^3 \right) \right].
  \label{Eq:hdecomp}
\end{equation}
Here $\breve{\chi}_{\mbox{\tiny{AB}}}$ and
$\breve{\psi}_{\mbox{\tiny{AB}}}$ are traceless, symmetric tensors
with respect to the unit sphere metric $\breve{h}_{\mbox{\tiny{AB}}}$.
The inverse is
\begin{equation}
  \tilde{h}^{\mbox{\tiny{AB}}} = \xi^2 \left[ \breve{h}^{\mbox{\tiny{AB}}} 
    + 2 \breve{\chi}^{\mbox{\tiny{AB}}} z + \left( \breve{\chi}^{\mbox{\tiny{CD}}} 
      \breve{\chi}_{\mbox{\tiny{CD}}} \breve{h}^{\mbox{\tiny{AB}}} + 
      \breve{\psi}^{\mbox{\tiny{AB}}} \right) z^2 + O\left( z^3 \right) \right].
  \label{Eq:Invhdecomp}
\end{equation}
It has been argued (\cite{lApC94}, \cite{pCmMdS95}) that in a general
treatment of radiation in asymptotically flat spacetimes, there should
be polyhomogeneous terms in this expansion, i.e., $z \log{z}$ and/or
$z^2 \log{z}$ and higher order terms. Such terms are consistent with
the zero-shear condition, for instance, and do not prevent the
existence of $\ScriPlus$ as a null surface in the conformal spacetime.
As noted in Sec.~\ref{Sec-Intro}, we see no reason why such terms
should be present when dealing with isolated systems in an
astrophysical context. There may be polyhomogeneous terms present at
higher orders in the expansion of $\tilde{h}_{\mbox{\tiny{AB}}}$ arising from the
polyhomogeneous terms in the expansion of the extrinsic
curvature that are associated with the CMC gauge choice.

Physical interpretation of the asymptotic solutions is greatly
simplified if the 3D conformal gauge is such that 2-surfaces of
constant conformal radius correspond to 2-spheres of the asymptotic
flat spacetime approaching $\dot{\ScriPlus}$ on a CMC hypersurface.
Since the level surfaces of the 3D conformal factor $\Omega$ approach
$\dot{\ScriPlus}$ uniformly in the conformal radius coordinate, this
is accomplished by demanding that the intrinsic geometry of
$\dot{\ScriPlus}$ be that of a 2-sphere, i.e., that the 2D conformal
factor $\xi$ be a constant $\xi_0$, independent of angle, on
$\dot{\ScriPlus}$, or equivalently, that the 2D scalar curvature of
$\dot{\ScriPlus}$, $^2 R \doteq 2\xi_0\,^2$, be uniform.  Assuming
that the initial data satisfy this condition, it will be preserved
if $\partial_t \xi \doteq 0$.  As a result of our asymptotic analysis,
we shall see this requires a certain Dirichlet boundary condition on
the trace $\tilde{K}$ of the 3D conformal extrinsic curvature.  In the
context of our elliptic equations for preserving the Nester triad
gauge, this provides the ``natural'' boundary condition on
$\tilde{\alpha} \tilde{K} = -(3/2)\tilde{D}_a \zeta_a$.  The
same point was noted in the different context of Ref.~\cite{vMoR09}.

The Gaussian normal coordinates are not suitable as global
coordinates, since they break down somewhere in the interior as the
normal spacelike geodesic congruence develops caustics, but they are
guaranteed to be valid in a neighborhood of $\dot{\ScriPlus}$,
assuming a regular conformal geometry.  Any breakdown of smoothness at
$\ScriPlus$ cannot be attributed to the Gaussian normal choice of {\em
  spatial} coordinates.

The non-zero Christoffel symbols for the metric of
Eq.~(\ref{Eq:ConformalSpatialMetric}) are just
\[
\tilde{\Gamma}^{\mbox{\tiny{A}}}{}_{{\mbox{\tiny{B}}} z} = \frac{1} {2}\tilde
h^{\mbox{\tiny{AC}}} \, \partial_z \tilde h_{\mbox{\tiny{CB}}} = -
\tilde \kappa^{\mbox{\tiny{A}}}{}_{\mbox{\tiny{B}}} , \qquad
\tilde{\Gamma}^z{}_{\mbox{\tiny{AB}}} = - \frac{1} {2} \partial_z \tilde
h_{\mbox{\tiny{AB}}} = \tilde \kappa _{\mbox{\tiny{AB}}},\qquad
\tilde{\Gamma}^{\mbox{\tiny{A}}}{}_{\mbox{\tiny{BC}}} 
 = \,^2\tilde{\Gamma}^{\mbox{\tiny{A}}}{}_{\mbox{\tiny{BC}}} ,
\]	 
where $\tilde{\kappa}_{\mbox{\tiny{AB}}}$ are the coordinate
components of the extrinsic curvature of the constant-$z$ 2-surfaces
based on the outward normal, which is in the minus $z$-direction, and
where $^2\tilde{\Gamma}^{\mbox{\tiny{A}}}{}_{\mbox{\tiny{BC}}}$ are
the Christoffel symbols associated with the 2-metric $\tilde{h}_{\mbox{\tiny{AB}}}$.
Let $\hat{\tilde{\kappa}}_{\mbox{\tiny{AB}}}$ denote the traceless
part of $\tilde \kappa _{\mbox{\tiny{AB}}}$ and $\tilde{\kappa}$ the
trace.  Calculating the conformal 3D Ricci tensor, we get
\begin{eqnarray}
\label{Eq:Conformal3DRicci}
\tilde R^z{}_z &=& \partial _z \tilde \kappa - \tilde \kappa
_{\mbox{\tiny{AB}}} \tilde \kappa^{\mbox{\tiny{AB}}} = \partial _z
\tilde \kappa - \hat{\tilde \kappa} _{\mbox{\tiny{AB}}} \hat{\tilde
  \kappa} ^{\mbox{\tiny{AB}}} - \frac{1}
{2}\tilde \kappa ^2 , \nonumber \\
\tilde R^z{}_{\mbox{\tiny{A}}} &=& \tilde \kappa _{|{\mbox{\tiny{A}}}} -
\tilde \kappa^{\mbox{\tiny{C}}}{}_{{\mbox{\tiny{A}}}|{\mbox{\tiny{C}}}} 
= \frac{1}{2}\tilde \kappa _{|{\mbox{\tiny{A}}}}  
 - \hat{\tilde \kappa}^{\mbox{\tiny{C}}}{}_{{\mbox{\tiny{A}}}|{\mbox{\tiny{C}}}} ,  \\
\tilde R^{\mbox{\tiny{A}}}{}_{\mbox{\tiny{B}}} &=&\,
^2\tilde{R}^{\mbox{\tiny{A}}}{}_{\mbox{\tiny{B}}} 
 + \partial _z \tilde\kappa^{\mbox{\tiny{A}}}{}_{\mbox{\tiny{B}}} 
 - \tilde \kappa \,\tilde\kappa^{\mbox{\tiny{A}}}{}_{\mbox{\tiny{B}}} 
 = \partial_z\hat{\tilde \kappa}^{\mbox{\tiny{A}}}{}_{\mbox{\tiny{B}}} 
 - \tilde\kappa \,\hat{\tilde \kappa}^{\mbox{\tiny{A}}}{}_{\mbox{\tiny{B}}} +
 \frac{1}{2}\left( {\,^2 \tilde R + \partial _z \tilde \kappa  - \tilde
    \kappa ^2 } \right)\delta^{\mbox{\tiny{A}}}{}_{\mbox{\tiny{B}}}.
  \nonumber
\end{eqnarray} 
The $|$ symbol indicates a covariant derivative in the constant-$z$
2-space with respect to the metric $\tilde{h}_{\mbox{\tiny{AB}}}$.
The scalar curvature is
\begin{equation}
\label{Eq:ConformalRicciScalar}
  \tilde R = \,^2 \tilde R + 2\partial _z \tilde \kappa - \frac{3}
  {2}\tilde \kappa ^2 - \hat{\tilde \kappa} _{\mbox{\tiny{AB}}}
  \hat{\tilde \kappa} ^{\mbox{\tiny{AB}}} .
\end{equation}

In terms of our expansion (\ref{Eq:hdecomp}) of the angular metric, it
is easy to verify that
\begin{equation}
  \hat{\tilde{\kappa}}^{\mbox{\tiny{A}}}{}_{\mbox{\tiny{B}}} = 
  \breve{\chi}^{\mbox{\tiny{A}}}{}_{\mbox{\tiny{B}}}+ 
  \breve{\psi}^{\mbox{\tiny{A}}}{}_{\mbox{\tiny{B}}} z + O\left( z^2 \right),
 \nonumber
\end{equation}
so $\hat{\tilde{\kappa}}^{\mbox{\tiny{A}}}{}_{\mbox{\tiny{B}}} \doteq
\breve{\chi}^{\mbox{\tiny{A}}}{}_{\mbox{\tiny{B}}}$ and
$\partial_z\hat{\tilde{\kappa}}^{\mbox{\tiny{A}}}{}_{\mbox{\tiny{B}}}
\doteq \breve{\psi}^{\mbox{\tiny{A}}}{}_{\mbox{\tiny{B}}}$. Also,
$\tilde{\kappa} = (2/\xi) \partial_z \xi$.  Furthermore, 
assuming $\xi \doteq \xi_0$ is uniform on $\ScriPlus$,
\begin{equation}
  ^2\tilde{\Gamma}^{\mbox{\tiny{A}}}{}_{\mbox{\tiny{BC}}}
  = \breve{\Gamma}^{\mbox{\tiny{A}}}{}_{\mbox{\tiny{BC}}}
  - \left[ 2\breve{\chi}^{\mbox{\tiny{A}}}{}_{(\mbox{\tiny{B}}
      \breve{\scriptscriptstyle |}\mbox{\tiny{C}})}
    - \breve{\chi}_{\mbox{\tiny{B}}\mbox{\tiny{C}}}{}^{\breve{\scriptscriptstyle |}
      \mbox{\tiny{A}}} + \delta^{\mbox{\tiny{A}}}{}_{(\mbox{\tiny{B}}}
    \partial_{\mbox{\tiny{C}})}\tilde{\kappa}_0  - \frac{1}{2}\breve{h}_{\mbox{\tiny{B}}\mbox{\tiny{C}}}
    \breve{h}^{\mbox{\tiny{A}}\mbox{\tiny{D}}}\partial_{\mbox{\tiny{D}}}\tilde{\kappa}_0
  \right] z + O\left( z^2 \right)
  \label{Eq:Gamma2D}
\end{equation}
and 
\begin{equation} 
  \,^2 \tilde{R} = 2\xi_0^2\,\left[ 1 + \left( \tilde{\kappa}_0 
      + \frac{1}{2}\tilde{\kappa}_0{}^{\breve{\scriptscriptstyle |}
        \mbox{\tiny{A}}}{}_{\breve{\scriptscriptstyle |}\mbox{\tiny{A}}}
      - \breve{\chi}^{\mbox{\tiny{AB}}}{}_{\breve{\scriptscriptstyle |}\mbox{\tiny{AB}}}
    \right) z + O\left( z^2 \right) \right].
\label{Eq:2DCurvature}
\end{equation}
The $\breve{\scriptscriptstyle |}$ symbol denotes a covariant
derivative with respect to the unit sphere metric.

Our definition of the conformal factor is not quite the same as in
Ref.~\cite{vMoR09} or~\cite{lApC94}, but like them, we can solve for
the conformal factor to third order in the expansion away from
$\dot{\ScriPlus}$.  Write the expansion of $\Omega$ as
\[
\Omega = \frac{K}{3}z\left[ {1 + c_1 \left( {x^{\mbox{\tiny{A}}} }
    \right)z + c_2 \left( {x^{\mbox{\tiny{A}}} } \right)z^2 + O\left(
      {z^3 } \right)} \right],
\]
the expansion of $\tilde \kappa$ as 
\[
\tilde \kappa = \tilde \kappa _0 + \tilde \kappa _1 z + O\left( {z^2 }
\right),
\]
and substitute into the Hamiltonian constraint,
Eq.~(\ref{Eq:HamRescaled}). The result is
\begin{equation}
   \Omega = \frac{K}{3}z\left\{ 1 - \frac{1}{4}\tilde{\kappa} _0 z
     + \frac{1}{12}\left[ \frac{1}{2}\tilde \kappa_0 ^2 - 2\tilde \kappa_1
     - \left. \left( \hat{\tilde K}^{ab} \hat{\tilde K}_{ab} 
     + \hat{\tilde \kappa} _{\mbox{\tiny{AB}}} \hat{\tilde \kappa}^{\mbox{\tiny{AB}}} 
     - \,^2 \tilde R \right)\right|_{\ScriPlus} \right] z^2 + O\left( {z^3} \right) \right\}. 
\label{Eq:OmegaExpansion}
\end{equation} 
The expansion of $\Omega$ becomes locally indeterminate, and
generically requires a polyhomogeneous $z^3 \log z$ term, at the next
order.

In the Gaussian normal coordinates, the outward normal to the
constant-$z$ 2-surfaces is just $\tilde{s}^z = - 1$,
$\tilde{s}^{\mbox{\tiny{A}}} = 0$.  Denote the traceless part of the
3D extrinsic curvature as projected into these surfaces by
\begin{equation}
  \label{Eq:Sigmadef}
    \Sigma^{\mbox{\tiny{A}}}{}_{\mbox{\tiny{B}}} 
    \equiv \hat{\tilde K}^{\mbox{\tiny{A}}}{}_{\mbox{\tiny{B}}} 
   - \frac{1}{2}\delta^{\mbox{\tiny{A}}}{}_{\mbox{\tiny{B}}} 
   \hat{\tilde K}^{\mbox{\tiny{C}}}{}_{\mbox{\tiny{C}}} .
\end{equation}
The zero-shear condition derived in Sec.~\ref{Sec:RegularityConds}
requires $\Sigma^{\mbox{\tiny{A}}}{}_{\mbox{\tiny{B}}} \doteq
\hat{\tilde \kappa}^{\mbox{\tiny{A}}}{}_{\mbox{\tiny{B}}} \doteq
\breve{\chi}^{\mbox{\tiny{A}}}{}_{\mbox{\tiny{B}}}$.  Note that 
$\breve{\chi}^{\mbox{\tiny{A}}}{}_{\mbox{\tiny{B}}}$ can be considered 
the asymptotic amplitude of the gravitational waves, and the
zero-shear condition is a reflection of the fact that a derivative of
the spatial metric along the normal to an asymptotically null
hypersurface becomes equivalent to a radial derivative of the spatial
metric as the normal becomes tangent to the
hypersurface.

The momentum constraint equation~(\ref{Eq:MomRescaled}) allows us to
determine the leading asymptotic behavior of the rest of the extrinsic
curvature in terms of
$\breve{\chi}^{\mbox{\tiny{A}}}{}_{\mbox{\tiny{B}}}$.  In the Gaussian
normal coordinate basis, we have
\begin{equation}
  \Omega \tilde{\nabla}_j \left( \Omega^{-1} \hat{\tilde{K}}^j{}_i \right) = 
  \Omega^{-1} \partial_j \Omega \hat{\tilde{K}}^j{}_i - 8\pi G \Omega^2\tilde{j}_i. 
  \label{Eq:GaussNormMomEq}
\end{equation}
First consider the $z$-component, which through first-order in $z$ is 
\begin{equation}
  \frac{K}{3} z \left[ \partial_z \left( \Omega^{-1} \hat{\tilde{K}}^z{}_z 
    \right) + \left( \Omega^{-1} \hat{\tilde{K}}^{\mbox{\tiny{A}}}{}_z \right)_{|\mbox{\tiny{A}}}
  \right] + \hat{\tilde{\kappa}}^{\mbox{\tiny{B}}}{}_{\mbox{\tiny{A}}} 
  \Sigma^{\mbox{\tiny{A}}}{}_{\mbox{\tiny{B}}} - \frac{3}{2} \tilde{\kappa} 
  \hat{\tilde{K}}^z{}_z = \frac{K}{3} \left( 1 - \frac{1}{2} \tilde{\kappa} z \right) 
  \Omega^{-1} \hat{\tilde{K}}^z{}_z.  
  \label{Eq:RadialMom}
\end{equation}
Using the zero-shear condition, the zeroth-order result is
\begin{equation}
  \frac{K}{3\Omega} \hat{\tilde{K}}^z{}_z =  \frac{K}{3\Omega} \tilde{s}^a 
  \tilde{s}^b \hat{\tilde{K}}_{ab} \doteq 
  \hat{\tilde{\kappa}}^{\mbox{\tiny{A}}}{}_{\mbox{\tiny{B}}}
  \Sigma^{\mbox{\tiny{B}}}{}_{\mbox{\tiny{A}}} \doteq 
  \breve{\chi}^{\mbox{\tiny{A}}}{}_{\mbox{\tiny{B}}} 
  \breve{\chi}^{\mbox{\tiny{B}}}{}_{\mbox{\tiny{A}}}.
  \label{Eq:KzzatScri}
\end{equation}
The transverse components of the momentum constraint to first-order in
$z$ are
\begin{equation}
  \frac{K}{3} z \partial_z \left( \Omega^{-1} \hat{\tilde{K}}^z{}_{\mbox{\tiny{A}}} \right) 
    + \hat{\tilde{K}}^{\mbox{\tiny{B}}}{}_{\mbox{\tiny{A}}|\mbox{\tiny{B}}}
   -  \tilde{\kappa} \hat{\tilde{K}}^z{}_{\mbox{\tiny{A}}}
 = \frac{K}{3\Omega} 
  \left( 1 - \frac{1}{2}\tilde{\kappa} z \right) \hat{\tilde{K}}^z{}_{\mbox{\tiny{A}}}
   - \frac{z}{2} \tilde{\kappa}_{|\mbox{\tiny{B}}} 
  \hat{\tilde{K}}^{\mbox{\tiny{B}}}{}_{\mbox{\tiny{A}}}.
\nonumber
\end{equation}
Expand
$\hat{\tilde{K}}^{\mbox{\tiny{B}}}{}_{\mbox{\tiny{A}}|\mbox{\tiny{B}}} =
\Sigma^{\mbox{\tiny{B}}}{}_{\mbox{\tiny{A}}|\mbox{\tiny{B}}} -
\frac{1}{2} \left( \hat{\tilde{K}}^z{}_z \right)_{|\mbox{\tiny{A}}}$,
using Eqs.~(\ref{Eq:Gamma2D}) and~(\ref{Eq:KzzatScri}), to get
\begin{equation}
  \frac{K}{3} z\, \partial_z \left( \Omega^{-1} \hat{\tilde{K}}^z{}_{\mbox{\tiny{A}}} \right) 
  + \breve{\chi}^{\mbox{\tiny{B}}}{}_{\mbox{\tiny{A}}|\mbox{\tiny{B}}}
  + z \left( \partial_z \Sigma^{\mbox{\tiny{B}}}{}_{\mbox{\tiny{A}}} \right)_{|\mbox{\tiny{B}}}
  - \frac{1}{2} \tilde{\kappa} \hat{\tilde{K}}^{}_{\mbox{\tiny{A}}}
  - \frac{1}{2} \tilde{\kappa}_{|\mbox{\tiny{B}}} 
  \breve{\chi}^{\mbox{\tiny{B}}}{}_{\mbox{\tiny{A}}} 
  = \frac{K}{3\Omega} \hat{\tilde{K}}^z{}_{\mbox{\tiny{A}}}.
     \label{Eq:TransverseMom}
\end{equation}
At zeroth order, 
\begin{equation}
  \frac{K}{3\Omega} \hat{\tilde{K}}^z{}_{\mbox{\tiny{A}}}
 \doteq \breve{\chi}^{\mbox{\tiny{B}}}{}_{\mbox{\tiny{A}}\breve{\scriptscriptstyle |}\mbox{\tiny{B}}}.
  \label{Eq:KAzatScri}
\end{equation}

Using these zeroth-order results, we can calculate the first-order
corrections. However, the equations cannot be satisfied by a simple
power series expansion. Instead, let
\begin{equation}
  \frac{K}{3\Omega} \hat{\tilde{K}}^z{}_z
 =  \breve{\chi}^{\mbox{\tiny{A}}}{}_{\mbox{\tiny{B}}}
  \breve{\chi}^{\mbox{\tiny{B}}}{}_{\mbox{\tiny{A}}} + d_1 z + d_1^\prime z \log{z} 
  + O \left( z^2 \right).
\end{equation}
Upon substitution into Eq.~(\ref{Eq:RadialMom}), the $d_1$'s cancel,
leaving $d_1$ indeterminate locally, but
\begin{equation}
  d_1^\prime = - \xi_0^2\breve{\chi}^{\mbox{\tiny{AB}}}{}_
  {\breve{\scriptscriptstyle |}\mbox{\tiny{AB}}}
  - \breve{\chi}^{\mbox{\tiny{A}}}{}_{\mbox{\tiny{B}}} 
  \left[ \breve{\psi}^{\mbox{\tiny{B}}}{}_{\mbox{\tiny{A}}} + 
    \left. \partial_z \left( \Sigma^{\mbox{\tiny{B}}}{}_{\mbox{\tiny{A}}} \right)
    \right|_{\ScriPlus} 
    - \tilde{\kappa}_0 \breve{\chi}^{\mbox{\tiny{B}}}{}_{\mbox{\tiny{A}}} \right].
   \label{Eq:RadialLogCoeff}
\end{equation}
Similarly, the coefficient of $z$ in the expansion of $(K/{3\Omega})
\hat{\tilde{K}}^z{}_{\mbox{\tiny{A}}}$ is indeterminate locally, and one has an 
expansion of the form
\begin{equation}
  \frac{K}{3\Omega} \hat{\tilde{K}}^z{}_{\mbox{\tiny{A}}}
 = \breve{\chi}^{\mbox{\tiny{B}}}{}_{\mbox{\tiny{A}}\breve{\scriptscriptstyle |}\mbox{\tiny{B}}}
 + e_{\mbox{\tiny{A}}} z + e_{\mbox{\tiny{A}}}^\prime z \log{z} + O \left( z^2 \right),
\end{equation}
where the coefficient of $z \log{z}$ is 
\begin{equation}
  \label{Eq:TransverseLogCoeff}
  e_{\mbox{\tiny{A}}}^\prime 
  = - \left( \partial_z \Sigma^{\mbox{\tiny{B}}}{}_{\mbox{\tiny{A}}} 
    - \frac{1}{2} \tilde{\kappa}_0 \breve{\chi}^{\mbox{\tiny{B}}}{}_{\mbox{\tiny{A}}} 
  \right)_{\breve{\scriptscriptstyle |}\mbox{\tiny{B}}}.
\end{equation}
It is also possible to calculate the coefficient $c_3^\prime$ of the
$z^3\log{z}$ term in the expansion for the conformal factor. The
result is $c_3^\prime = -d_1^\prime/8$.

The expansions above allow us, using L'H\^opital's rule, to directly evaluate at 
$\dot{\ScriPlus}$ the coordinate components of the apparently singular terms 
in the evolution equation for $\hat{\tilde K}_{ab}$,
\begin{displaymath}
S_{ij} \equiv \frac{1}{\Omega}\left( \tilde \nabla_i \tilde \nabla_j \Omega 
+ \frac{K} {3}\hat{\tilde K}_{ij} \right)^{TF} .
\end{displaymath}
The result is
\begin{eqnarray}
\label{Eq:RadialDerivHessianOmega}
S^z{}_z &\doteq& \frac{1}{3}
\left[ {\breve{\chi}^{\mbox{\tiny{C}}}{}_{\mbox{\tiny{D}}}
    \breve{\chi}^{\mbox{\tiny{D}}}{}_{\mbox{\tiny{C}}} + 2\xi_0^2 - \tilde \kappa _1 }
\right],
\nonumber \\
S^{\mbox{\tiny{A}}}{}_{\mbox{\tiny{B}}}
&\doteq& \partial _z \left( {\Sigma^{\mbox{\tiny{A}}}{}_{\mbox{\tiny{B}}} } 
\right) - \breve{\psi}^{\mbox{\tiny{A}}}{}_{\mbox{\tiny{B}}}  + 
\frac{1}{2} \tilde{\kappa}_0 \,\breve{\chi}^{\mbox{\tiny{A}}}{}_{\mbox{\tiny{B}}} 
- \frac{1}{6} \left( \breve{\chi}^{\mbox{\tiny{C}}}{}_{\mbox{\tiny{D}}}
  \breve{\chi}^{\mbox{\tiny{D}}}{}_{\mbox{\tiny{C}}}  + 2\xi_0^2 - \tilde{\kappa}_1 
\right)\delta^{\mbox{\tiny{A}}}{} _{\mbox{\tiny{B}}},
\\
S^z{}_{\mbox{\tiny{A}}} &\doteq&
\breve{\chi}^{\mbox{\tiny{C}}}{}_{\mbox{\tiny{A}}
  \breve{\scriptscriptstyle |}{\mbox{\tiny{C}}}}
- \frac{1}{2} \partial _{\mbox{\tiny{A}}} \tilde{\kappa}_0 .
\nonumber
\end{eqnarray}

Finally, we compute the expansion of the conformal lapse
$\tilde{\alpha}$ from the CMC slicing condition,
Eq.~(\ref{Eq:CMCRescaled}):
\begin{equation}
\tilde{\alpha} = \tilde{\alpha}_0\left[ 1 - \frac{1}{2}\tilde{\kappa}_0 z + \frac{1}{4}\left(
 \frac{1}{2}\tilde{\kappa}_0^2 - \tilde{\kappa}_1 
 - 3\breve{\chi}^{\mbox{\tiny{C}}}{}_{\mbox{\tiny{D}}}
  \breve{\chi}^{\mbox{\tiny{D}}}{}_{\mbox{\tiny{C}}}
  + 2\xi_0^2 \right) z^2 + O \left( z^3 \right) \right].
\label{Eq:ConfLapseExpansion}
\end{equation}
In deriving this expansion, we have assumed that $\tilde{\alpha}_0$ is uniform 
as stipulated in Sec.~\ref{Sec:Summary}.

\subsection{The Penrose Regularity Condition}

The apparently singular terms $S_{ij}$ are also related to the
electric part of the physical Weyl tensor evaluated at
$\ScriPlus$. The Penrose regularity condition asserts that the Weyl
tensor of the conformal spacetime should vanish at $\ScriPlus$.  The
conformal invariance of the Weyl tensor implies that the coordinate
components of the electric tensor are invariant,
\[
E_{ij} = C_{i\mu j\nu } \lambda _0^{\,\mu} \lambda _0^{\,\nu} = \tilde
C_{i\mu j\nu } \tilde \lambda _0^{\,\mu} \tilde \lambda _0^{\,\nu} =
\tilde E_{ij} .
\]
A well known expression for the physical Weyl electric tensor in a
vacuum spacetime (see Ref.~\cite{vMoR09}) is
\begin{equation}
  \label{Eq:ElectricTensor}
  E_{ij} = R_{ij} - K_i{}^k K_{kj} + KK_{ij} ,
\end{equation}
where the vacuum evolution equations have been used to eliminate time
derivatives.  Our assumptions about the fall-off of the
energy-momentum tensor at $\ScriPlus$ imply that
Eq. (\ref{Eq:ElectricTensor}) holds there.  In terms of the conformal
quantities, the traceless part of this gives
\[
\hat{\tilde E}_{ij} = \left\{ {R_{ij} - \hat K_i{}^k \hat K_{kj} }
\right\}^{TF} + \frac{K} {3}\hat K_{ij} = \frac{1} {\Omega }\left\{
  {\tilde \nabla _i \tilde \nabla _j \Omega + \frac{K} {3}\hat{\tilde
      K}_{ij} } \right\}^{TF} + \left\{ {\hat{\tilde R}_{ij} -
    \hat{\tilde K}_i{}^k \hat{\tilde K}_{kj} } \right\}^{TF} .
\]
Raising an index with the conformal metric in the Gaussian normal
coordinates, applying the results in
Eqs.~(\ref{Eq:Conformal3DRicci},~\ref{Eq:ConformalRicciScalar},~\ref{Eq:RadialDerivHessianOmega}),
and noting the identity
$\breve{\chi}^{\mbox{\tiny{A}}}{}_{\mbox{\tiny{C}}}
\breve{\chi}^{\mbox{\tiny{C}}}{} _{\mbox{\tiny{B}}} = \frac{1} {2}
\breve{\chi}^{\mbox{\tiny{C}}}{}_{\mbox{\tiny{D}}}
\breve{\chi}^{\mbox{\tiny{D}}}{}_{\mbox{\tiny{C}}}
\delta^{\mbox{\tiny{A}}}{}_{\mbox{\tiny{B}}}$,
\begin{eqnarray}
  \hat{\tilde E}^z{}_z  &\doteq& \frac{1}{3} \left[
      \breve{\chi}^{\mbox{\tiny{C}}}{}_{\mbox{\tiny{D}}} 
      \breve{\chi}^{\mbox{\tiny{D}}}{}_{\mbox{\tiny{C}}}  + 2\xi_0^2 - 
      \tilde \kappa_1 \right] + \frac{1}{3} \left[ \tilde{\kappa}_1  - 
      2\xi_0^2 - 2\breve{\chi}^{\mbox{\tiny{C}}}{}_{\mbox{\tiny{D}}} 
      \breve{\chi}^{\mbox{\tiny{D}}}{}_{\mbox{\tiny{C}}} \right] + \frac{1}{3}
  \breve{\chi}^{\mbox{\tiny{C}}}{}_{\mbox{\tiny{D}}} 
  \breve{\chi}^{\mbox{\tiny{D}}}{}_{\mbox{\tiny{C}}}  \doteq 0,
 \\ 
  \hat{\tilde E}^{\mbox{\tiny{A}}}{}_{\mbox{\tiny{B}}} &\doteq& 
  \partial_z \left( \Sigma^{\mbox{\tiny{A}}}{}_{\mbox{\tiny{B}}} \right)
  - \breve{\psi}^{\mbox{\tiny{A}}}{}_{\mbox{\tiny{B}}} + \frac{1}{2}
  \tilde{\kappa}_0 \,\breve{\chi}^{\mbox{\tiny{A}}}{}_{\mbox{\tiny{B}}}  - 
  \frac{1}{6}\left( \breve{\chi}^{\mbox{\tiny{C}}}{}_{\mbox{\tiny{D}}} 
    \breve{\chi}^{\mbox{\tiny{D}}}{}_{\mbox{\tiny{C}}}  + 2\xi_0^2 - 
    \tilde{\kappa}_1  \right) \delta^{\mbox{\tiny{A}}}{}_{\mbox{\tiny{B}}}
  \nonumber \\ 
  && +\, \breve{\psi}^{\mbox{\tiny{A}}}{}_{\mbox{\tiny{B}}}
  - \tilde{\kappa}_0 \,\breve{\chi}^{\mbox{\tiny{A}}}{}_{\mbox{\tiny{B}}} + 
  \frac{1}{6}\left( 2\xi_0^2 + 2\breve{\chi}^{\mbox{\tiny{C}}}{}_{\mbox{\tiny{D}}} 
    \breve{\chi}^{\mbox{\tiny{D}}}{}_{\mbox{\tiny{C}}} - \tilde{\kappa}_1 \right)
  \delta^{\mbox{\tiny{A}}}{}_{\mbox{\tiny{B}}}  - \frac{1}{6} 
  \breve{\chi}^{\mbox{\tiny{C}}}{}_{\mbox{\tiny{D}}} 
  \breve{\chi}^{\mbox{\tiny{D}}}{}_{\mbox{\tiny{C}}}
  \doteq \partial_z \left( \Sigma^{\mbox{\tiny{A}}}{}_{\mbox{\tiny{B}}} \right) - 
  \frac{1}{2} \tilde{\kappa}_0 \,\breve{\chi}^{\mbox{\tiny{A}}}{}_{\mbox{\tiny{B}}} ,
  \\ 
  \hat{\tilde E}^z{}_{\mbox{\tiny{A}}}  &\doteq& 
  \breve{\chi}^{\mbox{\tiny{C}}}{}_{\mbox{\tiny{A}}
    \breve{\scriptscriptstyle |}\mbox{\tiny{C}}}  - 
  \frac{1}{2} \partial_{\mbox{\tiny{A}}} \tilde{\kappa}_0  + \frac{1}{2} 
  \tilde{\kappa}_{|{\mbox{\tiny{A}}}}  - 
  \breve{\chi}^{\mbox{\tiny{C}}}{}_{\mbox{\tiny{A}}
    \breve{\scriptscriptstyle |}\mbox{\tiny{C}}} \doteq 0. 
\end{eqnarray} 

The Penrose regularity condition is not satisfied automatically once
the zero-shear condition is satisfied; it requires the additional
condition that
\begin{equation}
    \partial _z \left( \Sigma^{\mbox{\tiny{A}}}{}_{\mbox{\tiny{B}}} \right) 
  \doteq \frac{1}{2} \tilde{\kappa}_0 \,
    \breve{\chi}^{\mbox{\tiny{A}}}{}_{\mbox{\tiny{B}}}.
    \label{Eq:PenroseCondition}
\end{equation}
In the Bondi-Sachs asymptotic expansion on null 
hypersurfaces, the Penrose regularity condition is the vanishing 
of the traceless part of the $r^{-2}$ term in the expansion of the 
metric of the two-surfaces of constant Bondi radius $r$ 
(see~\cite{pCmMdS95}).  In that context 
the condition is preserved by the Einstein equations, once imposed 
in the initial data, and therefore  
the same must be true for Eq.~(\ref{Eq:PenroseCondition}).

The magnetic part of the Weyl tensor is
\[
B_{ij} = \tilde B_{ij} = \tilde \nabla_k \hat{\tilde K}_{\ell(i} \tilde
\varepsilon_{j)}{}^{k\ell}
\]
and using the asymptotic expansions above, one finds
\[
B^z{}_z \doteq 0,\qquad B^z{}_{\mbox{\tiny{A}}} \doteq 0,\qquad
B^{\mbox{\tiny{A}}}{}_{\mbox{\tiny{B}}} \doteq
-\breve{\varepsilon}^{\mbox{\tiny{A}}}{}_{\mbox{\tiny{C}}}
\left[ \partial_z \left( \Sigma^{\mbox{\tiny{C}}}{}_{\mbox{\tiny{B}}}
  \right) - \frac{1}{2} \tilde{\kappa}_0
  \,\breve{\chi}^{\mbox{\tiny{C}}}{}_{\mbox{\tiny{B}}} \right],
\]
with $B^{\mbox{\tiny{A}}}{}_{\mbox{\tiny{B}}} \doteq 0$ if and only if Eq.~(\ref{Eq:PenroseCondition})
holds. Therefore, the magnetic part of the Weyl tensor does not give
additional conditions.

Given that the Weyl tensor vanishes at $\ScriPlus$, much simpler
expressions can be obtained for the apparently singular terms in the
evolution equations than those of
Eq.~(\ref{Eq:RadialDerivHessianOmega}) directly in terms of tetrad
quantities in a general coordinate system, without the need to
propagate the normal geodesics away from $\dot{\ScriPlus}$ to
calculate $\tilde{\kappa}_{ab} $, etc.  One just uses the vanishing of
the tetrad components of the electric part of the Weyl tensor. The
regular form of the evolution equation at $\ScriPlus$ for
$\hat{\tilde{K}}_{ab}$ is given in Eq.~(\ref{Eq:PenroseBdaryEvol}).

\subsection{Metric Evolution Equations}

It is instructive to examine the asymptotic behavior of the evolution
equations for the conformal spatial metric in terms of the conformal
extrinsic curvature assuming the initial Gaussian normal coordinate
condition is preserved.  The general equations are
\[
\partial_t \tilde{h}_{ij} 
 = \beta^k \partial_k \tilde{h}_{ij} + \tilde{h}_{kj} \partial_i\beta^k 
 + \tilde{h}_{ik} \partial_j \beta^k 
 + 2 \tilde{\alpha} \left( \hat{\tilde{K}}_{ij} + \frac{1}{3} \tilde{h}_{ij} \tilde{K} \right).
\]
The evolution equation for $\tilde{h}_{zz}$ becomes an equation for $\beta^z$,
\[
    \partial_z {\beta}^z =  - \tilde{\alpha} \left( \hat{\tilde{K}}_{zz} +
    \frac{1}{3} \tilde{K} \right),
\]
from which
\begin{equation}
  \beta^z = \tilde{\alpha}_0 -  \left( \frac{\tilde{\alpha}\tilde{K}}{3} \right)_0 z - 
  \frac{1}{2} \left[ \tilde{\alpha}_0 \breve{\chi}^{\mbox{\tiny{C}}}{}_{\mbox{\tiny{D}}}  
    \breve{\chi}^{\mbox{\tiny{D}}}{}_{\mbox{\tiny{C}}}
    + \left( \frac{\tilde{\alpha} \tilde{K}}{3} \right)_1 \right] z^2 + O(z^3) .
     \label{Eq:betazshiftGNC}
\end{equation}
The condition for keeping $\tilde{h}_{\mbox{\tiny{A}}z} = 0$ is
\[
  \partial_z \beta^{\mbox{\tiny{A}}} + \tilde{h}^{\mbox{\tiny{AC}}} 
  \partial_{\mbox{\tiny{C}}} \beta^z = - 2 \tilde{\alpha} 
  \hat{\tilde{K}}^{\mbox{\tiny{A}}}{}_z,
\]
from which 
\begin{equation}
  \beta^{\mbox{\tiny{A}}} = \xi_0^2 \left[ 
    \breve{h}^{\mbox{\tiny{AB}}}\partial_{\mbox{\tiny{B}}}
    \left(\frac{\tilde{\alpha}\tilde{K}}{6} \right)_0
  - \tilde{\alpha}_0 
    \breve{\chi}^{\mbox{\tiny{AB}}}{}_{\breve{\scriptscriptstyle |}
      \mbox{\tiny{B}}} \right] z^2 + O(z^3). 
    \label{Eq:betaAshiftGNC}
\end{equation}
The boundary conditions on the shift and the lapse are from 
Sec.~\ref{Sec:Summary}.

Decompose the evolution equations for the angular part of the metric
as given by Eq.~(\ref{Eq:hdecomp}) into trace and traceless parts.
Defining the evolution of $\xi$ so Eq.~(\ref{Eq:hdeterminant}) is preserved,
the trace in zeroth order gives
\begin{equation}
  \frac{1}{\xi} \partial_t \xi \doteq \frac{\beta^z}{\xi} 
  \partial_z \xi - \frac{1}{3} \tilde{\alpha} \tilde{K} 
  \doteq \tilde{\alpha}_0 \left( \frac{1}{2} \tilde{\kappa}_0 - \frac{1}{3} \tilde{K} \right) ,
  \label{Eq:xievolution}
\end{equation}
using $\tilde{\kappa} = 2 \partial_z \log{\xi}$. The traceless part in
zeroth order, given the zero-shear condition, confirms that
$\partial_t \breve{h}_{\mbox{\tiny{AB}}} \doteq 0$, as assumed.

The condition for keeping the intrinsic geometry of
$\ScriPlus$ a true sphere, i.e., keeping $\xi$ uniform on $\ScriPlus$,
 is $\partial_t \xi \doteq 0$.  From Eq.~(\ref{Eq:xievolution}), 
 we see this requires the boundary condition on $\tilde{K}$ to be 
\begin{equation}
  \frac{2\tilde{\alpha} \tilde{K}}{3} \doteq \tilde{\alpha} \tilde{\kappa} ,
  \label{Eq:BCKtilde}
\end{equation}
which can be implemented in the elliptic Nester gauge equations of 
Sec.~\ref{SubSec:Ktilde}.  A significant consequence of this condition 
is that $\tilde{\alpha} - \beta^z = O(z^2)$.

At first order in $z$, and using the Gaussian-normal-coordinate shift,
one finds from the trace
\begin{equation}
  \tilde{D}_0 \tilde{\kappa} \doteq \frac{1}{\tilde{\alpha}_0} 
  \left(  \partial_t \tilde{\kappa}_0 - \tilde{\alpha}_0 \tilde{\kappa}_1 \right)
  \doteq  - \frac{1}{2} \tilde{\kappa}_0^2 + 
  \breve{\chi}^{\mbox{\tiny{CD}}} \breve{\chi}_{\mbox{\tiny{CD}}} 
  - \frac{2}{3\tilde{\alpha}_0} \partial_z (\tilde{\alpha} \tilde{K}) 
\label{Eq:coordkappaevol}
\end{equation}
and from the traceless part 
\begin{equation}
\partial_t\breve{\chi}^{\mbox{\tiny{A}}}{}_{\mbox{\tiny{B}}}
   \doteq \tilde{\alpha}_0 \left( \breve{\psi}^{\mbox{\tiny{A}}}{}_{\mbox{\tiny{B}}}
    - \partial_z \Sigma^{\mbox{\tiny{A}}}{}_{\mbox{\tiny{B}}} \right).
\label{Eq:coordchiEvolScri}
\end{equation}
After imposing the Penrose regularity condition, Eq.~(\ref{Eq:PenroseCondition}),
\begin{equation}
\partial_t\breve{\chi}^{\mbox{\tiny{A}}}{}_{\mbox{\tiny{B}}} 
\doteq \tilde{\alpha}_0 \left( \breve{\psi}^{\mbox{\tiny{A}}}{}_{\mbox{\tiny{B}}} 
    - \frac{1}{2} \tilde \kappa_0 \breve{\chi}^{\mbox{\tiny{A}}}{}_{\mbox{\tiny{B}}} \right) .
  \label{Eq:ChiEvolScri}
\end{equation}

Next, we examine the evolution equation for the conformally rescaled
extrinsic curvature,
\[
\partial_t\hat{\tilde K}^i{}_j = \beta^k \partial_k\hat{\tilde
  K}^i{}_j - \hat{\tilde K}^k{}_j\partial_k\beta^i + \hat{\tilde
  K}^i{}_k \partial_j\beta^k + \tilde\alpha \left( -\tilde R^i{}_j +
  \frac{1}{\tilde\alpha}\tilde\nabla^i\tilde\nabla_j\tilde \alpha -
  \frac{\tilde K}{3}\hat{\tilde K}^i{}_j - 2 S^i{}_j + 8\pi
  G\Omega^3\tilde{\sigma}^i{}_j \right)^{TF}.
\]
Evaluating the right-hand side of this equation at $\ScriPlus$ we
obtain, using
Eqs.~(\ref{Eq:Conformal3DRicci},~\ref{Eq:KzzatScri},~\ref{Eq:KAzatScri},~\ref{Eq:RadialDerivHessianOmega},~\ref{Eq:ConfLapseExpansion},~\ref{Eq:betazshiftGNC},~\ref{Eq:BCKtilde}),
\begin{equation}
  \partial_t\hat{\tilde K}^z{}_z \doteq 0,\qquad
  \partial_t\hat{\tilde K}^z{}_{\mbox{\tiny{A}}} \doteq 0,\qquad 
  \partial_t\hat{\tilde K}^{\mbox{\tiny{A}}}{}_{\mbox{\tiny{B}}} 
  \doteq \tilde\alpha_0\left(
    \breve{\psi}^{\mbox{\tiny{A}}}{}_{\mbox{\tiny{B}}} 
    - \partial_z \Sigma^{\mbox{\tiny{A}}}{}_{\mbox{\tiny{B}}}
  \right).
\end{equation}
In view of the Eq.~(\ref{Eq:coordchiEvolScri}), this shows that the
regularity conditions $\hat{\tilde K}^z{}_z \doteq 0$, $\hat{\tilde
  K}^z{}_{\mbox{\tiny{A}}} \doteq 0$, $\hat{\tilde
  K}^{\mbox{\tiny{A}}}{}_{\mbox{\tiny{B}}} \doteq \hat{\tilde
  \kappa}^{\mbox{\tiny{A}}}{}_{\mbox{\tiny{B}}}$ are automatically
preserved in the time evolution.

Now return to the question of the apparent breakdown in smoothness at
$\ScriPlus$ in the asymptotic expansion of the extrinsic curvature.
Note that from Eq.~(\ref{Eq:TransverseLogCoeff}), the condition for
the coefficient $e_1^\prime$ of the leading log term in
$\hat{\tilde{K}}^z{}_{\mbox{\tiny{A}}}$ to vanish is identical to the
Penrose regularity condition, Eq.~(\ref{Eq:PenroseCondition}).
However, applying the Penrose
regularity condition to Eq.~(\ref{Eq:RadialLogCoeff}) for the
coefficient $d_1^\prime$ of the leading log term in
$\hat{\tilde{K}}^z{}_z$ leaves
\begin{eqnarray}
  d_1^\prime &=& -\xi_0^2\breve{\chi}^{\mbox{\tiny{AB}}}{}_
  {\breve{\scriptscriptstyle |}\mbox{\tiny{AB}}}
  - \breve{\chi}^{\mbox{\tiny{A}}}{}_{\mbox{\tiny{B}}} \left[ 
    \breve{\psi}^{\mbox{\tiny{B}}}{}_{\mbox{\tiny{A}}} - \frac{1}{2}
    \tilde{\kappa}_0 \breve{\chi}^{\mbox{\tiny{B}}}{}_{\mbox{\tiny{A}}} \right]
  \nonumber\\
  &=& -\xi_0^2\breve{\chi}^{\mbox{\tiny{AB}}}{}
  _{\breve{\scriptscriptstyle |}\mbox{\tiny{AB}}} 
  - \frac{1}{2 \tilde{\alpha}} \partial_t \left( 
    \breve{\chi}^{\mbox{\tiny{A}}}{}_{\mbox{\tiny{B}}} 
    \breve{\chi}^{\mbox{\tiny{B}}}{}_{\mbox{\tiny{A}}} \right) ,
\end{eqnarray}
which does not vanish if any {\em outgoing} waves are present at
$\ScriPlus$. Therefore, the lack of smoothness of
$\hat{\tilde{K}}^z{}_z$ at $\ScriPlus$ cannot be prevented by a
regularity condition and must be dealt with in
any calculation involving emission of gravitational
radiation. Friedrich's proofs of smoothness at $\ScriPlus$ given
smooth initial data~\cite{hF83, hF86a} do not apply because they are
based on solving a complete symmetric hyperbolic system of evolution
equations. Our assumption of CMC hypersurfaces requires an elliptic
equation for the lapse, and it is this elliptic equation which
generates a breakdown in smoothness in the presence of outgoing
radiation. The elliptic equations which enforce the Nester gauge are
not relevant because they do not affect the evolution of the
coordinate metric. Our mathematical results on the presence of
polyhomogeneous terms in the asymptotic expansion of the extrinsic
curvature and conformal factor, after sorting through differences in
notation, are equivalent to those of Sec.~4 of Andersson and
Chru\'sciel~\cite{lApC94}.

We do not expect the lack of smoothness at $\ScriPlus$ to be a serious
impediment to the accuracy and/or stability of numerical calculations,
even though in principle it could prevent rapid convergence of
spectral-based numerical methods.

\section{Constraint Propagation}
\label{App-ConstraintPropagation}

In this appendix, we derive the constraint propagation system for our
evolution scheme. This system describes the evolution of constraint
violations which could be present in the initial data or, more
generally, could be triggered by truncation errors in a numerical
scheme.

Since the constraint equation~(\ref{Eq:CurlBRescaled}) is required for
vanishing torsion, we should allow for a connection $\nabla$ with
non-zero torsion tensor $T$ in the constraint propagation system. An
efficient method for deriving this system \cite{hFgN99} is based on
the Riemann and Bianchi identities for the curvature tensor $R$,
\begin{eqnarray}
\sum\limits_{(\beta\gamma\delta)} R^\alpha{}_{\beta\gamma\delta} 
 &=& \sum\limits_{(\beta\gamma\delta)} 
\left( \nabla_\beta T^\alpha{}_{\gamma\delta} 
 + T^\alpha{}_{\epsilon\delta} T^\epsilon{}_{\beta\gamma} \right),
\label{Eq:FirstBianchiID}\\
\sum\limits_{(\beta\gamma\delta)} \nabla_\beta R^\alpha{}_{\epsilon\gamma\delta} 
 &=& -\sum\limits_{(\beta\gamma\delta)} 
 R^\alpha{}_{\epsilon\kappa\delta} T^\kappa{}_{\beta\gamma}\, ,
\label{Eq:SecondBianchiID}
\end{eqnarray}
where $\sum_{(\beta\gamma\delta)}$ denotes the cyclic sum over
$(\beta\gamma\delta)$.  Expressed in terms of an arbitrary basis ${\bf
  e}_\alpha$ of vector fields, the torsion and curvature tensors are
defined by
\begin{eqnarray}
T^\alpha{}_{\gamma\delta} 
 &\equiv& -[{\bf e}_\gamma, {\bf e}_\delta]^\alpha 
 - 2\Gamma^\alpha{}_{[\gamma\delta]}\, ,
\nonumber\\
R^\alpha{}_{\beta\gamma\delta}
 &\equiv& e_\gamma(\Gamma^\alpha{}_{\beta\delta}) 
      - e_\delta(\Gamma^\alpha{}_{\beta\gamma})
   + \Gamma^\epsilon{}_{\beta\delta}\Gamma^\alpha{}_{\epsilon\gamma}
   - \Gamma^\epsilon{}_{\beta\gamma}\Gamma^\alpha{}_{\epsilon\delta}
   - \Gamma^\alpha{}_{\beta\epsilon}[{\bf e}_\gamma,{\bf e}_\delta]^\epsilon,
\nonumber
\end{eqnarray}
respectively. If the frame ${\bf e}_\alpha$ is orthonormal, Einstein's
field equations are equivalent to the requirements that the connection
coefficients $\Gamma_{\alpha\beta\gamma} =
\Gamma_{[\alpha\beta]\gamma}$ are antisymmetric in $\alpha\beta$, that
the torsion tensor is zero, and that the Ricci tensor
$R_{\alpha\beta}\equiv R^\gamma{}_{\alpha\gamma\beta}$ is related to
the stress-energy tensor $\tau_{\alpha\beta}$ according to
$R_{\alpha\beta} = 8\pi G\tau_{\alpha\beta} - 4\pi
G\eta_{\alpha\beta}\eta^{\gamma\delta}\tau_{\gamma\delta}$. An
equivalent way of stating this is the satisfaction of the equations
\begin{equation}
T^\alpha{}_{\gamma\delta} = 0,\qquad
\Delta_{\alpha\beta\gamma\delta} = 0,
\label{Eq:FieldEquations}
\end{equation}
where
\begin{displaymath}
\Delta_{\alpha\beta\gamma\delta} 
 \equiv R_{\alpha\beta\gamma\delta} 
 - \Gamma_{\alpha\beta\epsilon} T^\epsilon{}_{\gamma\delta} 
 - C_{\alpha\beta\gamma\delta}
 - S_{\alpha\beta\gamma\delta},\qquad
S_{\alpha\beta\gamma\delta} \equiv 8\pi G\left( \eta_{\alpha[\gamma}\tau_{\delta]\beta} 
- \eta_{\beta[\gamma}\tau_{\delta]\alpha} 
- \frac{2}{3} \eta_{\alpha[\gamma}\eta_{\delta]\beta}
 \eta^{\epsilon\kappa}\tau_{\epsilon\kappa} \right),
\end{displaymath}
and $C_{\alpha\beta\gamma\delta}$ is a tensor field satisfying the
algebraic symmetries $C_{[\alpha\beta\gamma]\delta} = 0$,
$C_{[\alpha\beta][\gamma\delta]} = C_{\alpha\beta\gamma\delta} =
C_{\gamma\delta\alpha\beta}$, $\eta^{\alpha\gamma}
C_{\alpha\beta\gamma\delta} = 0$. For a solution of the field
equations~(\ref{Eq:FieldEquations}), $C_{\alpha\beta\gamma\delta}$ is
the Weyl curvature tensor associated with the geometry.

In order to relate Eq.~(\ref{Eq:FieldEquations}) to our evolution and
constraint equations, we assume that the time-like leg ${\bf e}_0 =
{\bf n}$ coincides with the future-directed unit normal to the time
slices $\Sigma_t$. This implies that $T^0{}_{\gamma\delta} = 0$. We
decompose the remaining components of $T$ and $\Delta$ according to
\begin{eqnarray}
T_{\alpha\gamma\delta} &=& 2 F_{\alpha[\gamma} n_{\delta]}
 + G_{\alpha\beta}\varepsilon^\beta{}_{\gamma\delta},
\nonumber\\ 
\Delta_{\alpha\beta\gamma\delta} &=& -4 n_{[\alpha} {\cal E}_{\beta][\gamma} n_{\delta]} 
 - \varepsilon_{\alpha\beta}{}^{\iota} {\cal D}_{\iota\kappa} \varepsilon^\kappa{}_{\gamma\delta}
 - 2n_{[\alpha} {\cal B}_{\beta]\iota} \varepsilon^\iota{}_{\gamma\delta} 
 + 2\varepsilon_{\alpha\beta}{}^{\iota} {\cal H}_{\iota[\gamma} n_{\delta]},
\nonumber
\end{eqnarray}
where $\varepsilon_{\beta\gamma\delta}\equiv
n^\alpha\varepsilon_{\alpha\beta\gamma\delta}$ and the covariant
tensor fields ${\bf F}$, ${\bf G}$, ${\bf\cal E}$, ${\bf\cal D}$, ${\bf\cal B}$
and ${\bf\cal H}$ are orthogonal to ${\bf n}$. The left and right duals of
$\Delta$ are defined by
\begin{displaymath}
(*\Delta)_{\alpha\beta\gamma\delta} \equiv \frac{1}{2}\varepsilon_{\alpha\beta}{}^{\epsilon\kappa}\Delta_{\epsilon\kappa\gamma\delta},\qquad
(\Delta*)_{\alpha\beta\gamma\delta} \equiv \frac{1}{2}\Delta_{\alpha\beta\epsilon\kappa}
\varepsilon^{\epsilon\kappa}{}_{\gamma\delta},
\end{displaymath}
respectively, and induce the following transformations in the decomposition of $\Delta$:
\begin{eqnarray}
\Delta \mapsto *\Delta &:&
 {\bf\cal E}\mapsto {\bf\cal H},\quad
 {\bf\cal D}\mapsto {\bf\cal B},\quad
 {\bf\cal B}\mapsto -{\bf\cal D},\quad
 {\bf\cal H}\mapsto -{\bf\cal E},
\label{Eq:LeftDualTrans}\\
\Delta \mapsto \Delta* &:&
 {\bf\cal E}\mapsto {\bf\cal B},\quad
 {\bf\cal D}\mapsto {\bf\cal H},\quad
 {\bf\cal B}\mapsto -{\bf\cal E},\quad
 {\bf\cal H}\mapsto -{\bf\cal D}.
\label{Eq:RightDualTrans}
\end{eqnarray}
The trace of $\Delta$ is given by
\begin{eqnarray}
\Delta^{\gamma}{}_{\alpha\gamma\beta} 
 &=& R_{\alpha\beta} + \Gamma_\alpha{}^{\gamma\delta} T_{\delta\gamma\beta}
  - 8\pi G\left( \tau_{\alpha\beta} - \frac{1}{2}\eta_{\alpha\beta}\eta^{\gamma\delta}
  \tau_{\gamma\delta} \right)
\nonumber\\
 &=& n_\alpha n_\beta {\cal E}^\gamma{}_\gamma 
 - n_\alpha\varepsilon_\beta{}^{\gamma\delta} {\cal B}_{\gamma\delta}
 + \varepsilon_\alpha{}^{\gamma\delta} {\cal H}_{\gamma\delta} n_\beta
 - {\cal E}_{\alpha\beta} + {\cal D}_{\beta\alpha}
  - (\eta_{\alpha\beta} + n_\alpha n_\beta){\cal D}^\gamma{}_\gamma.
\label{Eq:TraceDelta}
\end{eqnarray}
The Weyl tensor has a similar decomposition to that of $\Delta$. However,
its algebraic symmetries imply that ${\bf\cal D}={\bf\cal E}$, ${\bf\cal H}={\bf\cal B}$, and that ${\bf\cal E}$ and ${\bf\cal B}$ are symmetric and traceless.

Our evolution system is obtained from the combinations of the field
equations~(\ref{Eq:FieldEquations}) which eliminate the Weyl
tensor. More precisely, the evolution
equations~(\ref{Eq:OrigBakEvoln},~\ref{Eq:OrigKabEvoln},~\ref{Eq:OrigNabEvoln})
are
\begin{eqnarray}
0 &=& F_{ab} B_a{}^k = -D_0 B_b{}^k + l.o.,
\nonumber\\
0 &=& \mu_{ab} \equiv {\cal E}_{ab} - {\cal D}_{ab} 
= -D_0 K_{ba} + \varepsilon_{bcd} D_c N_{da} + l.o.,
\nonumber\\
0 &=& \nu_{ab} \equiv {\cal B}_{ab} - {\cal H}_{ab}
 = -D_0 N_{ba} - \varepsilon_{bcd} D_c K_{da} + l.o.,
\nonumber
\end{eqnarray}
where``$l.o.$'' stands for lower order terms which do not contain
derivatives of $B_a{}^k$, $K_{ab}$, $N_{ab}$, $a_b$, $\omega_b$ or
$\tau_{\alpha\beta}$. The constraint
equations~(\ref{Eq:OrigBakConstraint},~\ref{Eq:OrigHamConstraint},~\ref{Eq:OrigMomConstraint},~\ref{Eq:OrigNabConstraint})
are equivalent to
\begin{eqnarray}
0 &=& G_{ab} B_a{}^k = -\varepsilon_{bcd} D_c B_d{}^k + l.o.,
\nonumber\\
0 &=& M \equiv -{\cal D}_{aa} = 2D_a n_a + l.o.,
\nonumber\\
0 &=& P_a \equiv \varepsilon_{abc} {\cal B}_{bc} =  D_b K_{ab} - D_a K + l.o.,
 \nonumber\\
0 &=& Q_a \equiv \varepsilon_{abc} {\cal D}_{bc} = D_b N_{ab} - D_a N + l.o.,
\nonumber
\end{eqnarray}
and the equation ${\cal B}_{aa} = 0$ is automatically satisfied as a
consequence of the hypersurface orthogonal gauge.

Now we use the
identities~(\ref{Eq:FirstBianchiID},~\ref{Eq:SecondBianchiID}) to
derive evolution equations for the constraint variables $G_{ab}$, $M$,
$P_a$ and $Q_a$. For this, we first rewrite them in terms of $T$ and
$\Delta$. Using the algebraic symmetries of
$C_{\alpha\beta\gamma\delta}$ and the fact that
$\sum_{(\beta\gamma\delta)} S_{\alpha\beta\gamma\delta} = 0$, this
yields
\begin{eqnarray}
\sum\limits_{(\beta\gamma\delta)} \nabla_\beta T_{\alpha\gamma\delta} 
&=& \sum\limits_{(\beta\gamma\delta)} \left[ \Delta_{\alpha\beta\gamma\delta} 
 + \left( {\bf e}_\alpha\cdot [{\bf e}_\epsilon,{\bf e}_\beta]
 + \Gamma_{\alpha\epsilon\beta} \right) T^\epsilon{}_{\gamma\delta} \right],
\label{Eq:FirstID}\\
\sum\limits_{(\beta\gamma\delta)} \nabla_\beta 
(\Delta_{\alpha\epsilon\gamma\delta} + C_{\alpha\epsilon\gamma\delta} 
 + S_{\alpha\epsilon\gamma\delta})
 &=& \sum\limits_{(\beta\gamma\delta)} \left[
 -\Gamma_{\alpha\epsilon\kappa}\Delta^{\kappa}{}_{\beta\gamma\delta}
 - {\bf e}_k(\Gamma_{\alpha\epsilon\beta})T^\kappa{}_{\gamma\delta} \right].
\label{Eq:SecondID}
\end{eqnarray}
We may further simplify these equations by noticing that
\begin{displaymath}
\sum\limits_{(\beta\gamma\delta)}\nabla_\beta S_{\alpha\epsilon\gamma\delta}
= \frac{1}{2}\varepsilon_{\beta\gamma\delta\iota}
   \varepsilon^{\iota\beta'\gamma'\delta'}\nabla_{\beta'} S_{\alpha\epsilon\gamma'\delta'}
 = \varepsilon_{\beta\gamma\delta}{}^{\iota}\nabla^\kappa (S*)_{\alpha\epsilon\iota\kappa},
\qquad
\sum\limits_{(\beta\gamma\delta)} \Delta^{\kappa}{}_{\beta\gamma\delta}
 = \varepsilon_{\beta\gamma\delta\alpha}(\Delta*)^{\kappa\epsilon\alpha}{}_\epsilon,
\end{displaymath}
where $(S*)_{\alpha\beta\gamma\delta} = 8\pi
G(\varepsilon_{\gamma\delta[\alpha}{}^\kappa\tau_{\beta]\kappa} -
\varepsilon_{\alpha\beta\gamma\delta}\eta^{\iota\kappa}\tau_{\iota\kappa}/3)$. Using
the expression~(\ref{Eq:TraceDelta}) for the trace of $\Delta$ and the
transformation properties (\ref{Eq:RightDualTrans}) of the right dual,
we find
\begin{displaymath}
(\Delta*)^{0\epsilon}{}_{0\epsilon} = 0,\quad
(\Delta*)^{0\epsilon}{}_{b\epsilon} = Q_b + \varepsilon_{bcd}\mu_{cd},\quad
(\Delta*)^{\epsilon}{}_{a\epsilon 0} = Q_a,\quad
(\Delta*)^{\epsilon}{}_{a\epsilon b} = -\varepsilon_{abc} P_c - \nu_{ba} 
+ \delta_{ab}\nu_{cc}.
\end{displaymath}
As a consequence, the right-hand sides of
Eqs.~(\ref{Eq:FirstID},~\ref{Eq:SecondID}) are a linear combination of
the constraint fields $G_{ab}$, $P_a$, $Q_a$ and the evolution
variables $F_{ab}$, $\mu_{ab}$, $\nu_{ab}$.

Eq.~(\ref{Eq:FirstID}) with $F_{ab}=0$ yields the following identities:
\begin{eqnarray}
&& \varepsilon_{bcd}\mu_{cd} + Q_b - a_a G_{ab} = 0,
\label{Eq:FirstId1}\\
&& \left[ (D_b - 2n_b)G_{ab} \right] D_a = Q_a D_a + G_{ab}[D_a,D_b],
\label{Eq:FirstId2}\\
&& D_0 G_{ab} = -\varepsilon_{abc} P_c - \nu_{ba} + \delta_{ab}\nu_{cc}
 + K_{ac} G_{cb} + G_{ac} K_{cb} - K G_{ab} 
 + \varepsilon_{acd} G_{cb}\omega_d + G_{ac}\varepsilon_{bcd}\omega_d.
\label{Eq:GabEvoln}
\end{eqnarray}
The first identity is an algebraic relation between the antisymmetric
part of the evolution equation $\mu_{ab}=0$ and the constraint fields
$Q_a$ and $G_{ab}$. It can also be obtained directly by computing the
antisymmetric part of Eq.~(\ref{Eq:OrigKabEvoln}). Upon using the
commutator relation $\varepsilon_{bcd} D_c D_d = -(G_{ab} + N_{ab} -
\delta_{ab} N) D_a$, the second identity yields a differential relation
between the constraint fields $Q_a$ and $G_{ab}$. The third identity
gives an evolution equation for $G_{ab}$.

In order to obtain evolution equations for the remaining constraint
fields $M$, $P_a$ and $Q_a$, we contract Eq.~(\ref{Eq:SecondID}) and
its left dual over $\alpha\beta$ and then over
$\delta\epsilon$. Assuming that the stress energy tensor is
divergence-free, this yields the twice contracted Bianchi identities
\begin{eqnarray}
&& \nabla_\alpha\Delta^{\gamma\alpha}{}_{\gamma\beta}
 - \frac{1}{2}\nabla_\beta\Delta^{\gamma\delta}{}_{\gamma\delta} 
 = *\Gamma_\beta{}^{\gamma\delta}(\Delta*)^\epsilon{}_{\delta\epsilon\gamma}
 - {\bf e}_\delta(\Gamma^{\alpha\gamma}{}_\gamma) T^\delta{}_{\alpha\beta}
 + \frac{1}{2}{\bf e}_\delta(\Gamma^{\alpha\gamma}{}_\beta) T^\delta{}_{\alpha\gamma},
\label{Eq:TwiceContractedBianchiID}\\
&& \nabla_\alpha(*\Delta)^{\gamma\alpha}{}_{\gamma\beta}
 - \frac{1}{2}\nabla_\beta(*\Delta)^{\gamma\delta}{}_{\gamma\delta} 
 = -\Gamma_\beta{}^{\gamma\delta}(\Delta*)^\epsilon{}_{\delta\epsilon\gamma}
 - {\bf e}_\delta(*\Gamma^{\alpha\gamma}{}_\gamma) T^\delta{}_{\alpha\beta}
 + \frac{1}{2}{\bf e}_\delta(*\Gamma^{\alpha\gamma}{}_\beta) T^\delta{}_{\alpha\gamma},
\label{Eq:TwiceContractedBianchiIDDual}
\end{eqnarray}
where $*\Gamma_{\alpha\beta\kappa} \equiv
\varepsilon_{\alpha\beta\gamma\delta}\Gamma^{\gamma\delta}{}_\kappa/2$. Using
the expression~(\ref{Eq:TraceDelta}) for the trace of $\Delta$ and its
left dual, we find, in terms of the constraint and evolution
variables,
\begin{eqnarray}
&& \Delta^{\gamma 0}{}_{\gamma0} = M - \mu_{aa},\quad
\Delta^{\gamma 0}{}_{\gamma b} = -P_b,\quad
\Delta^{\gamma}{}_{a\gamma 0} = -P_a + \varepsilon_{acd}\nu_{cd},\quad
\Delta^{\gamma}{}_{a\gamma b} = -\varepsilon_{abc} Q_c + \delta_{ab} M - \mu_{ab},
\nonumber\\
&& (*\Delta)^{\gamma 0}{}_{\gamma0} = \nu_{aa},\quad
(*\Delta)^{\gamma 0}{}_{\gamma b} = Q_b,\quad
(*\Delta)^{\gamma}{}_{a\gamma 0} = Q_a + \varepsilon_{acd}\mu_{cd},\quad
(*\Delta)^{\gamma}{}_{a\gamma b} = -\varepsilon_{abc} P_c + \nu_{ab}.
\nonumber
\end{eqnarray}
By virtue of Eqs.~(\ref{Eq:FirstId1},~\ref{Eq:FirstId2}) it turns out
that the zero component of Eq.~(\ref{Eq:TwiceContractedBianchiIDDual})
is satisfied identically, while
Eq.~(\ref{Eq:TwiceContractedBianchiID}) and the spatial components of
Eq.~(\ref{Eq:TwiceContractedBianchiIDDual}) give
\begin{eqnarray}
D_0 M + D_a P_a &=& \varepsilon_{bcd} D_b\nu_{cd} + l.o.,
\label{Eq:HamEvoln}\\
D_0 P_b - \varepsilon_{bcd} D_c Q_d + D_b M &=& -D_a\mu_{ab} + D_b\mu_{aa} + l.o.
\label{Eq:MomEvoln}\\
D_0 Q_b + \varepsilon_{bcd} D_c P_d &=& -D_a\nu_{ab} + D_b\nu_{aa} + l.o.
\label{Eq:QbEvoln}
\end{eqnarray}
The constraint propagation system for the original evolution
system, Eqs.~(\ref{Eq:OrigBakEvoln},~\ref{Eq:OrigKabEvoln},~\ref{Eq:OrigNabEvoln}),
is given by
Eqs.~(\ref{Eq:GabEvoln},~\ref{Eq:HamEvoln},~\ref{Eq:MomEvoln},~\ref{Eq:QbEvoln})
where the evolution equations $\mu_{ab} = \nu_{ab} = 0$ are
imposed. The lower order terms ``$l.o.$'' depend linearly on the
constraint fields $G_{ab}$, $M$, $P_b$ and $Q_b$, but not on their
derivatives. Therefore, the constraint propagation equations form a
linear, first-order symmetric hyperbolic system, and it follows that
the constraints are preserved by the evolution
system~(\ref{Eq:OrigBakEvoln},~\ref{Eq:OrigKabEvoln},~\ref{Eq:OrigNabEvoln}).

In the CMC and 3D Nester gauges, the evolution equations we impose are
slightly different. In order to see this, we decompose
\begin{displaymath}
\mu_{ab} = \hat{\mu}_{ab} + \varepsilon_{abc}\mu_c + \frac{\delta_{ab}}{3}\mu,\qquad
\nu_{ab} = \hat{\nu}_{ab} + \varepsilon_{abc}\nu_c + \frac{\delta_{ab}}{3}\nu,
\end{displaymath}
where $\hat{\mu}_{ab}$ and $\hat{\nu}_{ab}$ denote the symmetric,
trace-free parts of $\mu_{ab}$ and $\nu_{ab}$, respectively. Then, the
equations $\hat{\mu}_{ab} = \hat{\nu}_{ab} = 0$ yield the evolution
equations
(\ref{Eq:EvolnHatKab},~\ref{Eq:EvolnHatNab}). Eq.~(\ref{Eq:EvolnK}),
which yields the elliptic equation for the lapse, is equivalent to
$\mu = M$. However, when writing this equation in terms of the
conformally rescaled fields, the Hamiltonian constraint is used again to
eliminate the most singular terms at $\ScriPlus$. This amounts to
setting $\mu = 0$. Next, 
Eqs.~(\ref{Eq:EvolnNSubstHamCon},~\ref{Eq:EvolnLittlenSubstMomCon}),
which are used to derive the elliptic system for preserving the 3D
Nester gauge, are obtained from $\nu = 0$ and $2\nu_a = P_a$,
respectively. Using this together with $2\mu_b = -Q_b + a_a G_{ab}$
and Eq.~(\ref{Eq:FirstId2}), the constraint propagation system yields
\begin{eqnarray}
D_0 G_{ab} &=& l.o.,
\label{Eq:GabEvolnBis}\\
D_0 M  &=& l.o.,
\label{Eq:HamEvolnBis}\\
D_0 P_b - \frac{1}{2}\varepsilon_{bcd} D_c\bar{Q}_d + D_b M &=&  l.o.,
\label{Eq:MomEvolnBis}\\
D_0\bar{Q}_b + \frac{1}{2}\varepsilon_{bcd} D_c P_d &=& l.o.,
\label{Eq:QbEvolnBis}
\end{eqnarray}
where $\bar{Q}_b \equiv Q_b + a_a G_{ab}$. This system is weakly
hyperbolic. However, if the Hamiltonian constraint $M=0$ is imposed,
then the constraint evolution system is described by the
Eqs.~(\ref{Eq:GabEvolnBis},~\ref{Eq:MomEvolnBis},~\ref{Eq:QbEvolnBis})
which yield a linear, first-order symmetric hyperbolic system with
propagation speeds relative to normal observers equal to zero and one half the speed of light.

Depending on the election for the shift vector in
Sec.~\ref{SubSec:ShiftCond}, there might be additional constraints or
small adjustments to be made in the constraint propagation system. For
the spatial harmonic gauge, an additional constraint is $\tilde{V}^k =
0$, whose propagation equation is of the form $\tilde{D}_0\tilde{V}^k
= l.o.~$  For the minimal conformal strain and distortion
conditions~(\ref{Eq:MinimalConformalStrain})
and~(\ref{Eq:EllipticBeta}), the triad vectors ${\bf B}_a$ are
propagated along the time evolution vector field $\partial_t$ instead
of $\tilde{D}_0$, see Eq.~(\ref{Eq:partialtB}), which corresponds to
setting the combination $\alpha F_{ab} + \varepsilon_{bcd}
G_{ac}\beta_d$ to zero instead of $F_{ab} = 0$.

\section{The relation between the 3D Nester gauge condition and the 3D Dirac equation}
\label{App-NesterDirac}

In this appendix, we first review the relation found in
Ref.~\cite{aDfM89} between orthonormal three-frames satisfying the 3D
Nester gauge and nowhere vanishing spinor fields satisfying the 3D
Dirac equation. Then, we reconsider the equations in
Sec.~\ref{SubSec:System} which are responsible for preserving
the 3D Nester gauge and our choice of conformal factor throughout
evolution, and show that they can be rewritten as an inhomogeneous 3D
Dirac equation. Based on these observations, we show how to recast
these equations into second-order elliptic form.

The Dirac equation over a three-dimensional Riemannian manifold
$(\Sigma,h)$ with triad field ${\bf B}_a$ and corresponding connection
coefficients $\Gamma_{abc}$ is
\begin{equation}
  i\sigma_a\left( D_a + \frac{1}{2}\Gamma_{cda}\Sigma_{cd} \right)\xi = m\xi,
\label{Eq:3DDirac}
\end{equation}
where $\xi: \Sigma\to\Complex^2$ is a $SU(2)$-spinor field on $\Sigma$,
\begin{displaymath}
\sigma_1\equiv \left( \begin{array}{cc} 0 & 1 \\ 1 & 0 \end{array} \right),
\qquad
\sigma_2 \equiv \left( \begin{array}{cc} 0 & -i \\ i & 0 \end{array} \right),
\qquad
\sigma_3 \equiv \left( \begin{array}{cc} 1 & 0 \\ 0 & -1 \end{array} \right),
\end{displaymath}
are the Pauli spin matrices, $\Sigma_{cd} \equiv
\frac{1}{4}[\sigma_c,\sigma_d]$ and $m$ is the mass of the
field. Eq.~(\ref{Eq:3DDirac}) is invariant with respect to local
rotations
\begin{equation}
\xi = U\bar{\xi},\qquad {\bf B}_a = R_{ab}\bar{\bf B}_b,
\end{equation}
with $U: \Sigma \to SU(2)$ and $R: \Sigma \to SO(3)$ the corresponding
rotation defined by $U\sigma_a U^* = \sigma_b R_{ba}$. In terms of the
notation introduced in Sec.~\ref{Sec:3+1}, Eq.~(\ref{Eq:3DDirac}) can
be rewritten in the following way,
\begin{equation}
i\sigma_a(D_a - n_a)\xi = \left( m + \frac{1}{2} N \right)\xi.
\label{Eq:3DDiracBis}
\end{equation}

Now, suppose the frame ${\bf B}_a$ satisfies the integrated 3D Nester
gauge conditions~(\ref{Eq:Integrated3DNesterCondition}), such that
$n_a = D_a\Phi$ and $N = {\rm const}$. Setting $m \equiv -N/2$, 
Eq.~(\ref{Eq:3DDiracBis}) reduces to $i\sigma_a D_a(e^{-\Phi}\xi)
= 0$ with the solution $\xi = e^\Phi\xi_0$, $\xi_0\in\Complex^2$ being
a constant spinor.  Conversely, suppose $\xi = (\xi^0,\xi^1)^T$ is a
solution of Eq.~(\ref{Eq:3DDirac}) which is nowhere vanishing. Then,
the rotation
\begin{equation}
U \equiv \frac{1}{p}\left( \begin{array}{rr} 
\xi^0 & -(\xi^1)^* \\
\xi^1 & (\xi^0)^*
\end{array} \right),\qquad
p \equiv |\xi|,
\label{Eq:URotation}
\end{equation}
leads to the constant direction spinor solution $\bar{\xi} = p(1,0)^T$
of Eq.~(\ref{Eq:3DDiracBis}). It follows that the new frame $\bar{\bf
  B}_a$ satisfies $\bar{n}_a = \bar{D}_a(\log p)$ and $\bar{N} = -2m$,
and hence the integrated 3D Nester gauge.

Therefore, the existence and uniqueness of a three-frame satisfying
the integrated 3D Nester gauge condition~(\ref{Eq:3DDiracBis}) with $N
= N_0 = {\rm const}$ is equivalent to the existence and uniqueness of
a nowhere vanishing spinor field satisfying the three-dimensional
Dirac equation~(\ref{Eq:3DDirac}) with mass $m = -N_0/2$. The latter
being a linear, elliptic equation, it is amenable to an analysis
using standard tools from elliptic theory. Existence and uniqueness
theorems for similar equations have been obtained in
Refs.~\cite{oR82,oRkT84} in the context of the positivity of the ADM
and Bondi masses.

While the direction of the spinor $\xi$ is related to rotations of the
triad, its norm is related to conformal transformations
thereof. Indeed, the Dirac equation~(\ref{Eq:3DDirac}) is invariant
with respect to the conformal transformation $\xi =
\Omega\tilde{\xi}$, ${\bf B}_a = \Omega\tilde{\bf B}_a$, where $\Omega
> 0$ is a strictly positive function on $\Sigma$. This feature may be
used in order to fix the conformal factor (up to a positive
multiplicative constant) such that the rescaled quantities satisfy
$\tilde{n}_a = 0$. This corresponds to the choice for the conformal
factor in Sec.~\ref{Sec:Conformal}.

Finally, consider Eqs.~(\ref{Eq:GradD0Phi},~\ref{Eq:DivOmega})
for the quantities $v_0\equiv 2\alpha D_0\Phi$ and $v_a\equiv
\alpha\omega_a$ which are responsible for preserving the 3D Nester
gauge and the choice of the conformal factor throughout evolution. It
turns out that these equations are equivalent to the inhomogeneous,
3D Dirac equation
\begin{equation}
i\sigma_a D_a\xi = F,\qquad
F \equiv -\alpha\left( \begin{array}{ll} 
  \gamma_0 - i\gamma_3 \\ \gamma_2 - i\gamma_1 \end{array} \right),
\label{Eq:3DDiracInhomogeneous}
\end{equation}
for the spinor
\begin{equation}
\xi = \left( \begin{array}{ll} v_0 - i v_3 \\ v_2 - i v_1 \end{array} \right),
\label{Eq:3DDiracInhomogeneousSpinor}
\end{equation}
where $\gamma_0 \equiv \hat{N}_{ab}\hat{K}_{ab} + \frac{1}{3} K N +
D_0 N$ and $\gamma_a \equiv -\varepsilon_{abc}\hat{K}_{bd}\hat{N}_{dc}
+ \hat{K}_{ab}(3n_b + a_b) - \frac{2}{3} K a_a - 8\pi G j_a$. This can
be partially understood by considering the rotations defined in
Eq.~(\ref{Eq:URotation}) which map the orthonormal three-frame to one
that satisfies the 3D Nester gauge. Parametrizing $U$ by its rotation
angle $\Theta$ and axis $e_a$, $e_a e_a = 1$, we have
\begin{displaymath}
U = \exp\left( \frac{\Theta}{2i} e_a\sigma_a \right)
 = \cos\left( \frac{\Theta}{2} \right) - i\sin\left( \frac{\Theta}{2} \right) e_a\sigma_a.
\end{displaymath}
Hence, the corresponding spinor field is
\begin{displaymath}
\xi = p\left( \begin{array}{c}
 \cos\left( \frac{\Theta}{2} \right) - i\sin\left( \frac{\Theta}{2} \right) e_3 \\
 (e_2 - i e_1)\sin\left( \frac{\Theta}{2} \right) \end{array} \right).
\end{displaymath}
For an infinitesimal rotation and change $p = p_0 + \delta p$ in the
conformal factor, like the one required to preserve the 3D Nester
gauge and our condition on the conformal factor in an infinitesimal
time step, this yields precisely a spinor of the
form~(\ref{Eq:3DDiracInhomogeneousSpinor}) with $v_0 \equiv \delta p$
and $v_a \equiv p_0\delta\Theta e_a/2$.

Eq.~(\ref{Eq:3DDiracInhomogeneous}) constitutes an elliptic system of
equations for the fields $v_0$ and $v_b$ which can be solved subject to
appropriate boundary conditions. A second-order elliptic system can be
obtained from this by the ansatz $\xi = i\sigma_b D_b\eta$, with
$\eta$ a potential spinor field. Introduced into
Eq.~(\ref{Eq:3DDiracInhomogeneous}) this yields
\begin{equation}
-D_a D_a\eta + i N_{ab} \sigma_b D_a\eta - i N\sigma_a D_a\eta = F,
\end{equation}
where we have used the commutation relation $\varepsilon_{cab} D_a D_b
= N_{bc} D_b - N D_c$. In terms of the potentials $\psi$ and $\zeta_a$
introduced in Sec.~\ref{SubSec:Ktilde}, the potential spinor is given
by
\begin{equation}
  \eta = -\left( \begin{array}{ll} \psi - i\zeta_3 \\ \zeta_2 - i\zeta_1 \end{array} \right).
\end{equation}

\bibliography{refs}

\end{document}